**FACULTY OF MANAGEMENT**
**Aberdeen Business School**

# VALUE RELEVANCE OF THE COMPONENTS OF OIL AND GAS RESERVE QUANTITY CHANGE DISCLOSURES OF UPSTREAM OIL AND GAS COMPANIES IN THE LSE.

***ANIGHORO TEGA (1811048)***

**October 9th, 2019**

**Dissertation Supervisor: Dr. Ayodele Asekomeh**

**A dissertation submitted in partial fulfilment of the requirement for the Degree of Master of Science in Oil & Gas Accounting at Robert Gordon University, Aberdeen**

**Total word count**: 21654 (excluding abstract, acknowledgements, table of content diagrams, tables, references and appendices)

# ABSTRACT

*The high level of risk and uncertainty in harnessing oil and gas reserves poses an accounting dilemma in the reporting of reserves quantity information; information which is critical and relied on by investors for decision making.*

*Different studies have indicated that reserves disclosure information is fundamental to understanding the value of the firm. This study attempts to contribute to the growing value relevance literature on reserves disclosures by examining the value relevance of the components of oil and gas reserve quantity change disclosures of upstream oil and gas companies in the London Stock Exchange.*

*Particularly, it investigates the relationship between average historical share returns and changes in reserves from explorations, acquisitions, production, revisions and sale. It also examines the value relevance of the quality of these disclosures.*

*Using archival data from LSE, databases and annual reports, and applying a multifactor framework, the empirical results suggested that changes in reserves as well as the components of these changes where associated with share returns though insignificantly due to the significant impact of oil price and longitudinal effect posed by applying the measurement approach with utilizes historical returns. However, the quality of reserves disclosures has a positively significant relationship with share returns.*

*The volatility and decline in oil price is also reflected in both low average share returns at -0.4% and low average growth in reserves at 8.94% for the last 8years in the sector.*

*The sector is encouraged to diversify and also boost its reserves by complementing exploratory efforts with acquisitions of reserves especially given the volatility of oil prices. The IFRS is also encouraged to promote uniformity in reporting practices by formulating comprehensive accounting standards for reporting on reserves disclosures by upstream oil and gas firms in the UK.*

# ACKNOWLEDGEMENT

Firstly, my profound gratitude goes to God Almighty for his mercies, favour and wisdom which saw me through this phase of my educational pursuit.

My sincere appreciation also goes to my supervisor, Dr. Ayodele Asekomeh for his guidance, wise counsel, constructive suggestions and criticisms during my research which enabled me to adequately complete the work. My appreciation as well goes to all my lecturers at Robert Gordon University.

Special thanks go to my sponsors, Petroleum Technology and Development Fund (PTDF) Nigeria, for giving me an opportunity to study at this prestigious University and providing all the necessary funding requirements covering my tuition and upkeep.

Lastly, I would like to thank my family and friends. My appreciation goes to my beloved mother, Mrs. Jane Anighoro for her unwavering support all through these years and also sisters Garen, Elohor and Vwaire. To all my friends at Robert Gordon University, thank you for your companionship and support.

## Table of Contents

## LIST OF FIGURES

## LIST OF TABLES

## ACRONYMS AND ABBREVIATIONS

**UKCS** - UK Continental Shelf

**E&P** - Exploration and Production

**DDA** - Depreciation, Depletion and Amortization

**FTSE** - Financial Times Stock Exchange

**LSE** - London Stock Exchange

**AIM** - Alternative Investment Market

**FC** - Full Cost

**SE** - Successful Effort

**PSC** - Production Sharing Agreement

# 1 CHAPTER ONE: INTRODUCTION

## 1.1 INTRODUCTION

This chapter presents a general overview of the uniqueness of the oil and gas industry as it relates to the relevance of oil and reserves quantity in the determination of the value of the firm. The Chapter also discusses the aims and objectives of the study and gives a justification for its relevance.

## 1.2 BACKGROUND TO THE STUDY

The oil and gas industry possess various unique features, distinguishing it from every other industry (Power, Cleary and Donnelly, 2017). It has a complex mix of industry segments (upstream, midstream, downstream and services) as well as phased project activities (exploration, construction, production and decommissioning). A huge scale of investments is required in finding and developing oil and gas reserves and the cost of finding these reserves is unrelated to the future economic benefits generated. The reserves have a finite field life as well as high levels of risk and uncertainty in exploring for and commercially exploiting them.

These features have posed a number of accounting dilemmas in the reporting of E&P costs and reserves quantity disclosures. The accounting standard has however provided guidance in the treatment of E&P cost but there is a very limited framework regulating reserve quantity disclosures especially within the UK as this study will reveal.

Historical cost information and reserve disclosure information are both fundamental to understanding the firm's value (Berry and Wright, 2001; Boone, 2002; Asekomeh, Russell, Tarbert, & Lawal, 2010;). However, in this study, much preference is given to reserve quantity disclosure information.

This is because oil and gas reserves constitute the underlying asset and lifeblood of oil and gas exploration and production companies. But in line with the recognition criteria for assets according to IASB conceptual framework, they are not captured as assets in the financial statements due to their imprecise nature and difficulty in estimation and measurement. (IASB, 2018). The risk of reserve quantity estimation occurs from the point of exploration to the point of production. (McChlery et al, 2015).

Reserve quantity estimates are drawn from their recoverability which is a function of technology, geology and economics. Defining these parameters are problematic hence reserves cannot be estimated with complete objectivity. For instance, a field may have several reserve quantity estimates, increasing its uncertainty (Misund and Osmundsen, 2017).

Oil and gas reserves are therefore subjected to voluntary supplementary disclosures in the UK, as the accounting standards and regulations specify no mandatory disclosure requirements. The UK Oil Industry Accounting Committee (OIAC) only recommends best practices to guide the quality of reserve disclosures.

This "free rein" could pose the danger of reserve quantity information being taken for granted. Barry (1993) portrayed the exclusion of reserves from the statement of financial position as "odd" as the reserves' volume constituted the E&P firm's black hole, greatly impacting on all its facets but however attracting little attention.

Without doubts, oil and gas reserves are the most significant assets of E&P firms as they form a crucial element in an investor's decision on whether or not to proceed with investing in an oil and gas firm and the fair price at which to procure such company's shares (Rai, 2006).

Certain factors can lead to adjustments in reserves quantity resulting in a growth or a decline (Misund, 2018). A growth in reserves is attributed to acquisitions, successful exploratory and improved recovery activities and upward revisions while a decline is associated with sales of reserves, production and downward revisions. Information content of the changes in reserves is therefore a critical signal to the market.

As we shall see in the literature, past studies have shown the great importance of reserves to the value of the firm, in this case, the share price. Particularly, It has also revealed and demonstrated how Changes in reserves arising from acquisition announcement, exploratory activities and extension of reserves in place, revisions and production has impacted greatly on the share price (Clinch and Magliolo, 1992; Alciatore, 1993; Spear, 1994; Ohlson, 1995; Berry and Wright, 2001; Boyer and Filion, 2007; Bird, Grosse and Yeung, 2013; Sabet and Heaney, 2016; Edwin and Thompson, 2016; Misund, 2016;2018; Misund and Osmundsen, 2017).

This work focuses on the LSE market; one of the oldest and largest exchanges in the world with an age of 215 years, a market capitalization of $6187 billion and 3041 quoted companies comprising of both the Main Market for largest companies and the Alternative Investment Market (AIM) for smaller companies. (Jianu and

Jianu, 2018). Thus, focus is placed on both the FTSE All share and the FTSE AIM All Share index for Oil and Gas E&P Firms.

The ex-ante expectation for this work is that the oil and reserve quantity information is value relevant to investors and the dissected information content accounting for the changes in these reserves also hold value relevant information as observed with past researches.

## 1.3 STATEMENT OF THE RESEARCH PROBLEM

The value of upstream E&P firms is significantly driven by its oil and gas reserves. These reserves are not statutorily recorded in the reported financial statements but yet they serve as a major driver of economic activity (Taylor et al., 2012).

Amongst peculiar risk factors distinguishing the oil and gas industry as discussed in section 1.1, the major reason for this has been attributed to the imprecise estimates of oil and gas reserve quantities given the risk and uncertainty of exploration and development, making its recognition as an asset challenging.

According to Misund (2018), the uncertainty associated with reserves quantities and value is a source of confusion for investors whose investment decision is largely dependent on such information disclosures.

Therefore, the central theme for this research is contingent on the premise that should reserve disclosure requirement be value relevant, it therefore follows that the information on reserve quantity disclosures contained in the annual report should reflect on the share price. This is the cardinal factor necessitating this study which focuses on the value relevance of the components of oil and gas reserves quantity disclosures.

Past and contemporary studies on value relevance focused more on the evaluation of exploration and production historical cost expenditure disclosures and the FC and SE accounting policies than reserves quantity disclosures. (Power, Cleary and Donnelly, 2017; Misund, Osmunden and Sikveland 2015; Cortese and Irvine, 2010; Asekomeh et al, 2010, Bryant, 2003).

The value relevance literature suggests that an empirical relationship exists between share price and changes in oil and gas reserves (Bell 1981; Harris and Ohlson 1987; Magliolo, 1986; Spear, 1994; 1996; Ohlson, 1995; Berry, Hasan, & O'Bryan, 1998; Berry and wright, 2001; Boyer & Filion, 2007; Misund & Osmundsen, 2017).

However, only few studies have particularly addressed the importance of the information content accounting for the change in reserves quantity (Spear, 1994; Alciatore, 1993; Clinch and Magliolo, 1992; Cormier and Magnan 2002; Coleman, 2005; Olsen, Lee and Blasingame, 2011; Scholtens and Wagenaar, 2011; Costabile, Soltys and Spear, 2012; Bird, Grosse and Yeung, 2013; Sabet and Heaney, 2016; Edwin and Thompson, 2016; Misund, 2016; 2018; Misund and Osmundsen 2017; Gray, Hellman and Ivanova, 2019).

Within these narrow band of studies, most of the researches were focused on the United States, Canada and Australia, as these countries had more mandatory accounting standards and regulatory policies enforcing better reserves disclosure requirements, compared to what was obtainable in the United Kingdom (Gray, Hellman and Ivanova, 2019). There is therefore an apparent research gap in extensively exploring the area for which this research is undertaken.

The value relevance of reserves disclosure has been argued to be linked to the quality of accounting standards and regulations. The research explores how these guidelines have impacted on reserve disclosures within the UK and how the quality of these disclosures could affect share return.

Of the researches conducted with the UK oil and gas firms, focus was directed at the LSE AIM and Main market, having samples of firms at different levels of integration. However, this research adopts a specialized and focused approach, considering only the Exploration and Production firms who are the key players in the oil and gas industry.

Previous researchers have also focused on a shorter time period with a larger mix of oil and gas firms however this study adopts time series data of 8years (2011-2018) for a better insight.

In light of these analyses, this study is expected to contribute to the body of knowledge. Particularly, the outcome of this study will be useful for decision making to financial users, E&P companies, accounting standard setters and the petroleum industry in general by understanding the effect of each critical components influencing changes in oil and gas reserves and how this information can be utilized positively.

## 1.4 AIM AND OBJECTIVES

### 1.4.1 AIMS

This research is aimed at examining the relationship between the components that accounts for adjustment in oil and gas reserve quantity and share price in other to examine their individual value relevance for oil and gas upstream companies in the United Kingdom.

### 1.4.2 OBJECTIVES

- To examine the relationship between oil and gas reserves and share price.
- To investigate the value relevance of each individual component of oil and gas reserve quantity adjustments (acquisition announcements, revisions, exploration, production and sales) by analysing the information content of databases and the financial reports for selected oil and gas firms in the UK as well as share price data.
- To evaluate the impact of accounting policies and regulations for the UK jurisdiction on reserve disclosure requirements and assess the value relevance of reserve disclosure quality.
- To highlight the Implications of adjustments to components of oil and gas reserves for investors, E&P companies, accounting standard setters and the petroleum industry.

## 1.5 SYNOPSIS

The outline of the research paper is sectioned into several segments in the form of chapters. Chapters 1 and 2 provide the basis on which research is established in line with the topic under study. Chapter 3 prescribes the methodology considered applicable to achieve the aim and objectives of the research, Chapter 4 deals with data collection, analysis and interpretation and finally, chapter 5 provides critical discussions, conclusions and recommendations.

# 2 CHAPTER TWO: LITERATURE REVIEW

## 2.1 INTRODUCTION

According to Wentz (2014, p.81), "a literature review is a synthesis of prior research". This entails bringing together different ideas, articles, and sources, and giving it a new perspective (Wentz, 2014).

This chapter seeks to explore the extant literature on the relationship between the individual components affecting changes in oil and gas reserves and share price in other to examine their individual and aggregate value relevance. It assesses the determinants of firm value (share price) for oil and gas firm with reference to relevant theoretical models. It provides a critical discussion on the concept of oil and gas reserves and why they serve as a better measure of value relevance in the oil and gas industry compared to just earnings. It further explores the reserves ratio and how shareholders perceive the information content of these ratios as KPIs. Finally, it also discusses issues bothering on accounting policies and disclosures with regards to the UK spectrum and how these may influence the value relevance of reserves disclosures.

## 2.2 VALUE RELEVANCE ACCOUNTING

Value relevance of accounting information is described as a fundamental characteristic of accounting quality (Francis et al., 2004). It indicates the relationship between accounting data and investors' unanimous perception about the economic value of the company. It is the ability of accounting information to capture share value information, determined by a statistical test between both variables (Hellstrom, 2006).

A significant association between the share price and the accounting variable affirms that the accounting information contains the essential attribute of relevance as stated in the IASB's Conceptual Framework and vice versa (Power, Cleary and Donnelly, 2017).

There are two measurements of value relevance; the signalling perspective and the measurement perspective. The signalling perspective studies the market reaction to the announcement of accounting information while the measurement perspective measures the explicit relationship between value of the company and the accounting measures. The former is utilized in most value relevance studies (Hellstrom, 2006) however this present research adopts the measurement perspective.

The main accounting information considered in this study is oil and gas reserves while the measure of firm value considered is the share price.

### 2.2.1 SHARE PRICE AS A MEASURE OF FIRM VALUE

The value of a company could be defined as the present value of current and future cash flows (Ewing and Thompson 2016). The share price is an ideal indicator of a company's value as it represents the price at which present and potential investors are willing to pay for a unit of the company's shares.

Prior research (Power, Cleary and Donnelly, 2017; Ahmadi, Matteo, Mehdi, 2016; Shaeri, Adaoglu, Katircioglu, 2016, Cunado and De Gracia, 2014; Barth, Konchitchki and Landsman, 2013; Bushman et al, 2004; Francis et al, 2004), have established that the stock price captures the underlying economic value of the firm as perceived by investors.

Company-specific (idiosyncratic) and general market (systematic) information are the two principal factors affecting a company's share price (Jianu and Jianu, 2018). A fall or rise in share price provides the market with signals about changes in these information.

Oil and gas shares are regarded as one of the most stable and dependable shares on the London Stock Exchange (LSE) (Jianu and Jianu, 2018). They are sensitive to reflecting any key changes in oil and gas companies operations.

### 2.2.2 LIMITATIONS TO VALUE RELEVANCE STUDIES

Hellstrom (2006) has identified several limitations to value relevance studies. These include inefficiency of the market, quality of accounting regulations, business cycles and environment.

Most studies on value relevance are built on a framework modelling firm value as a function of equity and earnings, with the assumption of efficient markets. This may pose a limitation to the inferences derived from the value relevance tests because it is disputable if the market is truly efficient. However, Aboody et al. (2002) demonstrate that these challenges could be overcome by making future price changes a part of the research design to adjust for slowed market reactions.

Value relevance of accounting information is influenced by the quality of existing accounting standards, regulations and control mechanisms enforcing adequate disclosures to users of such information. Where these are weak or inexistent, it could affect the quality of the results.

The results of value relevance studies may also be affected by business cycles. Investors rather than base their valuation on a critical analysis of the accounting and market information, value firms higher in periods of economic boom than during recessions, regardless of their actual performance.
Finally, value relevance studies are impacted by the business environment. A centrally planned economy characterized by secretive practices, closure and restricted information flow due to the state's regulation of firms will have a reduced value relevance compared to a market economy which is more open with better information disclosure.

### 2.2.3 DETERMINANTS OF SHARE PRICE FOR E&P FIRMS

Company-specific (idiosyncratic) and general market (systematic) information are the two principal factors affecting a company's share price (Jianu and Jianu, 2018). Fama and French (1993) three-factor asset pricing model identified the share price of a firm as a measure of size (small versus big firms), value (high-value versus low-value firms) and market risk captured by beta. These factors were expanded by Carhart (1997) to include momentum. However according to the multifactor arbitrage pricing model by Stephen Ross (1976) more factors can be added to adjust these models for more relevance to reflect the uniqueness of the share price being assessed. For instance, in the oil and gas industry, these common risk factors include oil and gas reserves, oil and gas prices, exchange rates, firms' equity risk, interest rates, taxation policies due to location effect and economic boom/decline.

Ohlson (1995) modelled the market value of equity of a firm as function of profitability measure (earnings), book value of assets, dividend as well as any other additional items capturing future cash flows for the firm. Oil and gas reserves serve as this additional measure of share price, specific to the oil and gas industry as it represents future cash flows and profitability (Misund and Osmundsen, 2017).

Cormier and Magnan (2002) specified that earnings, cash flow and components of reserves determined the share price of E&P firms. Similarly, Boyer and Filion (2007) specified the determinants of share price in the oil and gas industry as cash flows, reserve changes and production. Misund (2018) draws from Ohlson (1995) and Boyer and Filion (2007) to state the determinants of E&P firms share price as function of change in reserves, profitability and the Fama, French and Cahart risk factors.

Oil and gas price is a major determinant of the share price of oil and gas firms. Contemporary studies have found a positive relationship between crude oil volatility and share price (Taamouti et al, 2017; Shaeri, Adaoglu and Katircioglu 2016; Cunado & de Gracia, 2014; Chang & Yu, 2013; Elyasiani, Mansur and Odusami, 2011; Ramos and Veiga, 2011; Cong et al., 2008). A rise in oil price results in a rise in the share price and vice versa. For instance between late 2014 to early 2016, there was a fall in the price of crude oil from about $100/barrel to less than $50/barrel due to increased production of US shale (Misund, 2018).

Kretzschmar and Kirchner (2009) explored the geographical location effect of oil and gas reserves on share price. They found share price to be sensitive to the concession or product sharing contractual arrangement of the countries where their reserves where located.

It can therefore be concluded that the determinant of oil and gas firm share price include change in oil and gas reserves, profitability measures (earning or cash flows), size of the firm, book value of the firm as well as other common risk factors such as oil and gas prices, equity risk, exchange rates, interest rates, location effect and level of economic activity.

## 2.3 OIL AND GAS RESERVES

### 2.3.1 MEANING

Oil and gas reserves constitute the major asset pivotal to an E&P firm's operations. The major aim of holding petroleum reserves is to generate future cash flows upon extraction from oil and gas reservoirs and subsequent monetization (Misund, 2018).

The SPE Petroleum Resources Classification and Definitions system categories petroleum resources into Reserves, Contingent Resources and prospective resources (SPE, 2011).

It defined Reserves as "quantity of petroleum which is anticipated to be commercially recovered from known accumulations from a given date forward"; Contingent resources as "quantities of petroleum which are estimated, on a given date, to be potentially recoverable from known accumulations, but which are not currently considered to be commercially recoverable"; and Prospective resources as "quantities of petroleum which are estimated, on a given date, to be potentially recoverable from undiscovered accumulations".

From the above definitions, it can be deduced that reserves have the highest discoverability, commercial recoverability and technical viability. Also, Reserves constitutes a small portion of total resources. This is because, oil resource is extensive, however, technical, economic and political limitations poses the onerous challenge in converting the known available resource into reserves.

According to Adelman and Watkins (2008), reserves is described as a "depletable" natural resource, finite in supply and subject to irreversible downturn given continuous extraction activities.

### 2.3.2 CLASSIFICATION

The SPE (2011) have classified commercial reserves according to maturity and probability of recoverability from underground reservoirs (See Figure 2.1). This category includes 1P (Proved), 2P (Proved plus Probable), and 3P (Proved plus Probable plus Possible) having probabilities of 90%, 50% and 10% of final recovered reserves respectively (Misund, 2018).

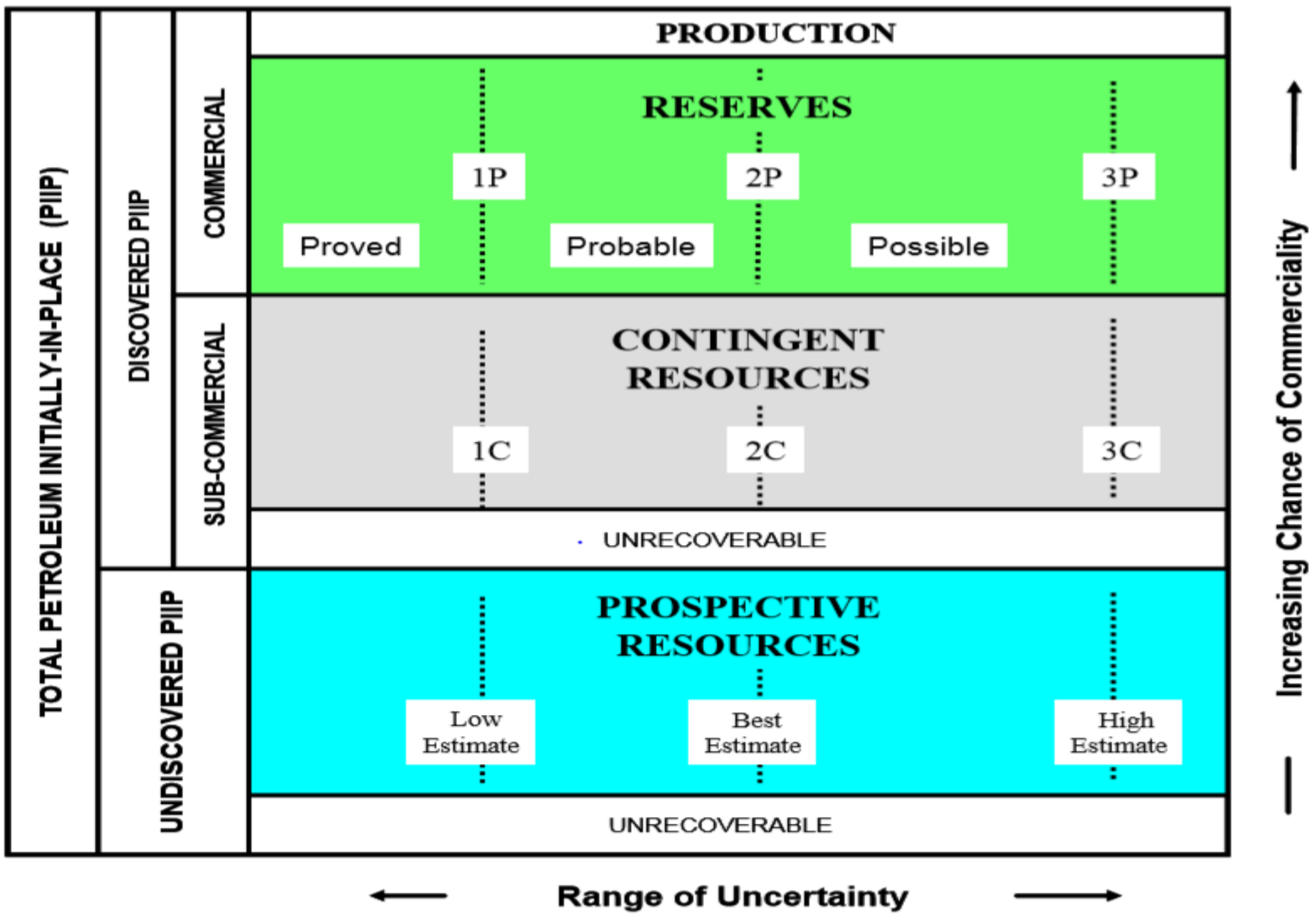

Figure 2. 1 SPE oil and gas reserves classification framework

*Source: Society of Petroleum Engineers (2011).*

These reserve estimates could be derived using probabilistic methods or deterministic methods (SPE, 2011). The probabilistic method is achieved using software such as the Monte Carlo analysis to generate a random sampling of distributions of the input factors and the relationship between them to derive a range of possible outcomes and associated statistical probabilities i.e. P90, P50 and P10).

The deterministic method employs professional judgment to estimate reserve quantities with a degree of reasonable certainty for each reserves division (Proved, Probable, and Possible) as single volumes. A single value is adopted for each input parameter based on known geological, engineering, and economic data to determine the reserve volume.

The best estimates is usually the 2P reserve because 1P and 3P reserves tend to converge at 2P in the long run (Owen, Inderwildi and King, 2010). The probabilistic method is common within Western Europe while the USA adopts the deterministic approach (Wright and Gallun, 2016). Application of different reserve estimation techniques makes comparability difficult, thus, entities are advised to disclose the method in use.

For clarity, SPE gives a definition of each classification of commercial reserves.

Proved reserves according to SPE (2011), could be described as amounts of oil and gas reserves deemed economically recoverable with a reasonable level of certainty, and estimated with geotechnical and engineering analysis at a given time, given current the operating techniques, level of economic activity and government regulation.

Proved developed and proved undeveloped reserves are the two sub divisions of proved reserves. Proved developed reserves constitute the most advanced reserve having the highest probability of being recovered. They may be extracted from existing wells either through the utilization of existing equipment and operating techniques or purchasing new ones where the cost of procurement is relatively negligible in comparison with the cost of drilling for a new well. (Misund and Osmundsen, 2017). On the other hand, proved undeveloped reserves are reserves which can be obtained either by drilling new wells on undrilled acreage with recoverable reserves or from existing wells requiring a relatively major expenditure for recompletion.

Unproved reserves are associated with the lowest probability of recoverability and can be subdivided into probable and possible reserves. Probable reserves are unproven reserves which are more likely to be recoverable given geological and engineering data analysis. There is a 50% probability that probable reserves recovered is equal or above the addition of estimated proved plus probable reserves. Possible reserves are, on the other hand, less likely to be recoverable than probable reserves given geoscience and engineering data analysis (PWC, 2017).

Disclosure requirement for UK, USA, Canada and Australia include one or more of these categories of reserves as the study will reveal in section 2.4.1. Several empirical studies have employed proved reserves to investigate the association between share returns and reserves changes (Misund, 2018). Misund & Osmundsen (2017) argues that investors have limited reliance on less mature reserves as they are less value relevant. This research paper considers it pertinent to utilize proved reserves.

### 2.3.3 IMPORTANCE OF OIL AND GAS RESERVES

Reserves information is highly meaningful to various stakeholders. They have key influence on political and socio-economic decisions made by E&P firms, shareholders, regulators and the government (Olsen, Lee and Blasingame, 2011). Upstream oil and gas firms' market value is largely dependent on its reserve estimates. For certain governments, a large proportion of its national income revenue may be derived from petroleum reserves or its economy could be majorly dependent on cheap imported oil (Mitchell, 2004).

Internally, information on the quantum of oil and gas reserve is a major consideration for mergers, acquisition and divestiture decisions, and security for principal and interest when seeking funding and lending decisions.

Reserves serve as indispensable instruments, easing access into the capital markets. Reserve volume is commonly accepted as collateral and also contribute to credit rating of E&P firms seeking funding from the bank (McChlery et al., 2015).

E&P firm's financial health is dependent largely on its booked oil and gas reserves. Financial analysis on key reserve ratios like the reserve life ratio, reserve replacement ratio, finding and development ratio, as well as costs such as DDA, are based on reserve values. These ratios are considered by investors as indicators of the firm's KPIs.

Clearly, the criticality of reserve quantum information reporting cannot be overemphasized as its usefulness spans across different participants in the industry.

### 2.3.4 LIMITATIONS OF PROVED/BOOKED RESERVES

The literature has identified several shortcomings of the information value of proved reserves. Firstly, investment decision has been continually complicated by uncertainty in reserves quantities estimates. Accurate and clear-cut estimates are elusive given adjustments to geological assumptions combined with continual change of reserve quantum figures from acquisitions, explorations, production and revisions (Sorrel et al, 2010). This may also affect the volume of oil and gas reserves which is a function of the reserve quantity and the price of oil and gas (Osmundsen, 2010).

As observed in section 2.3.2, proved reserves is just one of the various reserves categories. As will be discussed in section 2.4.1, for the US, proved reserves disclosure is mandatory but probable reserves voluntary. In the UK, proved and probable reserve disclosure is recommended but voluntary. Proved reserves alone, possesses incomplete information about the company's future growth as it excludes less mature reserves thus skipping conceivably vital information for investors and financial analysts (Misund and Osmundsen, 2017; Misund, 2018). This is supported by Boone, Luther and Raman (1998) who finds that by reporting unproved reserves (probable and possible), information gap between the market participants is reduced.

Finally, entitlement to reserves is conditioned by contractual arrangements such as concessions and production sharing agreements (Kretzschmar and Kirchner, 2009). E&P firms will have ownership of more reserves under a concessionary contract than a PSC (Misund, 2018).

### 2.3.5 OIL AND GAS RESERVES AS A MAJOR DETERMINANT OF SHARE PRICE IN THE PETROLEUM INDUSTRY

The value of an E&P firm is a function of its oil and gas reserves available for production and the ability to finance its operations and capital expenditure (Ewing and Thompson, 2016). This is supported by McChlery et al (2015) who opines that energy firms' market value is significantly derived from its petroleum reserves; assets not captured on its financial statements.

According to the Ohlson (1995), share returns, a measure of firm value is determined by the profitability, assets and other value-relevant information which is an indicator of future earnings but is not captured by prevailing measures of profitability. Oil and gas reserves serve as this additional value as it represents future cash flows and profitability (Misund and Osmundsen, 2017).

Investors and financial analysts, therefore, keenly monitor information released from E&P firms concerning reserve changes and the components contributing to or reducing these reserves (Misund, 2018). This is because such reserve provides them with statistics on the prospect of cash inflows, enabling them to determine a fair price for the shares of the firm (McChlery et al, 2015).

### 2.3.6 OIL AND GAS RESERVES VERSUS EARNINGS AS A MEASURE OF VALUE RELEVANCE IN THE PETROLEUM INDUSTRY

Reserves quantity has been regarded as a more stable and relevant measure to estimate the future cash flow that will accrue to an E&P firm compared to earnings (Teall, 1992; Cormier and Magnan, 2002; Misund and Osmundsen, 2017). The earnings in the statement of profit or loss for E&P firms is affected by the flexibility of using either the full cost or the successful effort methods. By capitalizing the cost of unsuccessful exploratory efforts, and spreading same via amortization, Firms using the FC would record more profits compared to firms using SE who expense such costs immediately (Noël, Ajayi and Blum, 2010; Cortese, Irvine and Kaidonis, 2009). This may pose confusion for investors, triggering reduced confidence in accounting earnings as compared to reserves information (Misund and Osmundsen, 2017).

Rai (2006) corroborates that proved and probable reserves are jointly more significant than earnings of a company when explaining abnormal share returns of Canadian upstream companies.

However, for more relevance in determining the value of a firm. A combined approach including a measure of reserves as well as a measure of profitability is recommended for a more rounded result.

Therefore, rather than including earnings as a measure of profitability, the literature has suggested that cashflows from operations is a more appropriate measure. (Cormier & Magnan, 2002; Misund, Asche and Osmundsen, 2008; Misund & Osmundsen, 2015; Misund, Osmundsen and Sikveland, 2015, Misund, 2018).

## 2.4 OIL AND GAS RESERVES DISCLOSURE

Information on the quantities of oil and gas reserves is fundamental to aiding users' understanding and comparability of E&P firms companies' financial position and performance (PWC, 2017).

Stakeholders view data acquired from disclosure behaviour research as critical given the high level of reliance on such data to make informed decisions. This applies for oil and gas reserve quantum data (Slack et al 2010; McChlery et al, 2015). Optimum disclosure is therefore a key requirement for accurately determining the value relevance of oil and gas reserves.

The disclosure requirements in the oil and gas industry is subjected to the degree of flexibility permitted by the accounting policies and regulation and these vary from country to country.

### 2.4.1 ACCOUNTING POLICIES AND REGULATIONS

The level of disclosure of oil and gas reserves is greatly determined by the accounting policies and regulations regulating the jurisdiction for which such disclosure is required. The discretion to disclose reserve information dictated by these policies could be voluntary or mandatory.

From the extant literature, the accounting regulations for the oil and gas industry have majorly focused on the disclosure of the historical cost of exploration expenditure, captured directly on the financial statement. The two prevalent accounting policies in this regard are the Full Cost and Successful Efforts methods (Power, Cleary and Donnelly, 2017; Abdo, 2016; Cortese and Irvine 2010)

However, given the uncertainty of oil and gas reserves quantity estimates and the extent to which disclosed information in this area is crucial to decision making of shareholders which in turn determines the value of firm, an examination of accounting policies and regulations influencing these disclosures is therefore critical.

Contemporary studies demonstrate that supplementary reserves disclosures are equally relevant as historical cost disclosures on exploration activities (Asekomeh, et al 2010; Boone, 2002).

A comparison is made between the United States, Canada, Australia and the United Kingdom disclosure requirements. For the US listed E&P companies, they are mandated by the Financial Accounting Standards Board (FASB) and the Securities and Exchange Commission (SEC) to report a considerable amount of supplementary reserve quantity information, to augment the typical financial statements (FASB, 2010; SEC, 2009). For reserve type, such firms are mandated to disclose proved developed and undeveloped reserves and are also permitted to voluntarily disclose probable reserves across oil, gas and nonconventional resources for its geographical locations (Misund and Osmundsen, 2017). Furthermore, they need to show the standardized measure; an estimation of the net present value of its proved reserves at end of the year. Lastly, they are required to report changes in this standardized measure for the period (Misund, 2018).

Canada improved on the disclose requirement in the US. Canadian Securities Administrators (CSA) NI 51-101 standards of disclosure for oil and gas activities necessitates mandatory disclosures of both proved and probable reserves, allowing for voluntary disclosure of contingent reserves (CSA 2015).

Australia has the most extensive disclosure policy. Australia's Joint Ore Reserve Committee (JORC) requires a compulsory disclosure of all categories of reserves, specifically, 100% of proved reserves, 60% of the probable reserves, and 30% of the possible reserves which must be endorsed by certified personnel (Rai, 2006; Gray, Hellman and Ivanova, 2019).

The UK standpoint differs from that of other jurisdictions. The UK oil and gas sector operates a voluntary system of reserves disclosures as backed up by its national GAAP, the IASB and FRC (formally ASB) which regulates the operations for both the listed and non-listed E&P firms.

Recommendations on accounting best practices on oil and gas reserve disclosures have been issued by UK OIAC Statements of Recommended Practice (SORPs) as well as the Operating and Financial Review (ASB 2006). SORP oversees Accounting for Oil and Gas Exploration, Development, Production and Decommissioning Activities (SORP 2001; PWC, 2017)

The SORP (s246-s251) and OFR (p77) specifically recommended the following disclosures:

- Statement of the source of the estimates.
- The name and qualification of an independent expert who audited the reserve data.
- The basis for arriving at the net reserve quantities.
- The oil and gas reserve balances.
- The accepted practice employed in defining reserve quantity: proved and probable reserves (probabilistic approach) or proved developed and undeveloped (deterministic approach).
- Movement in the net quantities of reserves.
- Reserve balances by geographic region.
- Reserve ratios/KPIs.

Consequent to the enactment of the new UK standards, FRS 100–102 in 2015 set by the Accounting Council of the FRC, the OAIC is no longer charged with regulation of the SORP. Notwithstanding, SORP remains highly fundamental in setting the industry's best practices regarding reserves disclosures as neither the IFRS nor the UK accounting standards currently provides any direction overseeing this area (Gray, Hellman and Ivanova, 2019).

Since 2005, listed firms within the UK have been mandated to make their financial statements IFRS complaint. Currently, no standardized reserve disclosure requirements exist under any IASB's IFRS standards (PWC, 2017).

IAS 1 on *Presentation of Financial Statement*, however, encourages that sources of estimations and key assumptions should be disclosed. It also specifies that supplementary information should be provided by a firm to complement its financial statements especially when the standard IFRS requirements are insufficient to accord such a firm fair representation (PWC, 2017).

The IASB's IFRS 6 on *Exploration for and Evaluation of Mineral Resources*, however, serves as the key standard related to the oil and gas sector (IFRS 2004). It is regarded as an interim standard having limited scope and thus, gives companies discretion to report reserve in their preferred mode resulting in varied accounting practices (Cortese, Irvine and Kaidonis, 2010, Mcchlery et al, 2015).

IASB in 2010 launched a research project, releasing a working paper focusing on extractive industry. This paper was still futile in regularising reserve quantity reporting and disclosures (Gray, Hellman and Ivanova, 2019). In fact, in July 2016, the research was tagged by the Board as a 'Pipeline Project', making it inactive (IASB, 2016).

However, following the IASB 2015 agenda consultation feedback, the board decided to start a new research project on the extractive industry with a view to gathering evidence to help decide whether to develop accounting requirements to either amend or replace IFRS 6. Deliberations are still ongoing till date (Gray, Hellman and Ivanova, 2019).

The reserve disclosure practices of 20 UK oil and gas firms with focus on the level of compliance with the OIAC recommendation was studied by Odo et al. (2016) in 2009. Their findings revealed that 65% of the firms sampled met or exceeded the recommendations, firms who had limited disclosure constituted 25%, and 10% did not disclose any reserves information.

On the contrary, McChlery et al. (2015) argue that the UK's voluntary disclosure approach is ineffective. Of the 86 upstream companies in the LSE observed in their study for the year 2007, reserve quantities was disclosed by a majority of them in some form but only a minority actually reported reserve quantity compliant with the UK SORP/OFR recommended practices.

According to McChlery et al (2015), the major driver for reserve disclosure was attributed to the risk and cost related to disclosing such information which was more for oil and gas firms at the exploratory stage than those at the producing stage given the imprecise nature associated with reserves estimates. Thus, firms were driven to disclosure reserves information only when the perceived benefits of doing so exceeded the cost.

The level of disclosure is also influenced by the firm size and quality of reserve audit. Larger firms can absorb more cost of disclosure relative to smaller firms thus providing more information on reserves while firms utilizing high-quality audit firms, signalled to the market a high quality of reserves information.

### 2.4.2 ISSUES WITH RESERVES DISCLOSURE

A full and accurate reserve disclosure is arguably more value relevant, however, this is beset with several challenges. As discussed in section 2.3.4, para 1, the pioneering issue associated with reserve disclosure stems from the uncertainty inherent in the estimation of reserve quantity making it less reliable.

According to Cormier and Magnan (2002), the information content of reserve being value relevant makes it vulnerable to the tendency of earnings management. This is exacerbated by reserve uncertainty, the level of market efficiency and the ability of regulators to monitor such disclosures. This is in line with the signalling theory and obfuscation hypothesis explored in section 2.7.

Varied accounting practices have opened a wide range of discretionary reporting choices of oil reserve estimates (Cortese, Irvine and Kaidonis, 2010). This has created difficulty for investors to compare two E&P's firm reserves on the same basis.

The UK E&P firms require only voluntary disclosure of proved and probable reserves. The result is a systematic distortion of the true resource value for oil companies (Misund and Osmundsen, 2017). As also discussed in section 2.3.4, para 2, this could lead to information asymmetry for the investor.

Reserves, although disclosed in the annual reports, are not part of the mandatory financial statements but rather are supplementary. Thus, they are not subjected to mandatory audits unless the E&P firm filing the report elects to use third-party engineering firm (McChlery et al, 2015; PWC, 2017) This raise the question of transparency and relevance of disclosures for investors.

Finally, the value relevance of accounting information might be low even when high quality accounting standards exists but are not complied with. Slack et al. (2010) affirm that disclosure quality affects stock returns. McChlery et al, (2015) reveal that only few UK E&P firms follow full disclosures in line with SORP. Therefore, regulation and control mechanisms must exist to enforce total adhesion by firms to reserve reporting requirements (Hellstrom, 2006).

## 2.5 OIL AND GAS RESERVES QUANTITY ADJUSTMENTS

Adjustments to oil and gas reserve quantity can be linked to several factors resulting in either a positive reserve growth or a fall in reserve quantity. A positive reserves growth is ascribed to exploration, extension and improved recovery, acquisitions and upward revisions. In reverse, negative reserves growth is due to production, sales and downward revisions (Coleman, 2005; Sabet and Heaney, 2016; Misund, 2018).

As will be discussed in section 2.8, the extant empirical literature has shown that disaggregated reserve quantity information possesses value relevance besides the cumulative figure alone. These researches established that unless separated into individual components, changes in proved reserves quantity have no additional information content. Alciatore, 1993; Spear, 1994; 1996; Sabet and Heaney, 2016; Misund, 2018). These components are evaluated individually below.

### 2.5.1 ACQUISITIONS AND SALES

E&P firms carry out reserve replacement by the acquisition of undeveloped acreage or directly purchasing reserves in the open market. (Coleman, 2005; Heaney 2016). These transactions are usually very significant to aggregate changes in reserves.

Global Upstream oil and gas transactions accounted for US$172.2b with about 970 deals with Europe accounting for about $25b in 2018 (EY, 2019). According to the Oil and Gas Uk business outlook (2019), the following LSE listed E&P companies operating in the UKCS recorded significant acquisition announcements in exploration prospect, predevelopment opportunities and producing fields for 2018:

- Cairn energy secured a farm-in to the Azinor Catalyst-operated Agar-Plantain exploration prospect.
- Shell acquired a stake in the Cambo area from Siccar Point.
- BP enhanced its stake in the Clair field, acquiring the interest held by ConocoPhillips.
- EnQuest acquired the remaining interest in the Magnus field from BP.

The procurement of reserves and acreage serve as a costly signal for future prospects of the E&P firm to the market and is therefore value relevant. (Sabet and Heaney, 2016; Misund, 2018). On the other hand, sales serve as a reduction to oil and gas reserves. The sale of an oil and gas asset could be necessitated when such asset has attained maturity and the cost of extraction and production exceeds the benefits. A sale could be in the form of whole or partial disposal of net interests in a joint venture. These could lead to a depletion of a company's reserves and is viewed negatively by the market.

### 2.5.2 EXPLORATION

Exploration is the technique of searching for accumulations of hydrocarbons trapped beneath the earth crust. Successful exploration is regarded as additions to reserves achieved organically (Misund, 2018). The discovery of reserves is a pivotal event used as a yardstick to value E&P firms (Berry and wright, 2001).

Magliolo (1986) advocated that there is market anticipation for the discovery of new reserves. It was also discovered by Shaw and Wier (1993) that the level of exploration had a significant positive relationship with the market value of an E&P firm and that deduced levels of exploration expenditures signalled inability of the firm to harness growth opportunities.

For example, the extensive Johan Sverdrup field discovery announced on 30th of September 2011, on the Norwegian continental shelf by Lundin, a Swedish oil company resulting to a 30% substantial stock price increase in the share price of the firm that same day (Misund and Osmundsen, 2017; Misund 2018).

However, studies have suggested that exploration efforts may negatively affect market valuation for certain period of time, for instance, in the period of plunging oil prices. (Misund, 2018; McConnell & Muscarella, 1985; Picchi, 1985).

In comparison to other components of reserve changes, Spear (1994) and Cormier and Magnan (2002) found explorations to be more value relevant compared to production. Their result is refuted by the study of Clinch and Magliolo (1992). Coleman, (2005) opined that E&P firms who encountered little or no successful with exploratory activities tend to change their replacement strategy to focus more on acquisition of reserves.

### 2.5.3 REVISIONS

Due to imprecise nature of petroleum reserves and the complex process of their estimation, reserve estimates are subject to continuous revisions over the field's life span (PWC, 2017).

Reserves can be revised up or written down. Investors respond positively to upward revisions (Costabile, Soltys and Spear, 2012; Gray, Hellman and Ivanova, 2019). A negative revision to reserves results in a write-down. A write-down is necessitated when it is observed that previous reserves estimates are higher than it ought to be. Reserve overstatements are worsened by the fact that reserves are not mandatorily subject to audits.

Write-downs can be due to uncontrollable factors such as fall in commodity prices or new technical data regarding the reservoir performance. It could also be induced by wilful neglect of reserve recognition procedures (Olsen, Lee and Blasingame, 2011).

McLane and Rose (2001) also identified several causes of reserve overstatement leading to write-downs: Defective technical practice leading to poor reserve estimation procedure fuelled by inadequate internal controls, pressure by managers to increase stock value in other to compete for investors, misguided incentives directed towards achieving an aggressive level of reserves, lack of professionalism and finally, human biases affecting judgement under uncertainty such as anchoring, overconfidence and availability.

Olsen, Lee and Blasingame (2011) examined the consequence associated with downward revisions arising from reserves overbooking for E&P companies. He identified that it could reduce the confidence of the market, creating nearly immediate destruction of share value. Shareholders could consequently file lawsuits against the firm involved. Internally, it could create tensions with the organisation such as civil penalties, dismissal of staffs involved and restructuring of management teams.

Similarly, Scholtens and Wagenaar (2011) analysed 100 revisions in several countries from 2000-2010 to analyse the impact of revisions of petroleum reserves on E&P firms' shares. They found revisions to have a significant impact on shareholder values with downward revisions having more effect on share returns in comparison with upward revisions.

Several instances of the negative effect of downward revisions on share price have been recorded. Olsen et al (2011) also observed that in recent years, certain reserves volumes have allegedly been overstated and written down by a lot of E&P firms. A negative revision exceeding 9.3 billion net BOE was recorded between 2003 and 2008.

Particularly, Shell declared a 28% (a fall of 3.9 billion BOE) write down of its proved oil and gas reserve in January 2004, resulting in a reduction of 12% in their share price in the following 4 weeks of the announcement (Misund, 2018). These overstatements were more predominant within the proved undeveloped reserves (Olsen, Lee and Blasingame, 2011).

Similarly, between 2001 -2006, Louisiana based Stone Energy Corporation had series of restatement of its reported assets in the Gulf of Mexico as a result of fraudulent activities. This caused its share price to drop by 30% (Olsen, Lee and Blasingame, 2011).

Likewise, Repsol YPF in Spain reduced its reserves in 2005 by 1.25million BOE owing to a flawed internal control system and unrealistic optimism for the technical and commercial performance of its fields. This resulted in an instantaneous decrease of about 7% in the price of the firm's shares on the day of the revision announcement.

It is therefore gathered that for companies with a history of large revisions, the reserve quantity disclosures of such companies are not considered value relevant by investors (Gray, Hellman and Ivanova, 2019)

### 2.5.4 PRODUCTION

Oil and gas reserves by their nature are depleting asset. The production of oil and gas causes an irreversible decline in the stock tank oil initially in place (STOIIP), that is, the volume of oil in a reservoir prior to production. However, more production signals generation of more cash flows to investors.

Clinch and Magliolo (1992) found production quantities were the most significant source of value relevant information of all other reserve components. Boyer and Filion (2007) however, revealed production to be a weak measure of value relevance as he finds a negative relationship between oil stock returns and changes in production of oil and gas. This he attributed to production been a contemporaneous measure and share returns, forward-looking.

Edwin and Thompson (2016) suggested market capitalization to be a function of the current production of hydrocarbons and the ability of the company to replace reserves at a rate that maintains future production. The authors employed the OLS multivariate framework to analyse the effect of the relationship between reserves and production on the firm value of E&P companies using a cross-section of 46 oil and gas companies ended 2012. They find a positive relationship between reserves and production, which were both significant indicators of firm value. This is supported by Misund (2018) who found share returns to be positively associated with increased oil production, and insignificantly impacted by rise in the production of natural gas as a result of the shale revolution experienced in 2008.

## 2.6 RESERVES RELATED KPI/RATIOS

Reserve ratios are regarded as key performance indicators (KPIs) that are value relevant to both the firm and its investors (Spear and Lee, 1999; McChlery et al, 2015).

While firms can utilize these ratios as a benchmark against similar firms in the industry to identify it comparative strength and weakness, investors can employ multiple-years analysis of these ratios to assess the current trends across several E&P firms and identify those firms having great prospects for future performance and consequently make well-informed decisions (Wright and Gallun. 2008, p. 704). The common reserve ratios are reserve replacement ratio, finding ratio, reserve life ratio, reserve cost ratio, reserve value ratio, average daily production per well and %net interest of wells. However the first three which are fundamental are explored below.

### 2.6.1 RESERVES REPLACEMENT RATIO (RRR):

The RRR captures the ability of the firm to replace its reserves used up through production. This ratio signifies the firm's ability to continue operating in the future. According to Kaiser (2013), in the absence of additional reserves, the net income of an E&P firm will fall given current prices, technology and operating cost.

Specifically, the RRR measures the total reserves additions via discoveries, extensions, improved recovery, upward revisions and acquisitions of reserves in place as a ratio to the total production within a financial year. A ratio greater than 1 signifies the firm is adequately replacing its reserves and vice versa (Wright and Gallun, 2008, p. 705).

### 2.6.2 FINDING RATIO (FD)

The FD ratio is derived from the RRR. It captures the exploratory success rate by measuring the level of reserves added organically through extension of existing reserves and new discoveries as these sources of additions are the most core to the operations of an E&P firm. According to Sabet and Heaney (2016), high FD ratio sends a positive market signal as it indicates that the E&P firm is financially stable and capable of investing in risky exploratory projects and that there is a high level of success associated with such projects.

### 2.6.3 RESERVES PRODUCTION/RESERVE LIFE RATIO (RPR)

This ratio approximates the number of years the current levels of oil and gas reserves owned by firm is expected to last given current rate of production. It is expressed as a ratio of the total proved reserves at the beginning of the year to the production (wright and Gallun, 2008, p. 707).

Reserve life is a crucial element in determining the value of an E&P firm (Ewing and Thompson, 2016). High RPR positively signals future production and profits for a company (Kaiser, 2013) as it implies that the firm can provide cash flows to meet its financial obligations for a longer period of time from existing reserves (Sabet and Heaney, 2016).

However, in a study by Ewing and Thompson (2016), they discovered that the market may penalize higher reserves to production ratios and prefer companies that front-load cash flows as a high ratio could also be an indication of production difficulty and viewed negatively by the market.

## 2.7 THEORETICAL MODELS FOR CONSIDERATION

### 2.7.1 INFORMATION ASYMMETRY THEORY

One assumption of the stock market is market efficiency where the prices of securities fully capture available information, thus all investors possessed the same degree of knowledge about the company and could make investment decision without altering the share price (Armstrong, Guay and Weber, 2010).

However, in reality, information asymmetry sometimes exists, making the market to be imperfect. This is evidenced by certain market participants being more or less informed than others about the firm’s prospect (Copeland, Weston and Shastri, 2005).

Managers of the firm tend to know more about its prospects compared to investors. They may not want to disclose certain reserve information if it exposes them to the risk of competitive disadvantage or if doing so impacts negatively on its profile, for example, information regarding challenges with successful exploratory and production activities. Investors, therefore, may find it difficult to distinguish between projects that are value-adding or detrimental and ultimately overprice the firm.

Inadequate disclosures of oil and gas reserves information hinder investors from making well-informed decisions and adequately price the firm. For instance, Boone, Luther and Raman (1998) find disclosing unproved reserves reduced information asymmetry between market participants.

### 2.7.2 AGENCY THEORY

Agency theory has been evaluated as a valid framework to examine reserve reporting in E&P firm's annual reports (Taylor et al., 2012). The company ownership is separate from its control, creating a principal-agent relationship. This theory postulates that the managers, who are the agents controlling the firm's operations are expected to act in the interest of the shareholders; the principals, commanding ownership of the firm's resources (Jensen and Meckling, 1976).

However, this may not be the case resulting in agency costs (Copeland et al, 2005). There are three kinds of these costs: cost of monitoring managers' actions, bonding cost of committing managers to their contractual obligations and residual loss due to inefficient decisions of managers. (Panda and Leepsa, 2017).

In the oil and gas industry, these costs are evidenced by significant monitoring and data verification cost using qualified geologists to reduce data uncertainty, in order to provide quality reserve information. This agency benefits as a result of this cost is a reduction in the organisation's cost of capital through a reduction in risk and contracting costs with agents. (Mirza and Zimmer, 2001).

This theory therefore suggests that firms are driven to disclosure reserves information only when perceived benefits of doing so exceeded the cost. According to McChlery et al (2015), this cost is more for firms at exploratory stage than those at the producing stage given the higher risk and uncertainty for the former.

### 2.7.3 LEGITIMACY THEORY

Legitimacy theory is dependent on the premise of a 'social contract' between the firm and the society within which it operates. A social contract refers to the expectations of the society about the mode of the firm's operations (Guthrie et al 2007).

Legitimacy theory can be applied to explain oil and gas firms' behaviour in implementing and developing voluntary disclosures of reserve information to fulfil its social contract in order to achieve its objectives (Schiopoiu and Popa, 2013). According to Dowling and Pfeffer (1975), legitimacy is a resource which the organisation dependents on for survival. Where society does not deem the firm's operations to be legitimately satisfactory, it will revoke the firm's 'contract' to continue its operations.

Therefore, in order for an organisation to be accorded legitimacy for its operations, it would voluntarily report on certain activities if it perceives that these activities are expected by the communities in which it operates (Guthrie et al 2006). For these reports to be effective, they must be accessible in publicised disclosures such as annual reports or other documents (Cormier and Gordon 2001). This is apposite to information on oil and gas reserves as this information is relied upon by investors as well as other stakeholders to adequately value the firm.

### 2.7.4 SIGNALLING THEORY

The signalling theory is attributed to the works of Arrow (1972) and Spence (1973). Signalling theory links the agency theory to describe firm behaviour aimed at reducing information asymmetry and adverse selection mechanism between itself and stakeholders through an increased level of discretionary disclosure in annual reports, also surmounting the limitations of inadequate mandatory disclosure regulations by so doing. (Connelly et al 2011).

The signalling theory advocates for a positive relationship between firm performance and the level of private information communicated to market participants. An implication of this is companies signalling their competitive strength through the communication of more and better information to the market. It is important to note that such signals must be private to the firm and observable by the market to be deemed efficacious (Dianelli, Bini and Giunta, 2013).

One core E&P firm's communication is reserves disclosures via annual reports as supplementary information. This signal could be negative or positive. Signal of improved reserves via successful exploration, acquisition and production constitute a positive signal as it indicates future cash inflows while signals of reduction to reserves via write-downs of overstated reserves or unsuccessful exploration and production challenges constitute a negative signal as it indicates reduced cash inflows and profitability.

The disputing nature of the principal-agent relationships may cause upstream oil and gas managers to adjust the signal they send to the market that is more favourable for the management. This is explored in the next theory discussed below.

### 2.7.5 OBFUSCATION HYPOTHESIS

Obfuscation is act of confounding the comprehension of financial statements for users of such information through reduced communicating clarity by concealing negative information behind cryptic and ambiguous wording and highlighting only positive information in a simplified manner (Bayerlein, 2010)

This obfuscation hypothesis is advanced by Courtis (1998). it assumes that managers are incentivised to opportunistically choose a presentation style and content on financial reporting which portrays a favourable impression of firm performance and therefore management is not neutral but biased in its presentation of accounting narratives. This results in management obscuring failures and negative outcomes while accentuating successes to manipulate market participants' perception of the firm and increase management compensations (Merkl-Davies, 2007)

As previously discussed in the signalling theory, regardless of the limitation of no mandatory reporting requirements for oil and gas reserves information, there is an incentive to voluntarily disclose more information by E&P firms to drive a positive reaction from the market and boost firm performance. However, the varied accounting practices have opened a wide range of discretionary reporting choices of oil reserve estimates (Cortese et al., 2010, p. 76).

This is fundamentally observed in the fact that the disputing nature of the principal-agent relationships may cause upstream oil and gas managers to adjust the signals they send to the market resulting in earnings management such that the managers communicate only better reserve position and withhold, conceal or postpone bad reserves news to manage the market reaction and improve the value of its shares and profitability (Cormier and Magnan, 2002). Discretionary reporting choice has also created difficulty for investors to compare two E&P's firm reserves on the same basis.

However, E&P firms may also have an incentive to report adverse reserve information to avoid litigation and reputational costs for failure to disclose and maintain the firms' equity value.

## 2.8 EMPIRICAL LITERATURE ON OIL AND GAS RESERVES AND FIRM VALUE

Early empirical studies examined the value relevance of aggregate oil and gas reserves. Bell (1983) found that initial Reserve Recognition Accounting (RRA) disclosures in the US elicited a positive response from the market. Conversely,

Dharan (1984) discovered these RRA disclosures had no additional impact on share prices. Similarly, Harris and Ohlson (1987) and Doran et al. (1988) research results were statistically insignificant to establish the value relevance of proved oil and gas reserve values. However, Boone (2002) attributed the absence of a significant relationship between reserve value changes and security returns observed in prior studies to model misspecification than to low value relevance of proved reserves disclosures or error in measurement. Boyer and Filion (2007) employed a multifactor framework to study the value relevance of cash flow, reserves changes and production on the quarterly returns of Canadian firms. The authors found a positive relationship between changes on oil and gas reserves and Canadian stock market return.

The foremost researches on the value relevance of the components of change in oil and gas reserve quantities are traced to studies by Alciatore (1993) and Spear (1994). Alciatore (1993) established that unless categorised into individual components, changes in the standardised measure had no additive information content. The six of the ten components considered to be value relevant were production, discoveries and reserve acquisitions, revisions, changes in income taxes and price changes. Spear (1994) corroborates in a similar study, revealing disaggregated reserve quantum data to possess significant information content such that exploration, improved recovery, production, revisions and acquisitions had value relevance besides just the aggregate figure. Furthermore, explorations were adjudged the most value relevant. This conclusion was again confirmed by Spear (1996).

Contemporary studies have examined the value relevance of the disaggregated components accounting for adjustments in oil and gas reserves either independently or collectively (Berry and Wright, 2001; Kretzschmar and Kirchner, 2009; Bird, Grosse and Yeung, 2013; Sabet and Heaney, 2016; Misund, 2016; Misund and Osmunden, 2017; Misund, 2018; Jianu and Jianu, 2018).

Berry and Wright (2001) assessed the value relevance of reserves disclosures by examining the relationship between the firm's effort and ability to discover new reserves by looking at 246 US public companies for 1989-1993 at the Arthur Andersen Oil & Gas Reserve Disclosure Database. They found that for firms replacing its reserve via future discoveries and extension of proved reserves, exploratory efforts alone are not enough but also the ability to make successful discoveries. Ability reflects on the relationship between firm effort and firm future cash flows. Hence efforts with commensurate successful discovery are value relevant while efforts yielding negative results are valued negatively by the firm. They suggested that firms with low ability to discover reserves should focus on alternatives like purchase of proven reserves rather than exploratory activities.

Kretzschmar and Kirchner (2009) explored the value relevance of location-specific oil field tax terms effect on E&P firms' reserve replacement by analysing 51 E&P SEC complaint companies holding 59% of global reserves in a PSC and Concession regime-mix for a 9year period (1998 to 2006). He found out that the location's regime for the reserve replacement for large E&P firms is a determinant of share price and thus shares of such companies with concession dominated assets outperform those of companies with high PSC reserves holdings especially with increase in the price of oil. This is because with the PSC regime, asset entitlement is limited for the firm

Bird, Grosse and Yeung (2013) analysed 307 Australian mining firms to study the impact of JORC compliant discovery announcements and reserve disclosures on share price. They found a positive market reaction to earning and exploration announcements. These disclosures were discovered to be made by larger and more mature firms.

An empirical study by Sabet and Heaney (2016) on the value relevance of announcement of the acquisition of oil and gas acreage and reserves of 1391 separate acreage or reserve acquisition announcements made by listed E&P firms in the United States equity market from 1992 to 2011 on the Herold Merger and Acquisition Database found that acquisition of reserves and acreage were value relevant, but consistent with information asymmetry, reserves had stronger positive signal to the market value of E&P firms than acreage as the former was more costly to acquire because of the high cost of identifying reserves existent within such acreage. Control variables where size, growth, leverage, RPR, and FD. However, share price was found to be higher for acreage acquisition than reserves acquisition for periods of high crude oil volatility. They also found acquisition of reserves as compared to acreage, benefitted firms with low successful track record of exploratory activities, while the share price of firms with low reserve production ratio improved upon acquisition of either acreage or reserves.

Misund (2016) revealed that E&P firms have high value relevance especially given the presence of supplementary estimates of oil and gas reserves hence mitigating the effect of high variability of earnings posed by the presence of high intangible assets whose value were affected by the volatility of oil and gas prices.

Misund (2018) also conducted an investigation to compare the value relevance of exploration and acquisition as replacement strategy of oil and gas reserves by separating total reserves into its subcomponents. Employing a sample of 4,218 firm-years for North American and international oil and gas companies, firstly, he finds stock returns to be positively related to changes in oil and gas reserves. His research disclosed that the market was indifferent about either method of reserves replenishment as all reserves additions and reductions were solely measured as proved reserves not including unproven reserves. Therefore, should the market distinguish, it could present arbitrage opportunity to investors. However, due to a structural shift caused by shale revolution for periods after 2008 and worsened by the world financial markets affected by the credit crises in the banking sector, there was a break in the oil and gas price link, plummeting the price of gas, thus, change in gas reserves had bigger impact on share return than changes in oil reserves. This consequently resulted in a positive significant relationship between the increase in oil production and share returns and less significant relationship between the increase in natural gas production and share price. Another explanation for the fall in gas prices was that oil reserves had an earlier cash flow in comparison with gas reserves, given the nature of their production sequence.

Jianu and Jianu (2018) adopting the OLS Ohlson share price model analysed how commitment to amassing reserves influence the share price of E&P Companies recording losses by examining 51 listed companies in the LSE for the period 2014-2016 marking significant fall in oil and gas prices. He discovered that increasing cost to acquire reserves especially during the period where losses are incurred has a positive significant effect on share price as this action implies future cash flows thus neutralizing the negative perceptions generated by the losses incurred

Extant literature has studied value relevance of oil and gas reserves disclosures as a function of its classification (Berry et al (1997), Rai 2006, Misund 2017). Berry et al (1997) found a positive relationship between aggregate proved reserves and share price. But by analysing the individual value relevance of proved developed and proved undeveloped reserves, they found only proved developed reserves to be value relevant. This was because proved undeveloped reserves were regarded as unreliable as they required sizable cost commitment to make them productive.

Similarly, Misund and Osmunden (2017) investigated the relationship between share return and the three classifications of reserves based on maturity levels: proved developed, proved undeveloped and probable reserves employing a multifactor framework and using a sample of 94 Canadian and International oil companies for the period 1993-2013 (455 firm years). He discovered that proved developed reserve had positive relationship with returns while the probable reserves, had no impact. Almost similar to probable reserves, proved undeveloped reserves had a weak significant relationship at 10% to share price. This he explained by the fact that investors viewed changes in less mature reserves as uncertain and hence information about them are not considered useful to forecast future cash flows. However, with shale revolution for periods after 2009, he observed that changes in proved developed gas reserves turned from positive to negative relationship with share price. Similarly, proved undeveloped reserves now had a negative relationship with share price, while changes in probable gas reserves moved to a positive but insignificant relationship with share price. This can be explained by the fall in natural gas price for the period of the shale revolution. His findings corroborate earlier research by Rai (2006) who conducted identical research for 30 Canadian E&P firms for the period 2002-2004, but finding both proved and probable reserves to be significant and value relevant.

## 2.9 CONCLUSION

From the foregoing, this study seeks to extend the works of the value relevance empirical literature for the oil and gas industry as reviewed in the previous section by examining the relationship between the changes in oil and gas quantity reserves as well as the informational contents such changes and the average share returns for UK upstream LSE listed firms for the period 2011 to 2018. In addition, it also seeks to analyse the quality of such disclosures and the implication for the firm value.

# 3 CHAPTER THREE: RESEARCH METHODOLOGY

## 3.1 INTRODUCTION

According to Kothari and Garg (2014), research is a voyage of discovery, the pursuit of truth through study, observation, comparison and experiment". Research methodology is a systematic plan to conduct research (Saunders, Lewis and Thornhill 2016). It provides a step-by-step process outlining the way research is to be undertaken.

This chapter describes the research philosophy, approach, methods, strategy and design used in this study to achieve the research aims and objectives. The chapter also explores the research samples, sources of data and the data collection techniques, highlighting limitations as well as ethical considerations. Issues on the credibility of findings are also explored. This chapter takes on a format of the research onion (Figure 3.1) prescribed by Saunders, Lewis and Thornhill (2016 p.164).

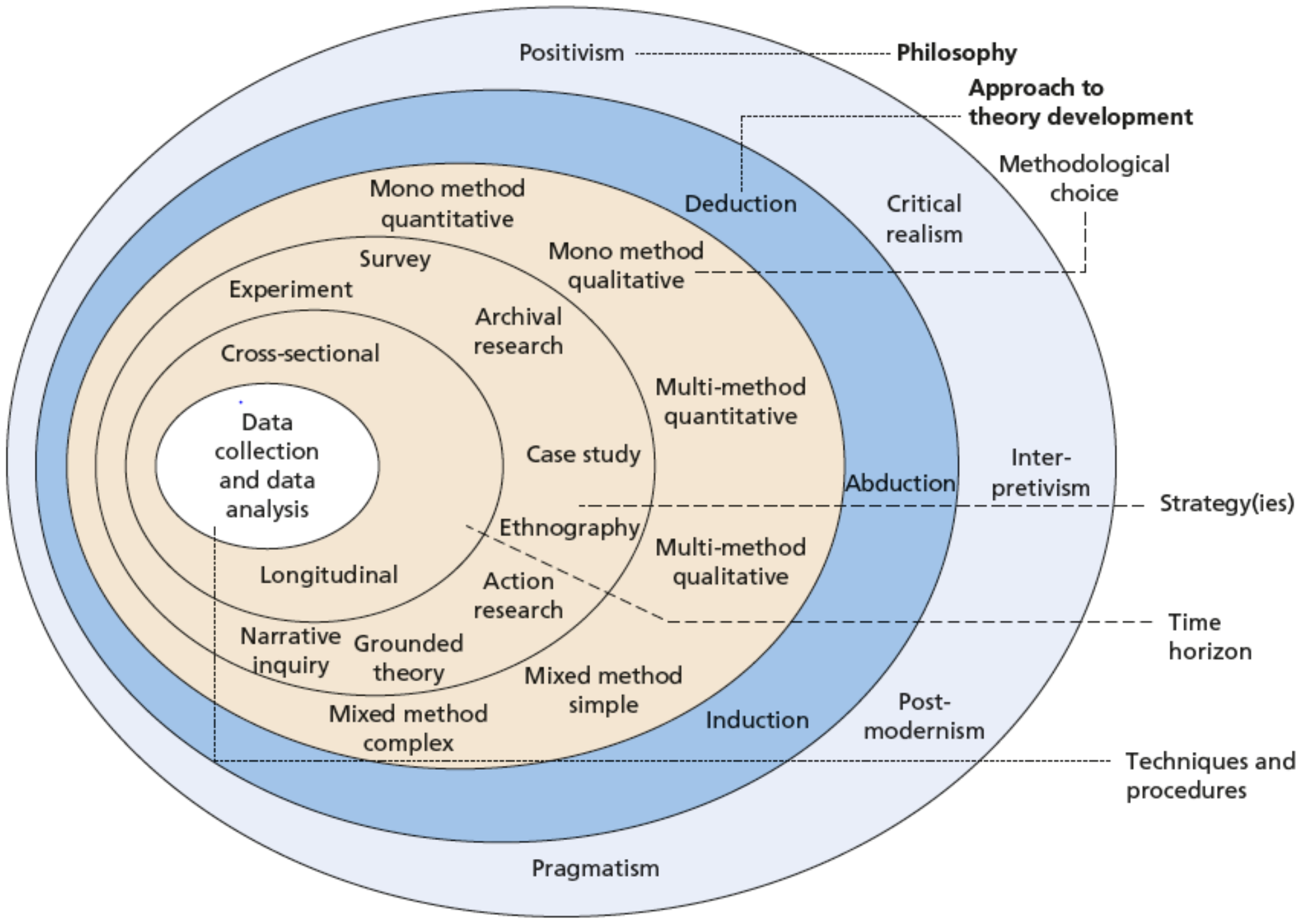

Figure 3. 1: Research onion (Saunders, Lewis and Thornhill 2016 p.164)

## 3.2 RESEARCH PHILOSOPHY

According to Saunders, Lewis and Thornhill (2016), research philosophy is defined as a system of beliefs and assumptions guiding the development of knowledge. It includes a set of epistemological, ontological and axiological assumptions which underpins the methodology and design of the research.

Epistemological assumptions are assumptions about what constitutes acceptable, valid and legitimate knowledge for the research undertaken and the contributions such research would make to knowledge. Ontological assumptions focus on the assumptions about the nature of reality which shapes how the researcher views and studies the research object while the axiological assumption relates to the researcher's values and ethics and how it affects the research process (Mackenzie and Knipe 2006).

According to Niglas (2010), the two continua of approaches to research assumptions are Objectivism and Subjectivism and the two major types of research philosophy; positivism and interpretivism. This research is established on an objectivism approach and a Positivism philosophy. The assumptions and justifications for this approach and philosophy is addressed in relation to the research topic.

In contrast to subjectivism which conceptualises social entities as a function of perceptions, share meaning and consequential actions of social actors, objectivism assumes organisations were like social entities, existing independent of social actors, thus, all social actors experienced only one true social reality. Therefore, social entities could be subject to scientific researches (Sanders, Lewis and Thornhill, 2016).

Ontologically, objectivism assumes that organisations are like independent physical entities of nature. Thus, this study assumes that management is an objective entity having formal structured management function similar to all business organisations, wherein subordinates are accountable to management who are in turn accountable to shareholders, necessitating a principal-agent relationship. Epistemologically, objectivism assumes the use of scientific methods, facts, numbers and observable phenomenon to make generalisations. The same has been assumed for this study. Axiologically, objectivism also assumes that the researcher's values are neutral and independent of the research to make the results bias-free. This is critical factor for this research (Burrel and morgan, 1979).

The positivist research philosophy has been favoured for this research. This philosophy is built on the objectivism approach to research assumptions. Positivists believe that human beings, their actions and society could be studied objectively. It rejects the metaphysical and subjective ideas and focuses on the tangible (Niglas 2010). Positivism assumes the philosophical stance of the natural scientist by analysing an observable social reality to produce law-like generalisations (Saunders, Lewis and Thornhill, 2016). It relies on scientific empirical methods, utilizing observable and quantifiable data to produce credible and meaningful results. Positivism philosophy uses existing theories to develop hypothesis which is then tested, confirmed or refuted. Data collected is analysed to detect causal relationship to make law like generations as contributions to the body knowledge (Collis and Hussey, 2014 .p 44).

Conversely, interpretivism argues that human beings and their social worlds cannot be studied scientifically as physical phenomena because social actors experienced different social realities (Saunders, Lewis and Thornhill, 2016). It is built on the subjectivist approach. With this perspective, managements were therefore seen as different groups of people experiencing different workplace realities. The researcher following this philosophy relies on qualitative analytical methods to understand and interpret the different participant perceptions to make judgements (Collis and Hussey 2014). Considering the extant literature on value relevance of oil and gas disclosures, positivism provides a better perspective to undertake this study.

Based on the foregoing, this research also adopts the functionalist paradigm of organisational operations which underpins the positivist research philosophy. It combines an objectivist and a regulation perspective. It assumes rational explanations for organisation actions and develops sets of recommendation for improvement within the current structure rather than radically challenging the existing framework (Burrel and Morgan, 1979).

## 3.3 RESEARCH APPROACH, METHOD, STRATEGY AND SAMPLING TECHNIQUES

### 3.3.1 RESEARCH APPROACH

Research approach is defined as a path of conscious scientific reasoning (Niglas 2010). The two fundamental approaches to scientific enquiry are the Deductive and the Inductive approach. Burney (2008) sets out the processes unique to each approach (Figure 3.2 and 3.3).

Deductive research is associated with positivism research philosophy and the quantitative method of research. Deductive research is a theory-testing process aimed at ascertaining if an existing theory is applicable to a certain set of data observations (Hyde 2000). Hypotheses are developed from theories and expressed as a relationship between variables. These hypotheses are then subjected to rigorous tests based on the observed data. The outcomes are then examined to confirm, modify or reject the hypothesis (Saunders, Lewis and Thornhill, 2016).

Caution must be exercised by introducing control variables when testing hypothesis. This will ensure that there are no outliers resulting in ambiguous results.

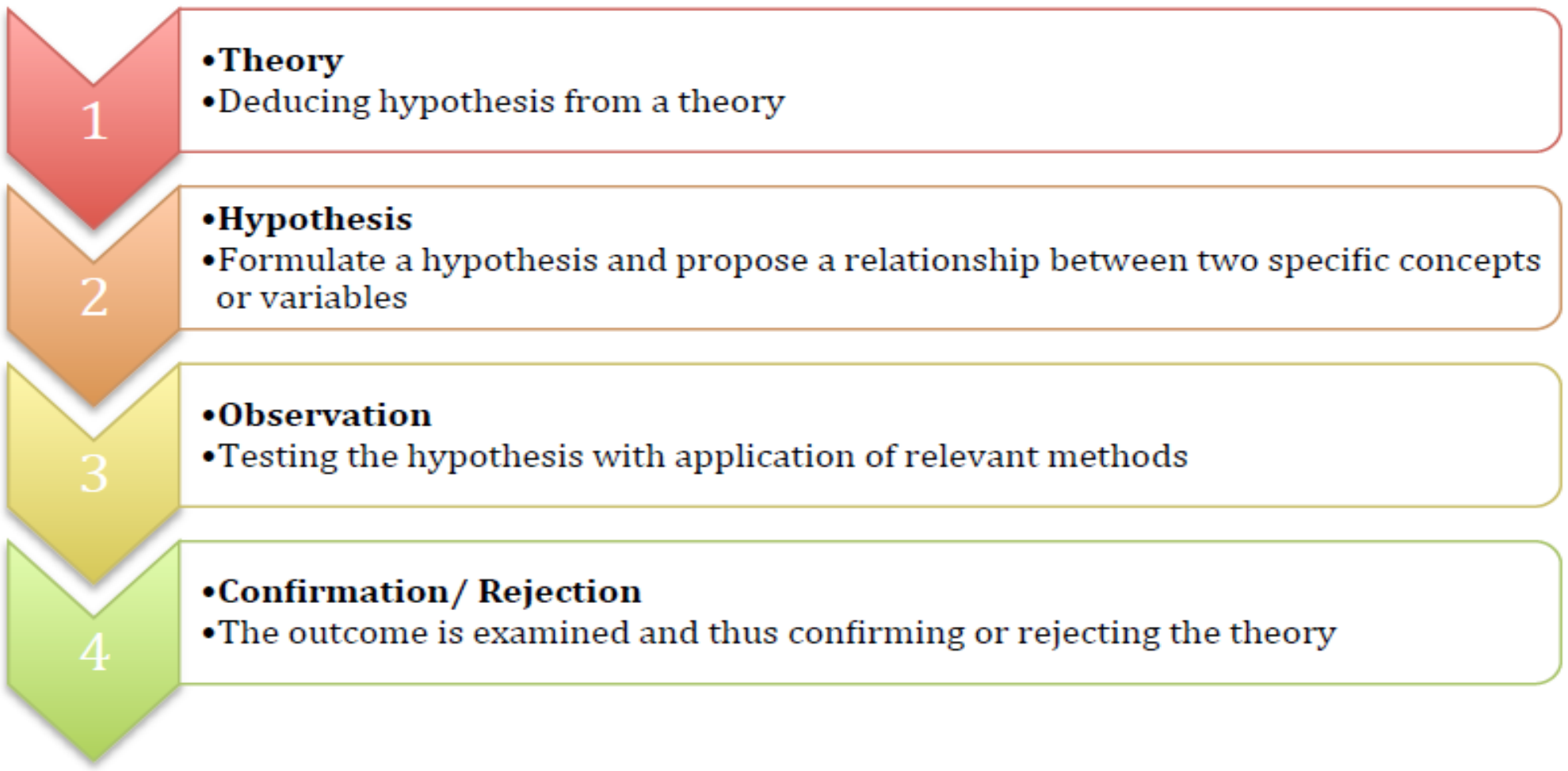

Figure 3. 2: Stages of Deductive Reasoning (Waterfall Approach) (Burney, 2008)

In reverse, inductive approach flows from specific observations to broad generalizations (Trochim, 2006). The researcher collates data to derive conclusions from it (Saunders, Lewis and Thornhill, 2016). The results of the data analysis are then applied to formulate a theory (Hyde 2000). The advantage with inductive approach is that it includes the real-world context in which certain events affecting the data has taken place. However there is the danger of an erroneous inductive conclusion and human bias effect.

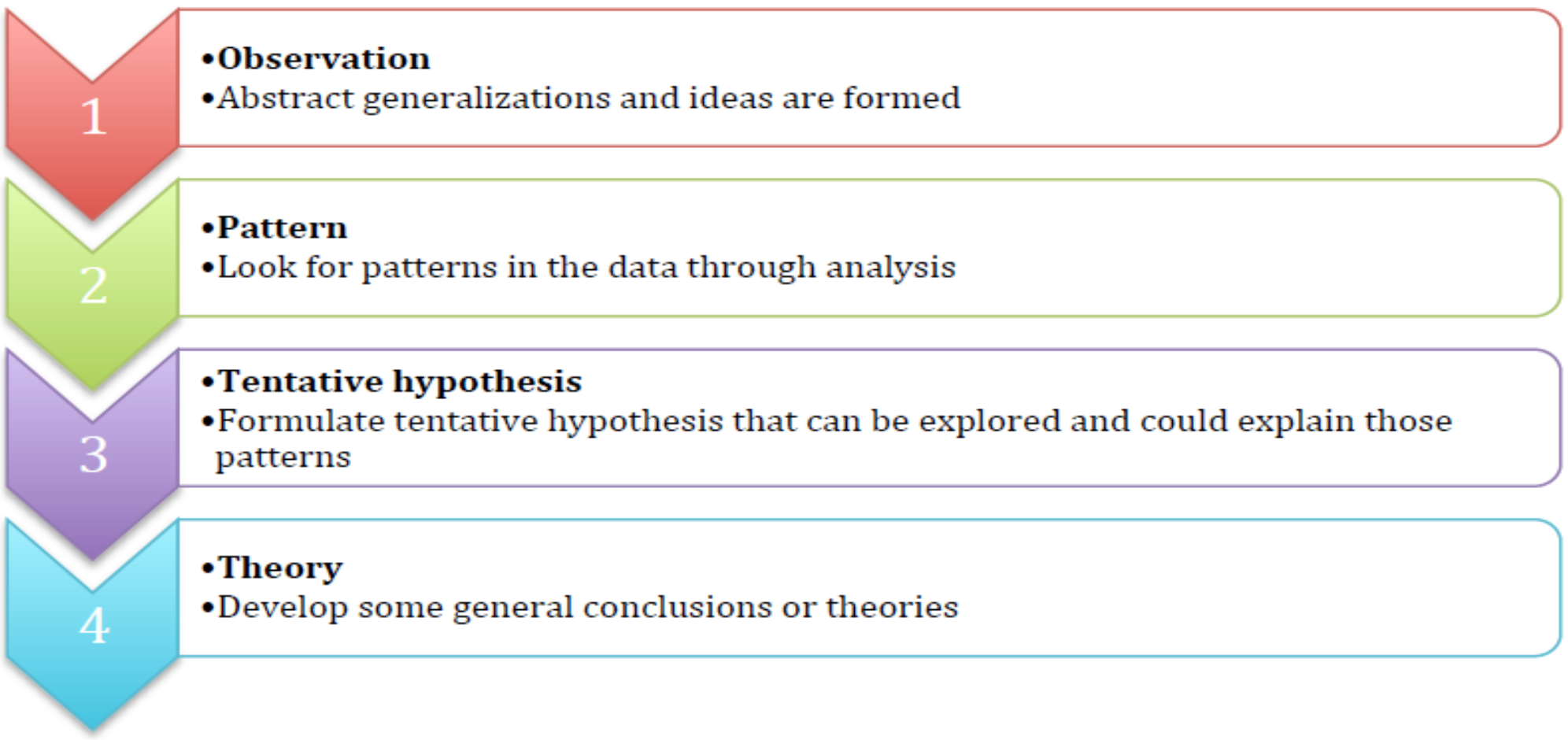

Figure 3. 3: Stages of inductive Reasoning (Hill climbing) (Burney, 2008)

From the foregoing, the deductive approach is adopted as the most appropriate based on the existing literature, research philosophy and the research design prescribed in Section 3.4. Also, as observed in section 2.7, there exist several theoretical models underpinning reserve disclosures. These theories have provided an avenue for the researcher to develop hypotheses to be tested on the observed data set with a deductive approach.

### 3.3.2 RESEARCH METHODS

The three major methods of research in management are quantitative, qualitative and mixed methods (Saunders, Lewis and Thornhill, 2016). There is no best method, however, the researcher's choice is determined by the appropriate method considered fit for the study. The appropriate method produces plausible evidence to best answer the research questions and achieves the research objectives (Scott, 2010). The choice of method also influences the data collection techniques and analysis procedures (Collis and Hussey, 2014).

Quantitative research applies positivism and deductive approach to examine and predict causal relationships between variables through statistical and graphical analytical techniques, employing the use of controls to ensure validity of the data (Saunders, Lewis and Thornhill, 2016). It is aimed at collecting and analysing data in numeric form.

In contrast, qualitative research utilizes the interpretive philosophy and inductive approach to study participants' meanings and the relationships between them in order to develop a conceptual framework and theoretical contribution (Saunders, Lewis and Thornhill 2016). A combination of quantitative and qualitative research results in a mixed method.

Given the nature of this study, its aims, objectives and the underpinning research philosophy and approach, a quantitative method has been selected. This is also in tune with the methods used in previous studies. This will provide a highly efficient and objective approach to finding outcomes, attributing causes, comparing and ranking of variables.

Quantitative research facilitates understanding patterns and answers the 'how' questions while qualitative research answers the 'why' questions (Biggam 2015). This choice method will enable the researcher understand how the aggregate and the components of oil and gas reserve disclosure impacts on share price for the span of time under study.

Conclusively, the mono-method quantitative technique has been adopted for this research, that is, applying one quantitative data collection technique and corresponding analytical procedure (Sanders, Lewis and Thornhill, 2016). Statistical figures on reserve information, share price, cashflows, equity market risk and oil price have been obtained from databases, annual reports and stock market websites.

### 3.3.3 RESEARCH STRATEGY

According to Saunders, Lewis and Thornhill (2016), research strategy refers to plan undertaken by a researcher to provide answers to proposed research questions. It links the research philosophy and methodology applied in data collection and analysis (Denzin and Lincoln 2011).

Saunders, Lewis and Thornhill (2016 p.177) provided different research strategies available to a researcher. These he grouped based on the choice of methodology as observed below (Figure 3.3). The choice of research strategy is dependent on its ability to answer the research questions, meet the objectives and is integrated with the research design, methodology, existing literature, underpinning research philosophy and resources such as time, data availability and finance (Sanders, Lewis and Thornhill, 2016)

Table 3. 1: Research strategies (Author, 2019)

| Research strategy | | |
|---|---|---|
| **Quantitative** | **Qualitative** | **Combination/mixed** |
| Experimental research | Action research | Case study |
| Survey | Ethnography | Archival research |
| | Grounded theory | |
| | Narrative research | |

Therefore, no single strategy is superior to another and neither are they mutually exclusive. In light of the analysis of the merits and demerits of each research strategies identified by Saunders, Lewis and Thornhill (2016, p. 179-199) as well as the considerations on the research design and data constraints, the archival research method has been adopted as the most appropriate for this research.

Archival research strategy employs administrative records and documents as a major data source (Saunders, Lewis and Thornhill 2016). These records are not limited to only historical documents but also current ones. Although these records were not initially collated for the purpose of research, they could serve as secondary data as they provide meaning information on the reality that is being studied and also saves resources that would have been utilized to collect a new data set.

Archival research is capable of answering explanatory, exploratory or descriptive research questions. It permits the researcher to answer research questions that focus on past trends and the changes that have occurred over time. It is easy to use if the research considers the use of quantifiable historical data which can be easily accessed digitally from a variety of sources. Section 3.4.5 below displays the variables being investigated which can all be easily extracted on the internet from database, published financial statements, finance and stock exchange web sites at no or very minimal cost. This strategy, therefore, allows the researcher achieve the objectives itemized regarding value relevance of the component of reserve disclosures of UK listed E&P firms.

## 3.4 RESEARCH DESIGN

This researcher adopts a descriptive and explanatory design to determine the cause and effect relationship between the E&P oil and gas firms' share price and the components of reserve quantity change.

As suggested by Ross (1976) and implemented by Alciatore (1993), Spear (1994), Ohlson (1995), Boyer and Filion (2007), Misund (2017; 2018) in their value relevance studies of oil and gas reserves, this study similarly adopts a multifactor empirical model in regressing the share return of oil and gas on a set of fundamental variables using SPSS.

The researcher builds majorly on the Ohlson (1995) and the Misund (2018) models which set a framework to analyse the information content of oil and gas disclosure information. Ohlson (1995) attributed the value of the firm to profitability (earnings), book value of equity and other value relevant information that relates to future earnings that were not yet captured as shown below in Eq. (1).

$$P_t = \beta_0 + \beta_1 BV_t + \beta_2 E_t + \beta_3 V_t + \varepsilon_t \qquad \text{(Eq. 1)}$$

Where t is the time period, $P_t$ is share returns, $BV_t$ is book value of equity, $E_t$ is earnings, $\varepsilon_t$ is the error term and $V_t$ is other value relevant information representing future earnings.

As with Misund (2018), this study uses change in oil and gas reserves to represent other value relevant information impacting on the future cashflows. In addition, this study decomposes the individual components of change in reserves quantity into exploratory activities, acquisitions, purchases and revisions as reported in the supplementary statements of the annual reports of companies. In other to achieve the research objectives, this research proposes three (3) models to study the relationship between these parameters and share price of the firm.

### 3.4.1 CHANGE IN OIL AND GAS RESERVES IMPACT ON SHARE PRICE

This first model is developed from Misund (2018), which is a modification of the Ohlson (1995) model in Eq. (1). His model is shown below as Eq. (2):

$$R_{it} = C + \beta_1 CF_{it}/P_{it-1} + \beta_2 \Delta CF_{it}/P_{it-1} + \sum_{j=1}^{6} \lambda_j R_t^j + \Delta TR_{it} + \varepsilon_{it} \qquad \text{(Eq. 2)}$$

The dependent variable $R_{it}$ is the share return for firm $i$ at time $t$. $\delta^{total}TR_{it}$ (Change in total oil and gas reserves) is the major explanatory variable for future related value relevant information. $\Delta TR_{it}$ is calculated as total oil and gas reserves at the year-end less the total oil and gas reserves at the beginning of the year ($\Delta TR_{it} = TR_t - TR_{t-1}$).

To prevent omitted variable bias, control variables are cashflows and risk factors. Misund (2018) includes cashflow from operations ($CF_{it}$) and changes in cashflows from operations ($\Delta CF_{it}$) as measures of profitability which was considered to more appropriate than earnings used by Ohlson (1995) as discussed in section 2.3.6. However, this study considers cashflows from operations only. Cashflows from operations is further adjusted using total assets as deflator which was found to be more suitable for this study compared to total number of outstanding shares used by Misund (2018). However the large values of total assets compared to cashflows from operations may pose a limitation to the statistical significance of cashflows as a determinant of share returns. $R_t^j$ represents the risk factors. Misund (2018) adopts six (6) risk factors which were explored in Section 2.2.3, par 1, but this study considers only two of them which are $X_{it;}$ equity risk and $\Delta OP_{t;}$ change in oil prices. A dummy variable, $D_{i1}$ is introduced into this study as a proxy for firm size to capture E&P firms listed either in the main or aim market. As identified in section 3.6.1, main listed companies are for large firms and the aim listed companies are for smaller firms. Finally, $C$ and $\varepsilon_{it}$ are the intercept and residual. The first model for this study is therefore coined as follows:

$$R_{it} = C + \beta_1 \Delta TR_{it} + \beta_2 CF_{it} + \beta_3 X_{it} + \beta_4 \Delta OP_t + \beta_5 D_{i1} + \varepsilon_{it} \qquad \text{(Eq. 3)}$$

Where,

$D_{i1}$ is set to 0 for FTSE AIM All-Share firms and 1 for FTSE All-Share firms

### 3.4.2 EXPLORATORY ACTIVITIES, ACQUISITIONS, PRODUCTION AND REVISIONS IMPACT ON SHARE PRICE.

The second model in this study is developed from this first in Eq.(3). It splits the change in reserve quantity into 5 major components (Eq. 4). Each of these components is considered a separate explanatory variable ($\Delta TR_{it} = TR_exp + TR_acq + TR_pro + TR_rev + TR_sal$).

$$R_{it} = C + \beta_1 TR_EXP_{it} + \beta_2 TR_ACQ_{it} + \beta_3 TR_PRO_{it} + \beta_4 TR_REV_{it} + \beta_5 TR_SAL_{it} + \beta_6 CF_{it} + \beta_7 X_{it} + \beta_8 \Delta OP_t + \beta_9 D_{i1} + \varepsilon_{it} \quad \text{(Eq. 4)}$$

$TR_EXP_{it}$ is exploratory activities including extensions and improved recovery, $TR_ACQ_{it}$ is acquisitions, $TR_PRO_{it}$ is production, $TR_REV_{it}$ is revisions and $TR_SAL_{it}$ is sales. This improves on Boyer and Filion (2007) by avoiding overlapping of these variables. All changes in reserves' components use total outstanding shares at the beginning of year as deflator in Cormier and Magnan (2002) and Berry and Wright (2001), however, like Misund (2018), this study uses barrels of oil equivalents at start of the year as the deflator.

### 3.4.3 DISCLOSURE EFFECT ON SHARE PRICE

To derive the third model, the researcher adjusts the first model by including two disclosure parameters. According to Beretta and Bozzolan (2004), quality of disclosure depends not only on the quantity of information disclosed but also on the richness of such information. As discussed in section 2.4.1, the IFRS has no mandatory disclosure requirements for UK listed oil and firms. However as suggested by Slack et al., (2010) and McChlery et al. (2015), the researcher tries to assess the disclosure quality given voluntary disclosures regulations by measuring how the level of disclosure of the individual reserve quantity components and reserve quantity KPIs (see section 2.6) as recommended by SORP/OFR in section 2.4.1 further impacts on the share price using two dummy variables shown below.

$$R_{it} = C + \beta_1 TR_EXP_{it} + \beta_2 TR_ACQ_{it} + \beta_3 TR_PRO_{it} + \beta_4 TR_REV_{it} + \beta_5 TR_SAL_{it} + \beta_6 CF_{it} + \beta_7 X_{it} + \beta_8 \Delta OP_t + \beta_9 D_{i1} + \beta_{10} D_{i1} + \beta_{11} D_{i2} + \varepsilon_{it} \quad (5)$$

Where,

$D_{i1}$ is set to 1 where the firm discloses more than 2 components accounting for changes in the reserve quantity or 0 otherwise

$D_{i2}$ is set to 1 where the firm disclosures one or more reserve KPI or 0 otherwise.

### 3.4.4 HYPOTHESIS FORMULATION

The researcher predicts that at 95% level of significance, there would be a positive and significant relationship between the change in oil and gas reserve and share price on one hand, as well as the a significant relationship between the components of change in oil and gas reserves and share price in line with previous empirical studies explored in the empirical review in Section 2.8.

### 3.4.5 CALCULATION OF OPERATIONAL VARIABLES

The independent and dependent variables used for the research design are summarized and calculated as shown in the table below. The raw data set and the computed data set due to these adjustments are shown in appendices 5 and 6.

Table 3. 2: Operational Variables (Author, 2019)

| OPERATIONAL VARIABLES | CALCULATIONS |
|---|---|
| **(a) Independent variables** | |
| Share return ($\boldsymbol{R_{it}}$) | This study utilizes contemporaneous returns<br>$\frac{\text{Current year share price} - \text{Preceding year price}}{\text{Preceding year price}} = \frac{P_{it} - P_{it-1}}{P_{it-1}}$<br>***Comments**: average monthly share price which was available on uk.investing.com is utililized for this study. This is used to calculate the average annual share prices. This measure is considered to better capture the annual performance of the firm as compared to using end of the year prices*. |
| **(b) Dependent Variables** | |
| Change in reserves ($\Delta\boldsymbol{TR_{it}}$) | $= \frac{\text{total reserves at the yr end} - \text{total reserves at the yr start}}{\text{total reserves at the yr start}}$<br>$= \frac{TRsv_{it} - TRsv_{it-1}}{TRsv_{it-1}}$ |
| Exploratory activities ($\boldsymbol{TR_EXP_{it}}$) | $= \frac{exploration + improved\ recovery + extensions}{\text{total reserves at the yr start}} = \frac{TR_EXP_{it}}{TR_{it-1}}$ |
| Acquisitions ($\boldsymbol{TR_ACQ_{it}}$) | $= \frac{Acquisitions}{\text{total reserves at the yr start}} = \frac{TR_ACQ_{it}}{TR_{it-1}}$ |
| Production ($\boldsymbol{TR_PRO_{it}}$) | $= \dfrac{Production}{\text{total reserves at the yr start}} = \dfrac{TR_PRO_{it}}{TR_{it-1}}$ |
| Revisions ($\boldsymbol{TR_REV_{it}}$) | $= \dfrac{Revisions}{\text{total reserves at the yr start}} = \dfrac{TR_REV_{it}}{TR_{it-1}}$ |
| Sales ($\boldsymbol{TR_SAL_{it}}$) | $= \dfrac{Sales}{\text{total reserves at the yr start}} = \dfrac{TR_SAL_{it}}{TR_{it-1}}$ |
| **(c) Control VariableS** | |
| Cashflows ($\boldsymbol{CF_{it}}$) | $= \dfrac{Cashflows\ from\ operations}{Total\ Assets} = \dfrac{CF_{it}}{TA_{it-1}}$ |
| equity risk ($\boldsymbol{X_{it}}$) | This is the expected risk premium, it is captured using variability in earnings (Farrelly et al, 1985)<br>$= \dfrac{\text{Current year earnings} - \text{Preceding year earnings}}{\text{Preceding year earnings}}$<br>$= \dfrac{X_{it} - X_{it-1}}{X_{it-1}}$<br>***Comments****: Earnings is measured using EBITDA (Earnings before Interest, Tax, Depreciation and Amortisation.* |
| Change in oil price ($\Delta\boldsymbol{OP_t}$) | $= \dfrac{\text{current yr oil price} - \text{preceding yr oil price}}{\text{preceding yr oil price}} = \dfrac{OP_{it} - OP_{it-1}}{OP_{it-1}}$<br>***Comments****: historical monthly UK brent oil price is obtained from www.eia.gov. This is used to calculate annual oil prices.* |

## 3.5 TIME HORIZON

Providing clarity on the time frame covered by a research is fundamental to providing a solution to the research problem. Some research problem may require a cross-sectional study and others, a longitudinal study. The former is a snapshot process of studying one or more phenomena at a particular point in time while the latter studies such phenomena over an extended period of time (Saunders et al, 2016). Longitudinal studies, however time-consuming, has the comparative advantage of looking at past periods and can therefore observe changes in variables and proffer a cause and effect relationship between them.

This study was conducted as a longitudinal research using time series data from 2011-2018 obtained from secondary sources. This was because some firms had no record of reserves or operational activity for earlier periods and some were not listed in the LSE for those earlier years. This ultimately resulted in data unavailability for earlier periods.

## 3.6 SAMPLING TECHNIQUE

For accuracy in findings, the data must be presentative of the population. This is dependent on the sampling technique adopted. The different sampling techniques generally fall into the probability and non-probability categories (Saunders, Lewis and Thornhill 2016, p. 276). Within the band of oil and gas firms in the UK, this study analyses only the E&P oil and gas firms listed on the LSE. Therefore, a non-probability sampling, particularly the purposive (homogenous) sampling technique has been adopted.

This technique permits the researcher to use his judgement to select elements that best fit the sample that will provide data to answer the research question and achieve the research objectives (Saunders, Lewis and Thornhill, 2016). It is critical for a small sample as in the case of this study. It is based on a particular subgroup where all sample members possess similar characteristics, enabling them to be studied in greater depth and revealing minor differences (Saunders, Lewis and Thornhill, 2016).

### 3.6.1 SAMPLE FRAME AND SAMPLE SIZE

The sample frame is precisely defined as the entire body of persons or objects under consideration. The consideration for the sample frame focused on E&P oil and gas companies trading at the London stock exchange. The LSE was selected due to its size as one of the two largest stock exchanges in Europe and its compliance with IFRS (Power, Cleary and Donnelly, 2017).

The FTSE All-Share Index and the FTSE AIM All-share Index within which oil and gas firms are listed in the LSE were considered for this study. The FTSE All-Share Index represents 98% of UK market capitalisation as it contains larger firms while the FTSE AIM All-share Index contains smaller firms with high growth potential (FTSE Russell All-Share Index fact sheet, 2019).

As of August 28, 2019, on the londonstockexchange.com, the FTSE All-Share Index had a total of 15 companies classified under the non-renewable energy industrial sector and the FTSE AIM All-Share Index had 90 companies under the oil and gas industrial sector. This summed up to a total of 105 oil and gas companies. The profile of these companies was examined individually from www.financialtimes.com and uk.investing.com to understand the nature of business. 30 Companies identified as service, equipment, logistics and management or downstream were eliminated at this point from leaving a sample frame of 75 strictly E&P companies.

Purposeful sampling technique was applied to further select only firms trading at £3 and above as at that date. This was done because it was observed from looking at the annual report of a random sample of these firms who traded below £3, that they were newly listed in the LSE thus had no historical data for the period considered for this study or they were still at license procurement and acreage acquisition stage and therefore had no reserve information. 49 E&P firms were found to trade above £3 however only 25 of these firms constituted the final sample size (Table 3.3) for the period 2011-2018, which cumulated into a total of 181 fam-years. This period was selected due to the unavailability of reserve and share price data for earlier periods generally. The firms eventually eliminated from the resulting 49 E&P firms are shown in table 3.4.

Table 3. 3: Study Sample Size

| S/N | CODE | NAME | MARKET | SHARE PRICE(£) |
|---|---|---|---|---|
| 1 | BP. | British Petroleum | main | 502.90 |
| 2 | CNE | Cairn Energy | main | 166.00 |
| 3 | ENQ | Enquest | main | 18.41 |
| 4 | PMO | Premier Oil | main | 79.46 |
| 5 | RDS'B' | Royal Dutch Shell, B | main | 2269.50 |
| 6 | SIA | Sia Soco INTL | main | 63.80 |
| 7 | TLW | Tullow Oil | main | 201.70 |
| 8 | AMER | Amerisur | aim | 17.88 |
| 9 | CASP | Caspian Sunrise | aim | 10.35 |
| 10 | ELA | Eland Oil and Gas | aim | 123.60 |
| 11 | EME | Empyrean | aim | 10.05 |
| 12 | HUR | Hurricane Energy | aim | 43.72 |
| 13 | IGAS | IGas Energy | aim | 50.65 |
| 14 | PMG | Parkmead Group PLC | aim | 45.10 |
| 15 | RPT | Regal Petroleum | aim | 32.90 |
| 16 | SDX | SDX Energy | aim | 19.50 |
| 17 | SQZ | Serica Energy PLC | aim | 111.60 |
| 18 | SEY | Sterling Energy | aim | 10.70 |
| 18 | SOU | Sound Energy | aim | 8.38 |
| 20 | TRIN | Trinity E&P PLC | aim | 10.75 |
| 21 | ZOL | Zoltav Resources | aim | 37.50 |
| 22 | NOG | Nostrum Oil and Gas | main | 19.00 |
| 23 | CABC | Cabot Energy | aim | 3.50 |
| 24 | EGRE | Egdon Resources | aim | 4.75 |
| 25 | PPTC | President Energy | aim | 4.65 |

Source: Londonstockexchange.com

Table 3. 4: Companies eliminated from sample

| S/N | CODE | NAME | Market | SHARE PRICE (£) | REMARKS |
|---|---|---|---|---|---|
| 1 | ECO | Eco Atlantic | aim | 156.00 | No reserves info. |
| 2 | FOG | Falcon Oil | aim | 13.62 | No reserves info. |
| 3 | I3E | I3 Energy | aim | 52.80 | Only 2years data on Osiris/FAME |
| 4 | IOG | Independent Oil&Gas | aim | 17.75 | No reserves info. |
| 5 | JSE | Jadestone Energy | aim | 51.50 | Only 2years data on Osiris/FAME |
| 6 | PANR | Pantheon Resources | aim | 18.26 | No reserves info. |
| 7 | RKH | Rockhopper Expl. | aim | 20.12 | No reserves info. |
| 8 | SAV | Savannah Petroleum | aim | 23.90 | No reserves info. |
| 9 | ENOG | Energean Oil | main | 997.00 | Only 2years data on Osiris/FAME |
| 10 | DGOC | Diversified Gas | aim | 105.00 | Listed in AIM 2017 |
| 11 | SENX | Serinus Energy | aim | 9.00 | Listed in AIM 2018 |
| 12 | TXP | Touchstone Energy | aim | 15.25 | Listed in AIM 2017 |
| 13 | BLVN | Bowleven | aim | 11.00 | Reserves undisclosed |
| 14 | SLE | San Leon | aim | 32.10 | Reserves undisclosed |
| 15 | PRR | Providence Resources | aim | 6.52 | No reserve info |
| 16 | BLOE | Block Energy | aim | 6.25 | Listed in AIM 2018 |
| 17 | LEK | Lek oil | aim | 6.14 | No reserve info |
| 18 | MATD | Petro matad | aim | 6.20 | No reserve info |
| 19 | AAOG | Anglo African oil&gas | aim | 3.85 | Listed 2017/no reserve info |
| 20 | BPCB | Bahamas Petrol | aim | 3.10 | Same as no. 19 |
| 21 | CERP | Columbus Energy | aim | 3.40 | No reserve info |
| 22 | TLOU | Tlou energy | aim | 4.70 | Listed 2017 |
| 23 | UOGU | United oil and gas | aim | 4.50 | relisted 2017 |
| 24 | CHARC | Chariot oil | aim | 3.81 | No reserve info |

Source: Londonstockexchange.com

## 3.7 DATA

### 3.7.1 SOURCE OF DATA

The data sources are majorly categorised as primary and secondary data. Secondary data is one which is adopted and not originally generated by the researcher as it is already in existence while primary data is created by the researcher for the purpose of the research utilizing tools like questionnaire, interview and observation (Tashakkori and Teddlie, 2003 p. 314). In line with the research design and strategy for this study, secondary rather than primary data has been used to source quantitative information.

This data source is time saving given the time constraints for this study. Also, it can be checked and verified to confirm its authenticity (Saunders, Lewis and Thornhill, 2016).

### 3.7.2 DATA COLLECTION

181 Annual reports were utilized for this study. The FAME and Osiris databases were employed to obtain annual reports for the firms for periods ranging from 2011-2018. For companies not found within these databases, their annual reports were sourced directly from their websites. The oil and gas reserve quantity information was then extracted individually from the annual reports. These databases were also employed to obtain financial information such as Earnings, cash flows from operations and total assets. Data on historical monthly share price for the firms was gotten from uk.investing.com. Finally, UK Brent historical monthly oil price data was obtained from the US Energy Information Administration website (www.eia.gov)

The analysis and interpretation of the collected data is provided in chapter four. Appendix 6 provides the sample data set. This data can be assessed publicly and it is possible for any subsequent researcher to replicate and test it under the set criteria. A possible limitation is that the data required further processing and computation to match the research design described in section 3.4. the adjusted data set is captured in Appendix 5.

## 3.8 CREDIBILITY OF RESEARCH FINDINGS

As suggested by Raimond (1993), to ensure the quality of the research design such that the data evidence gathered and the conclusions drawn are logical and can stand up to close scrutiny (Raimond, 1993), the validity and reliability of this research are assessed.

### 3.8.1 VALIDITY

According to Saunders, Lewis and Thornhill (2016, p. 202), validity assesses the level of appropriateness of measures applied, accuracy of results analysed and generalization of the findings. Research findings are invalid when such findings are arrived at falsely or the reported relationship is inaccurate.

Firstly, the value-free and objective nature of positivism philosophy and quantitative method applied in this study promotes certainty. Secondly, care has been taken to limit instability and inconsistency of measurements arising from biases or errors on the part of the researcher and for the variables under consideration. Also statistical packages such as SPSS for data analysis has been employed to further boost accuracy.

Thirdly, data analysed has been built upon the extant verified theoretical and empirical framework explored in chapter two. Finally, the statistical generalization and conclusions made entirely refer to the sample under study which is the FTSE ALL SHARE and FTSE AIM ALL SHARE E&P oil and Gas firms listed on the LSE.

The researcher will prove the validity of findings through an accurately defined empirical relationship between component of oil and gas quantity disclosures and share price.

### 3.8.2 RELIABILITY

Reliability refers to the ability of a research design to be replicated by a subsequent researcher and still achieve consistency with findings (Saunders, Lewis and Thornhill 2016, p. 202)

The sample information, published annual reports and other credible data from the stock exchange and financial databases can be publicly accessible and thus verified. In addition, the statistical analysis employed also permits the researcher to conduct the analysis repeatedly to ensure that results can be relied upon.

The certainty level preferred for this study is set at 95% level of significance. This implies that should the sample in section 3.6.1 be selected 100 times, 95 of the sample should certainly represent the characteristics of the sample frame under consideration.

## 3.9 ETHICAL CONSIDERATION

Generally ethical principles during research requires that the researcher should uphold high ethical standard in course of conducting the study. The study should not subject any group to any form of harm, embarrassment or any substantial damage or disadvantage. There should be privacy of respondents. Decision of a target sample to refuse consent to data collection should be respected and the research should be conducted within the strict university research ethical procedures. However, this research had negligible human interaction as data was pooled from secondary sources.

## 3.10 LIMITATIONS OF RESEARCH DESIGN

The researcher depended on companies' annual reports to access oil and gas reserve information. As addressed in section 2.4.2, discretionary accounting reporting policies of reserve quantity information may affect the level and quality of disclosures.

There is a limitation of sample size due to restriction to only E&P upstream firms given the nature of the study. This study utilized only 25 purely E&P firms listed within both the LSE FTSE All-Share index and FTSE AIM All-Share index. This has also affected the number of variables that can be utilized in the regression model, as a small sample extended over a large number of variables affects the validity and reliability of findings.

Finally, a limitation of data availability for certain firms in certain time periods was evident given the year they became listed on the LSE. This led to consideration of a time frame below 10 years. This also resulted in the exclusion of certain firms from the sample size.

During data collection, reserve information for certain firms was not explicitly disclosed in the supplementary or additional information sections. However, they were dispersed around the operational and strategic review sections. Also, some reserve component information were not aggregated properly but disclosed as daily rates across different types of reserves. The research therefore had to calculate the aggregate values.

# 4 CHAPTER FOUR: DATA ANALYSIS AND PRESENTATION

## 4.1 INTRODUCTION

The sample data in this study consists of E&P oil and gas companies listed on the London Stock Exchange Aim and Main Market for the period 2011-2018. On the londonstockexchange.com, the FTSE All-Share Index and the FTSE AIM All Share-Index both had a total of 105 oil and gas companies. The profile of these companies was examined individually from www.financialtimes.com and uk.investing.com to understand the nature of business. 30 Companies identified as service, equipment, logistics and management or downstream were eliminated at this point from the list, leaving a sample frame of 75 strictly E&P companies from which a sample of 25 companies was finally selected using the criteria identified in Section 3.6.1 for the period 2011-2018, constituting 181 firm-year observations.

The research strategy and design utilized secondary data from several sources identified in section 3.7.2 to empirically test the predicted variables on share returns. The relationships between the independent and dependent variables were expressed using statistical modelling. SPSS data analysis tool was used to determine trends, correlations and other relationships to either validate or invalidate the predicted relationships and provide a deductive assessment for the research problem. All the statistics were set at 95% significance level.

The rest of this chapter reveals the descriptive and correlation statistics, the trend analysis of the explanatory variables and regression results. The conclusions and recommendations of this study are discussed in chapter five.

## 4.2 DESCRIPTIVE AND CORRELATION STATISTICS

### Descriptive Statistics

The researcher conducted an initial descriptive statistics assessment on the computed data set in Appendix 5 and the results showed a high level of skewness and kurtosis indicating abnormality (Table 4.1).

Table 4. 1: Initial descriptive statistics

**Descriptive Statistics**

| | | Rit | ResC | Exp | Acq | Pro | Rev | Sal | OP | Risk | CF | D1 | D2 | D3 |
|---|---|---|---|---|---|---|---|---|---|---|---|---|---|---|
| N | Valid | 181 | 181 | 181 | 181 | 181 | 181 | 181 | 181 | 181 | 181 | 181 | 181 | 181 |
| | Missing | 0 | 0 | 0 | 0 | 0 | 0 | 0 | 0 | 0 | 0 | 0 | 0 | 0 |
| Mean | | .26321 | .3679 | .2481 | .1882 | -.087 | .0401 | -.025 | .0201 | .258 | .044 | .34 | .88 | .85 |
| Median | | -.0748 | .0000 | .0000 | .0000 | -.056 | .0000 | .0000 | -.0275 | -.08 | .039 | .00 | 1.00 | 1.00 |
| Std. Deviation | | 2.6500 | 2.430 | 1.834 | 1.435 | .1414 | .3682 | .2046 | .2697 | 2.65 | .112 | .474 | .328 | .357 |
| Variance | | 7.023 | 5.905 | 3.363 | 2.060 | .020 | .136 | .042 | .073 | 7.02 | .013 | .225 | .107 | .128 |
| Skewness | | 11.377 | 8.378 | 12.10 | 11.93 | -4.11 | 3.993 | .238 | -.311 | 11.4 | -.524 | .695 | -2.336 | -1.986 |
| Std. Error of ... | | .181 | .181 | .181 | .181 | .181 | .181 | .181 | .181 | .181 | .181 | .181 | .181 | .181 |
| Kurtosis | | 141.28 | 74.51 | 154.6 | 151.2 | 20.79 | 26.57 | 35.52 | -.799 | 141 | 3.72 | -1.534 | 3.494 | 1.966 |
| Std. Error of Kurtosis | | .359 | .359 | .359 | .359 | .359 | .359 | .359 | .359 | .359 | .359 | .359 | .359 | .359 |
| Minimum | | -.8836 | -1.00 | -.270 | .0000 | -1.00 | -1.000 | -1.00 | -.4714 | -.89 | -.51 | 0 | 0 | 0 |
| Maximum | | 33.596 | 23.42 | 23.84 | 18.55 | .0306 | 2.795 | 1.643 | .3976 | 33.6 | .349 | 1 | 1 | 1 |

***Rit**- is share return, **ResC** – is change in reserves, **Exp**- is exploration, **Acq**- is acquisition, **Pro**- is production, **Sal**- is sales, **OP**- is change in oil price, **Risk** – is the firm risk and **CF** – is cashflows from operations scaled to total assets. **D1, D2, D3** are the dummy variables. **D1** for market listed (main or aim), **D2** for no. of reserve change component disclosed (greater or equal to 2 or less) and **D3** for no. of reserve KPI disclosed (greater or equal to 1or less) All reserve variables have been scaled to reserves at the beginning of the year.*

For diagnostic purpose, the researcher performed an initial regression on each of the three models and discovered that in confirmation of the irregular initial descriptive statistics result, the Mahalanobis and Cook's maximum distance statistics showed a large number of multivariate outliers in the data sets greatly impacting on the results. The P values for each of mahal distances for each specific model using the cumulative chi-square was calculated to identify and remove the cases with outliers. 15 Cases with Mahal. p values less than 0.05 were excluded (see Appendix 4). Using this refined data, another descriptive statistic was conducted (Table 4.2) showing improved results.

Table 4. 2: Final descriptive statistics

**Descriptive Statistics**

| | | Rit | ResC | Exp | Acq | Pro | Rev | Sal | OP | risk | CF | D1 | D2 | D3 |
|---|---|---|---|---|---|---|---|---|---|---|---|---|---|---|
| N | Valid | 166 | 166 | 166 | 166 | 166 | 166 | 166 | 166 | 166 | 166 | 166 | 166 | 166 |
| | Missing | 0 | 0 | 0 | 0 | 0 | 0 | 0 | 0 | 0 | 0 | 0 | 0 | 0 |
| Mean | | -.004 | .089 | .0843 | .070 | -.068 | -.010 | -.008 | .0137 | -.011 | .048 | .35 | .90 | .87 |
| Median | | -.074 | .000 | .0000 | .000 | -.056 | .000 | .000 | -.028 | -.079 | .041 | .00 | 1.00 | 1.00 |
| Std. Deviation | | .512 | .518 | .2968 | .320 | .079 | .199 | .040 | .2727 | .512 | .100 | .478 | .296 | .340 |
| Skewness | | 2.08 | 3.14 | 6.907 | 7.74 | -3.09 | -1.52 | -6.96 | -.312 | 2.10 | .142 | .638 | -2.76 | -2.187 |
| Std. Error of Skewness | | .188 | .188 | .188 | .188 | .188 | .188 | .188 | .188 | .188 | .188 | .188 | .188 | .188 |
| Kurtosis | | 7.37 | 14.2 | 59.85 | 68.3 | 14.1 | 11.5 | 51.8 | -.817 | 7.46 | 1.39 | -1.6 | 5.687 | 2.818 |
| Std. Error of ... | | .375 | .375 | .375 | .375 | .375 | .375 | .375 | .375 | .375 | .375 | .375 | .375 | .375 |
| Minimum | | -.884 | -1.00 | .0000 | .000 | -.590 | -1.00 | -.329 | -.4714 | -.887 | -.309 | 0 | 0 | 0 |
| Maximum | | 2.65 | 2.94 | 3.013 | 3.26 | .031 | .867 | .0098 | .3976 | 2.65 | .349 | 1 | 1 | 1 |

***Rit**- is share return, **ResC** – is change in reserves, **Exp**- is exploration, **Acq**- is acquisition, **Pro**- is production, **Sal**- is sales, **OP**- is change in oil price, **Risk** – is the firm risk and **CF** – is cashflows from operations scaled to total assets. **D1, D2, D3** are the dummy variables. **D1** for market listed (main or aim), **D2** for no. of reserve change component disclosed (greater or equal to 2 or less) and **D3** for no. of reserve KPI disclosed(greater or equal to 1or less) All reserve variables have been scaled to reserves at the beginning of the year.*

Although skewness and kurtosis are still present in the variables, it is improved compared to the previous results. Also, the high standard deviation statistics observed from the previous result reduced drastically to further improve the normality of the data set.

For all the LSE listed E&P firms in the sample, the mean share return which captured the average firm performance for the sector was low at -0.4% for the period under study.

The mean change in reserves was 8.94% from 2011-2018. This implied that the companies in the sample experienced considerable growth in reserves within this period. Most of the increase was from exploration and acquisition with arithmetic mean of 8.43% and 7.0% respectively while reserves changes from production, revisions and sales constituted the decrease in reserves with averages of -6.8%, -1% and -0.8% respectively. This established exploration to be the major source of reserve replacement and production to be the major factor for downward changes in reserves for the companies within the period.
Change in oil price had a mean of 13.7% showing considerable volatility. Volatility in oil price has been attributed to be an impact factor for movement in reserve components (Misund, 2018). The average risk for the firms was -1.1% and the average cash flows from operations scaled-down total assets was 4.8%.

Table 4. 3: Descriptive statistics and charts of dummy variables.

| | | Frequency | % |
|---|---|---|---|
| **D1** | FTSE AIM All Share | 107 | 64.8 |
| | FTSE All Share | 59 | 35.2 |
| **D2** | Less than 2 Reserve Component | 16 | 9.1 |
| | Greater or equal to 2 Reserve Component | 150 | 90.9 |
| **D3** | No Reserve KPI | 22 | 12.7 |
| | Greater or equal to 1 Reserve KPI | 144 | 87.3 |

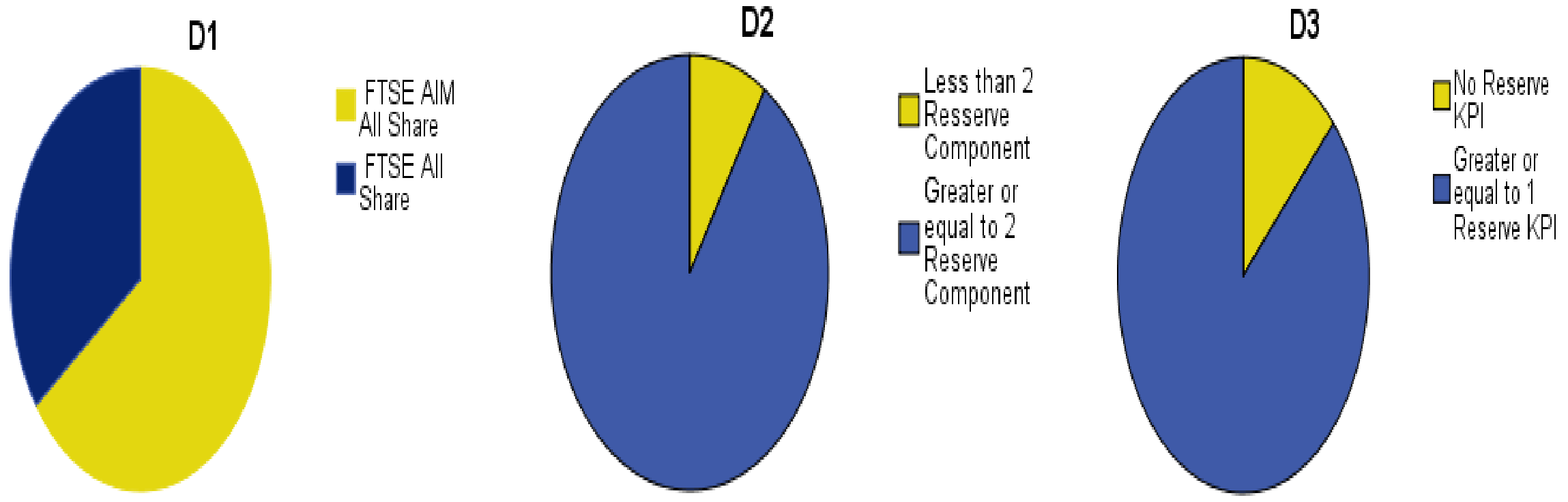

Table 4.3 above shows the descriptive statistics for the dummy variables. For D1, the number of E&P firms listed on the AIM market was more at 64.8% compared to those listed on the Main Market at 35.2%. The statistic for D2 revealed that 90.9% of LSE listed E&P firms disclosed at least two change in reserves component information. Similarly, the D3 results show a high number of E&P firms in the LSE (87.3%) also disclosed reserve KPI information. This result is contrary to the findings of McChelery et al. (2015) who found poor reserve disclosure for UK Oil and Gas firms. However, it is supported by the findings of Odo et al. (2016) who revealed that a high level of reserve disclosures existed for UK oil and gas firms.

## Correlation

Table 4.4 below, shows the correlation for the independent and dependent variables used in the statistical analysis shown by the correlation coefficient

Table 4. 4: Pearson Correlation Coefficients for the Variables

| Correlations | | | | | | | | | | | | |
|---|---|---|---|---|---|---|---|---|---|---|---|---|
| | **Rit** | **ResC** | **Exp** | **Acq** | **Pro** | **Rev** | **Sal** | **OP** | **risk** | **CF** | **D1** | **D2** |
| **Rit** | 1.000 | | | | | | | | | | | |
| **ResC** | 0.162 | | | | | | | | | | | |
| **Exp** | 0.171 | 0.552 | | | | | | | | | | |
| **Acq** | -0.016 | 0.576 | -0.024 | | | | | | | | | |
| **Pro** | -0.070 | -0.026 | -0.178 | 0.000 | | | | | | | | |
| **Rev** | 0.093 | 0.345 | 0.039 | -0.043 | -0.166 | | | | | | | |
| **Sal** | 0.008 | 0.042 | 0.023 | -0.030 | 0.020 | -0.102 | | | | | | |
| **OP** | 0.336 | 0.019 | 0.003 | 0.099 | 0.047 | -0.005 | 0.029 | | | | | |
| **risk** | 0.014 | 0.156 | 0.169 | -0.020 | -0.069 | 0.084 | 0.008 | 0.312 | | | | |
| **CF** | 0.017 | -0.038 | 0.198 | -0.115 | -0.092 | -0.016 | -0.083 | 0.066 | 0.013 | | | |
| **D1** | -0.089 | -0.113 | -0.087 | -0.086 | -0.061 | 0.144 | -0.273 | 0.009 | -0.073 | 0.437 | | |
| **D2** | -0.167 | -0.011 | 0.083 | 0.071 | -0.279 | -0.031 | -0.066 | -0.184 | -0.168 | 0.335 | 0.197 | |
| **D3** | -0.119 | -0.138 | 0.068 | -0.025 | -0.200 | -0.034 | -0.079 | -0.111 | -0.121 | 0.312 | 0.137 | 0.414 |

***Rit***- *is share return,* ***ResC*** *– is change in reserves,* ***Exp***- *is exploration,* ***Acq***- *is acquisition,* ***Pro***- *is production,* ***Sal***- *is sales,* ***OP***- *is change in oil price,* ***Risk*** *– is the firm risk and* ***CF*** *– is cashflows from operations scaled to total assets.* ***D1, D2, D3*** *are the dummy variables.* ***D1*** *for market listed (main or aim),* ***D2*** *for no. of reserve change component disclosed (greater or equal to 2 or less) and* ***D3*** *for no. of reserve KPI disclosed (greater or equal to 1or less) All reserve variables have been scaled to reserves at the beginning of the year.*

The above analysis shows small correlation between the dependent variable (Share return) and the independent variables. Amongst the explanatory variables, reserve change via exploration had the highest (positive) correlation with share return at 0.171. change in oil price was the control variable having the highest correlation with share return at 0.336. It was also observed that certain correlations were contrary to previous empirical predictions. For instance, Reserve change via Acquisitions, D2 and D3 had a negative correlation with share returns while Reserve change via Sales had a positive correlation with share returns. However, this will be further examined by applying rigorous statistical modelling and analytical tools and assessing the significance of the outcomes as will be revealed in section 4.4.

There is also an observable low correlation coefficient between the independent variables. The highest are the correlation between reserve change and acquisitions at 0.576 and reserve change and exploration at 0.552. The general low correlations among the independent variables are acceptable as it represents the absence of strong relationships between the independent variables resulting in multicollinearity. This could cause difficulty in separating the effect of the independent variables on the dependent variable (Saunders, Lewis and Thornhill 2016). This is corroborated by the collinearity tolerance values and variance inflation factor (VIF) for each of the models as shown in the SPSS output under Appendix 3. All the collinearity tolerance values are greater than 0.1 and the VIFs are within the acceptable value of less than 10 indicating no multicollinearity.

## 4.3 TREND ANALYSIS

To validate the statistical observations of the results in section 4.2, the arithmetic mean observations for each of the different reserve variable on a yearly basis across all the firms in the sample was determined as shown in Appendix 2a. The researcher generated times series graphs to further understand the movement of these variables for the period under study using change in oil price as the major control variable. All reserve variables have been scales to reserve quantity at the beginning of the year. Appendix 2b shows different trend graphs plotted individually for each reserve variable against returns and oil price for further clarity. However, a summary of these trends for all the reserve variables is shown below for ease of comparison.

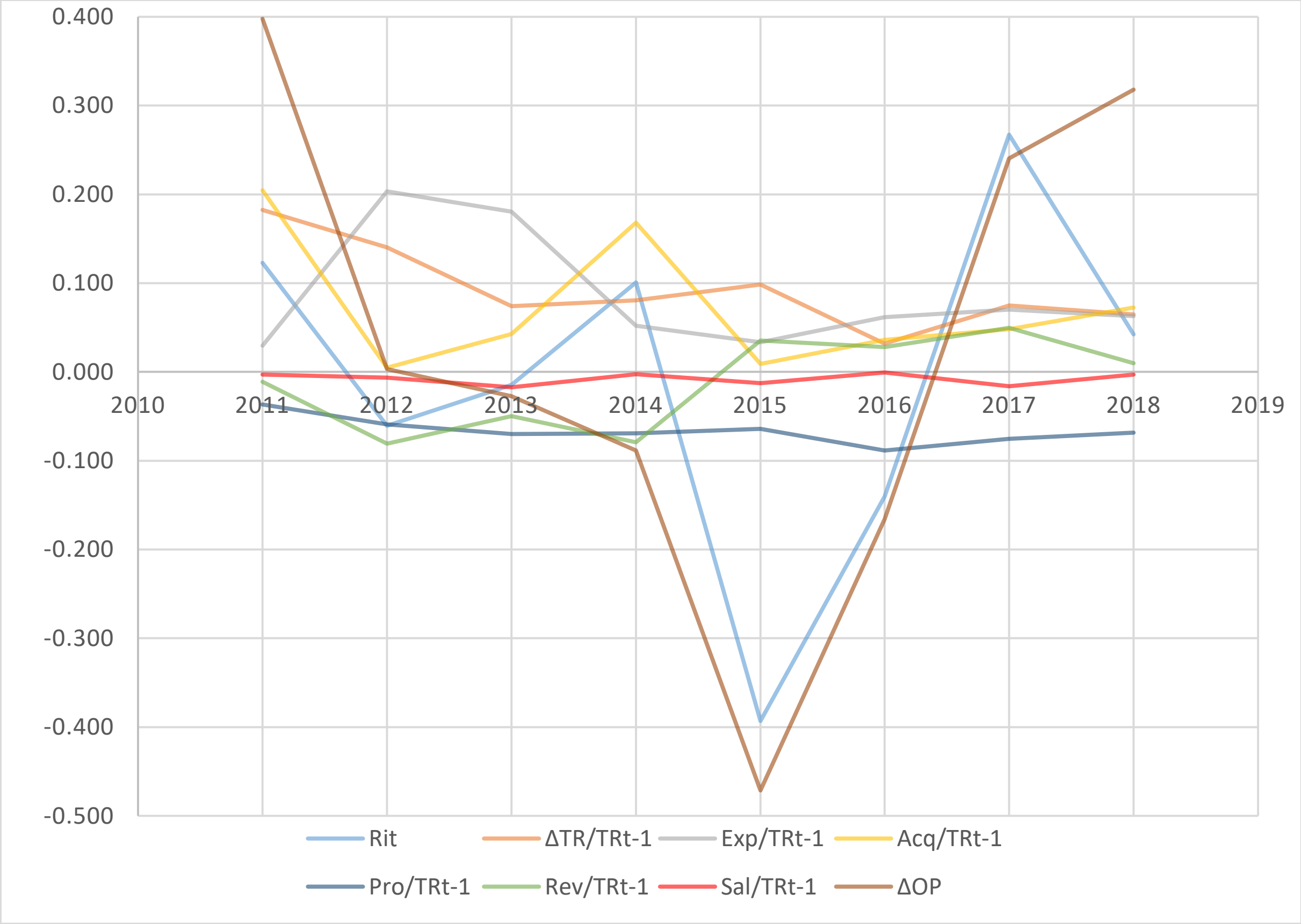

Figure 4. 1: Graph of Reserve variables, share returns and oil price

Observing the trend for share returns and changes in reserves, growth in reserve has been low and fairly consistent throughout the period, however it has fallen by more than 50% from 18.2% in 2011 to 6.5% in 2018. Average share returns have moved closely in relation to changes in the price of oil. The statistics reveal no improvement in sector performance for the period under study. A rapid decline in oil price and share returns is observed in 2012 and from 2014 to 2016. This decline was also reflected in the changes in reserves for those periods. The decline in oil price for those periods is attributed to supply glut caused by booming shale oil production and fall in the demand of oil from emerging economies.

Looking at the trend for share returns and changes in reserves due to exploratory activities, high reserve growth in exploration is experienced from 2011-2013, and a constant decline for periods thereafter. This can also be attributed to the movement in the price of crude, which is seen to be symmetrical with share returns. However, changes in reserves due to exploration has improved from 2.9% in 2011 to 6.3% in 2018.

Studying the movement for share returns and changes in reserves due to acquisitions, share returns have moved quite similarly to reserve acquisitions for the period under study. Between 2012 and 2013 marked a decline in reserve acquisitions. However, reserve acquisition was on the increase for the period thereafter up until 2014 which saw a steep decline hitting zero levels in 2015. Reserve acquisitions have been consistently low from 2015 till date. Generally, between 2011 and 2018 reserve acquisitions have fallen by about 60% from 20.5% in 2011 to 7.2% in 2018.

Examining the trend for share returns and changes in reserves due to production reveals that the levels of reserve reductions through production have been fairly consistent between 2011 and 2018. However, there is a slight decrease in production between 2014 and 2016 marking periods of low oil prices and share returns. On the average production has increased from 3.7% in 2011 to 6.8% in 2018.

Inspecting the trend for share returns and changes in reserves due to revisions, a constant and low negative revision to reserves between 2011 towards the start of 2015 is observed, while a positive but low revision is recorded between 2015 and 2018. Generally, revisions to reserves show a slight increase from -1.1% in 2011 to 1% in 2018

Finally, by analysing the trend for share returns and changes in reserves due to sales, it is revealed that the level of reserve sale has been consistently very low for the period 2011-2018. Change in reserves due to sales remains unchanged from 2011 to 2018 at 0.3%.

## 4.4 REGRESSION ANALYSIS

Based on the three models in the research design under section 3.4, three different OLS regression analyses were performed to test the relationship between share returns and changes in reserve quantity components. As pointed out in section 4.2, the regression has been carried using the refined data set after removing the identified multivariate outliers (see Appendix 4) from the computed data set which was significantly impacting on the analysis. The next set of tables (Table 4.5- 4.7) presents the results of the empirical estimation of the models, showing the impact of the explanatory variables on the dependent variable.

### 4.4.1 CHANGE IN OIL AND GAS RESERVES IMPACT ON SHARE PRICE

For the first model of the study, the regression equation to denote the relationship between share returns and changes in total reserves quantity was derived as:

$$R_{it} = C + \beta_1 \Delta TR_{it} + \beta_2 CF_{it} + \beta_3 X_{it} + \beta_4 \Delta OP_t + \beta_5 D_{i1} + \varepsilon_{it} \qquad \text{(Eq. 3)}$$

The hypothesis for this model is stated as follows:

$H_0$ : No statistically significant relationship between share returns and changes in reserves.

$H_1$ : Presence of a statistically significant relationship between share returns and changes in reserves.

The regression output in Table 4.5 below corresponds to Eqn. (3) above. The Durbin-Waston statistics is 2.002 which is close to the acceptable value of 2, hence there is no serial autocorrelation between the residuals of the variables used in the model.

At 95% confidence level, the model is statistically significant to predict the outcomes accurately as the p-value of the F-Statistics is less than 5%.

The adjusted $R^2$ is at 61.8% indicating that 61.8% of the variance in share returns is explained by the predictor variables of the model at a 95% confidence level. This suggested that the explanatory variable, Change in Reserves (**ResC**) has an effect on share returns.

Table 4. 5: Change in oil and gas reserves and share returns regression

**Model Summary**[b]

| Model | R | R Square | Adjusted R Square | Std. Error of the Estimate | Durbin-Watson |
|---|---|---|---|---|---|
| 1 | .796[a] | 0.634 | 0.618 | 2.6887232 | 2.002 |
| a. Predictors: (Constant), D1, ΔOP, ΔEarnings, ResC, CF | | | | | |
| b. Dependent Variable: Rit | | | | | |

**ANOVA**[a]

| Model | | Sum of Squares | df | Mean Square | F | Sig. |
|---|---|---|---|---|---|---|
| 1 | Regression | 42.820 | 5 | 8.564 | 1846.208 | .000[b] |
| | Residual | 0.499 | 160 | 0.003 | | |
| | Total | 43.319 | 165 | | | |
| a. Dependent Variable: Rit | | | | | | |
| b. Predictors: (Constant), D1, OP, ResC, risk, CF | | | | | | |

**Coefficients**[a]

| Model | | Unstandardized Coefficients | | Standardized Coefficients | | | 95.0% Confidence Interval for B | |
|---|---|---|---|---|---|---|---|---|
| | | B | Std. Error | Beta | t | sig. | Lower Bound | Upper Bound |
| 1 | (Constant) | 0.006 | 0.006 | | 1.144 | 0.254 | -0.005 | 0.017 |
| | ResC | 0.098 | 0.009 | 0.068 | 0.909 | 0.365 | -0.009 | 0.025 |
| | OP | 0.574 | 0.017 | 0.428 | 3.114 | **0.002** | 0.019 | 0.086 |
| | risk | 0.395 | 0.009 | 0.283 | 10.449 | **0.000** | 0.367 | 0.562 |
| | CF | 0.031 | 0.048 | 0.046 | 0.635 | 0.526 | -0.065 | 0.126 |
| | D1 | -0.107 | 0.010 | -0.105 | 2.720 | **0.042** | -0.075 | -0.433 |
| a. Dependent Variable: Rit | | | | | | | | |

*__Rit__- is share return, __ResC__ – is change in reserves, __Risk__ – is the firm risk and __CF__ – is cashflows from operations scaled to total assets. __D1__ is dummy variable for market listed (main or aim), All reserve variables have been scaled to reserves at the beginning of the year.*

ResC had a positive relationship with share returns with a coefficient of 0.098, implying that for one unit increase in reserves, share returns increases by 9.8%. However, it had a p value of 0.365 which was greater than the critical value of 5%, suggesting a weak and statistically insignificant relationship.

For the control variables, change in oil price had a positively significant relationship with share returns ($p<0.05$) with a coefficient of 0.574 indicating that for a one-unit increase in oil price, share returns increased by 0.574. Risk also had a positive relationship with share returns with a coefficient of 0.395, which was statistically significant indicating that one unit increase in risk resulted in share returns increasing by 0.395. D1 which is a proxy for market listing had a negatively significant relationship with share returns suggesting that E&P firms listed in the main market had their share returns further reduced by 10.7% compared to AIM-listed firms. Cashflows from operations had a positively insignificant relationship with share returns with a coefficient of 0.031.

Conclusively, the null hypothesis for this model is not rejected as the result is not statistically significant to accept the alternate hypothesis. This corroborates the findings of Harris and Ohlson (1987), Doran et al. (1988), Alciatore (1993) and Spear (1994).

## 4.4.2 EXPLORATORY ACTIVITIES, ACQUISITIONS, PRODUCTION AND REVISIONS IMPACT ON SHARE PRICE.

In accordance with the second model of the study, the regression equation to denote the relationship between share returns and changes in total reserves quantity components was derived as:

$$R_{it} = C + \beta_1 TR_EXP_{it} + \beta_2 TR_ACQ_{it} + \beta_3 TR_PRO_{it} + \beta_4 TR_REV_{it} + \beta_5 TR_SAL_{it} + \beta_6 CF_{it} + \beta_7 X_{it} + \beta_8 \Delta OP_t + \beta_9 D_{i1} + \varepsilon_{it} \quad \text{(Eqn. 4)}$$

The hypothesis for this model is stated as follows:

$H_0$ : No statistically significant relationship between Share returns and reserve from exploration, acquisition, production, revisions and sales

$H_1$ : Presence of a statistically significant relationship between share returns and reserve from exploration, acquisition, production, revisions and sales

The regression output in Table 4.6 below corresponds to Eqn. (4) above. The Durbin-Waston statistics is 2.055 reveals no serial autocorrelation between the residuals of the variables used in the model.

The model is statistically significant at 95% confidence level to accurately predict the outcomes as the p-value of the F-Statistics is less than 5% critical value.
The adjusted $R^2$ is at 66.8% indicating that 66.8% of the variance in share returns is explained by the predictor variables of the model.

The change in reserves variable (ResC) is disaggregated into components which constitutes the explanatory variables for this model.

Table 4. 6: Change in reserves components and share returns regression

**Model Summary**[b]

| Model | R | R Square | Adjusted R Square | Std. Error of the Estimate | Durbin-Watson |
|---|---|---|---|---|---|
| 2 | .818[a] | 0.669 | 0.668 | 0.0563440 | 2.055 |

a. Predictors: (Constant), D1, OP, Pro, Acq, Exp, Rev, Sal, risk, CF
b. Dependent Variable: Rit

**ANOVA**[a]

| Model | | Sum of Squares | df | Mean Square | F | Sig. |
|---|---|---|---|---|---|---|
| 2 | Regression | 42.824 | 9 | 4.758 | 898.822 | .000[b] |
| | Residual | 0.495 | 156 | 0.003 | | |
| | Total | 43.319 | 165 | | | |

a. Dependent Variable: Rit
b. Predictors: (Constant), D1, OP, Pro, Acq, Exp, Rev, Sal, risk, CF

**Coefficients**[a]

| Model | | Unstandardized Coefficients | | Standardized Coefficients | t | Sig. | 95.0% Confidence Interval for B | |
|---|---|---|---|---|---|---|---|---|
| | | B | Std. Error | Beta | | | Lower Bound | Upper Bound |
| 2 | (Constant) | 0.007 | 0.007 | | 1.014 | 0.312 | -0.007 | 0.020 |
| | Exp | 0.049 | 0.016 | 0.038 | 0.198 | 0.843 | -0.028 | 0.034 |
| | Acq | 0.028 | 0.014 | 0.021 | 0.131 | 0.896 | -0.026 | 0.029 |
| | Pro | -0.013 | 0.057 | -0.002 | -0.220 | 0.826 | -0.126 | 0.100 |
| | Rev | 0.029 | 0.023 | 0.011 | 1.259 | 0.210 | -0.016 | 0.074 |
| | Sal | -0.029 | 0.116 | -0.002 | -0.253 | 0.801 | -0.258 | 0.200 |
| | OP | 0.561 | 0.017 | 0.483 | 3.088 | **0.002** | 0.019 | 0.087 |
| | risk | 0.394 | 0.009 | 0.253 | 10.272 | **0.001** | 0.266 | 0.752 |
| | CF | 0.034 | 0.051 | 0.030 | 0.673 | 0.502 | -0.067 | 0.136 |
| | D1 | -0.101 | 0.011 | -0.090 | -2.796 | **0.047** | 0.173 | 0.331 |

a. Dependent Variable: Rit

***Rit**- is share return, **ResC** – is change in reserves, **Exp**- is exploration, **Acq**- is acquisition, **Pro**- is production, **Sal**- is sales, **OP**- is change in oil price, **Risk** – is the firm risk and **CF** – is cashflows from operations scaled to total assets. **D1** is dummy variable for market listed (main or aim), All reserve variables have been scaled to reserves at the beginning of the year.*

The coefficient of the explanatory variables conformed to previous empirical predictions. Exploration, acquisitions and revisions had a positive coefficient of 0.049, 0.028 and 0.029 respectively, suggesting a positive relationship with share returns. Productions and sales had a negative coefficient of -0.013 and -0.029 respectively, indicating a negative relationship with share returns. These results are consistent with the finding of Alciatore (1993), Spear (1994), Berry and wright (2001), Cormier and Magnan (2002), Bird et al. (2013), Sabet and Heaney (2016), Misund (2018). However, the p values of all the explanatory variables were statistically insignificant as they exceeded the critical value of 5%, thus an indication of a weak relationship between these variables and share returns.

For the control variables, the statistics still remained constant as with the first model. oil price and risk maintained a positive significant relationship, Cashflows, a positive insignificant relationship and D1, a negative significant relationship with share price.

The insignificance of the results of the reserve variables could be attributed to several factors. Firstly, according to Spear (1994), an insignificant relationship may be explained by the fact that oil price has the most significant effect on share returns and also drives the activities of individual components of changes to reserves. Therefore, the signals conveyed by the components of changes in reserve quantity to the market participants may be so small individually compared to the significant effect of oil price changes.

The second factor is due to the structure of the model. As pointed out in section 2.2, the study adopted the measurement perspective rather than the signalling perspective to determine the value relevance of reserves disclosures. Previous studies as reviewed in section 2.8 adopted the signalling perspective by using share returns accumulated the week following the release of the annual reports or by using the lagged returns which therefore significantly captured the reserve disclosure effects, however the researcher argues that such measure of share returns does not fully capture the average firm performance for the period.

This study has therefore employed the average annual share returns to measure the impact of reserve disclosures on the overall firm performance for the period. However, this approach may have resulted in the insignificance of the results.

From the foregoing, the null hypothesis of this model is accepted as the results are not statistically significant.

### 4.4.3 DISCLOSURE EFFECT ON SHARE PRICE

For the third model of the study, the regression of the equation is derived as

$$R_{it} = C + \beta_1 TR_EXP_{it} + \beta_2 TR_ACQ_{it} + \beta_3 TR_PRO_{it} + \beta_4 TR_REV_{it} + \beta_5 TR_SAL_{it} + \beta_6 CF_{it} + \beta_7 X_{it} + \beta_8 \Delta OP_t + \beta_9 D_{i1} + \beta_{10} D_{i2} + \beta_{11} D_{i3} + \varepsilon_{it} \quad (5)$$

Essentially, this third equation was derived by adjusting the second model of the study to measure the effects of the quality of reserve disclosures on share returns by introducing two additional dummy variables, D2 and D3:

Such that,

$D_{i2}$ is set to 1 where the firm discloses more than 2 components accounting for change in the reserve quantity or 0 otherwise

$D_{i3}$ is set to 1 where the firm disclosures one or more reserve KPI or 0 otherwise.

Table 4. 7: Disclosure effect and share returns regression

| Model Summary[b] | | | | | |
|---|---|---|---|---|---|
| Model | R | R Square | Adjusted R Square | Std. Error of the Estimate | Durbin-Watson |
| 3 | .824[a] | 0.679 | 0.678 | 0.0566746 | 2.051 |
| a. Predictors: (Constant), D3, Acq, Rev, Exp, Sal, OP, Pro, D1, risk, D2, CF | | | | | |
| b. Dependent Variable: Rit | | | | | |

| ANOVA[a] | | | | | | |
|---|---|---|---|---|---|---|
| Model | | Sum of Squares | df | Mean Square | F | Sig. |
| 3 | Regression | 42.825 | 11 | 3.893 | 689.062 | .000[b] |
| | Residual | 0.495 | 154 | 0.003 | | |
| | Total | 43.319 | 165 | | | |
| a. Dependent Variable: Rit | | | | | | |
| b. Predictors: (Constant), D3, Acq, Rev, Exp, Sal, OP, Pro, D1, risk, D2, CF | | | | | | |

| Coefficients[a] | | | | | | | | |
|---|---|---|---|---|---|---|---|---|
| Model | | Unstandardized Coefficients | | Standardized Coefficients | t | Sig. | 95.0% Confidence Interval for B | |
| | | B | Std. Error | Beta | | | Lower Bound | Upper Bound |
| 3 | (Constant) | 0.000 | 0.017 | | 0.025 | 0.980 | -0.033 | 0.033 |
| | Exp | 0.041 | 0.016 | 0.032 | 0.194 | 0.846 | -0.028 | 0.035 |
| | Acq | 0.021 | 0.014 | 0.018 | 0.077 | 0.939 | -0.027 | 0.029 |
| | Pro | -0.005 | 0.060 | -0.001 | -0.081 | 0.935 | -0.124 | 0.114 |
| | Rev | 0.029 | 0.023 | 0.020 | 1.282 | 0.202 | -0.016 | 0.075 |
| | Sal | -0.027 | 0.117 | -0.021 | -0.235 | 0.815 | -0.258 | 0.203 |
| | OP | 0.565 | 0.018 | 0.429 | 3.097 | **0.002** | 0.020 | 0.089 |
| | risk | 0.384 | 0.009 | 0.283 | 9.882 | **0.001** | 0.266 | 0.726 |
| | CF | 0.027 | 0.055 | 0.019 | 0.482 | 0.630 | -0.082 | 0.135 |
| | D1 | -0.124 | 0.011 | -0.104 | -2.614 | **0.046** | -0.033 | -0.011 |
| | D2 | 0.106 | 0.018 | 0.092 | 2.549 | **0.044** | 0.100 | 0.142 |
| | D3 | 0.115 | 0.015 | 0.094 | -2.191 | **0.043** | 0.140 | 0.814 |
| a. Dependent Variable: Rit | | | | | | | | |

***Rit***- *is share return,* ***ResC*** *– is change in reserves,* ***Exp***- *is exploration,* ***Acq***- *is acquisition,* ***Pro***- *is production,* ***Sal***- *is sales,* ***OP***- *is change in oil price,* ***Risk*** *– is the firm risk and* ***CF*** *– is cashflows from operations scaled to total assets.* ***D1, D2, D3*** *are the dummy variables.* ***D1*** *for market listed (main or aim),* ***D2*** *for no. of reserve change component disclosed (greater or equal to 2 or less) and* ***D3*** *for no. of reserve KPI disclosed (greater or equal to 1or less) All reserve variables have been scaled to reserves at the beginning of the year.*

The regression output in Table 4.7 above corresponds to Eqn. (5) above. The Durbin-Waston statistics is 2.051 reveals no serial autocorrelation between the residuals of the variables used in the model.

The model is statistically significant at 95% confidence level to accurately predict the outcomes as the p-value of the F-Statistics is less than 5% critical value.
The adjusted $R^2$ is at 67.8% indicating that 67.8% of the variance in share returns is explained by the predictor variables of the model.

The result revealed that D2 had a positive significant relationship with share returns. This means that where a firm disclosed above two reserve change components, it boosted share returns by 10.6%. D3 also had a positive significant relationship with share returns indicating that where a firm disclosed at least one reserves KPI, share return was boosted by 11.5%. This is supported by the findings of Slack et al. (2010), McChelery et al. (2015) and Odo et al. (2016). These results therefore suggested a significant relationship between the quality of reserves disclosures and share returns.

It is also observed that the reserve component variables and the control variables still maintained the same relationship with share returns as with the second model. From the foregoing, the null hypothesis of this model is rejected as the results on the quality of disclosures is statistically significant.

Conclusively from all the statistical observations in all three models, there is no evidence to support the hypothesis that the changes in reserves as well as the components of reserve changes, have a significant impact on the share returns. It can therefore be implied that although reserves disclosures may influence investors decision at a particular point in time as observed in previous studies, it is not significant enough to impact on the average share returns of LSE listed E&P firms over a longer time interval, especially given the more significant impact of oil prices.

The next chapter provides a critical discussion of the research findings in trying to achieve the research aims and objective. It draws relevant conclusions in the light of underpinning theories and opinions of previous empirical studies.

# 5 CHAPTER FIVE: CRITICAL DISCUSSION, SUMMARY AND CONCLUSION

## 5.1 INTRODUCTION

This is the concluding chapter and the most crucial part of the study. It aims to critically discuss the results in light of previous researches and theories. To present the findings, the discussion will be divided into two main sections. The first section is the general discussion integral to the overall report and the second section provides for a summary of finding and critical analysis, linking the results to extant studies and theories. Conclusions are subsequently drawn and recommendations proffered.

## 5.2 GENERAL DISCUSSION

The oil and gas industry possess several unique characteristics distinguishing it from every other industry (Power et al, 2017). A notable feature is the high levels of risk and uncertainty in exploring for and commercially exploiting oil and gas reserves. This feature has posed an accounting dilemma in the reporting of reserves quantity disclosures (McChelery et al., 2015).

Reserve disclosure information is fundamental to understanding the value of a firm (Wright, 2001; Boone, 2002; Asekomeh, Russell, Tarbert, & Lawal, 2010) as oil and gas reserves constitute the underlying asset and lifeblood of oil and gas E&P companies. However, there is a very limited framework regulating reserve quantity disclosures, most especially within the UK. In line with the IASB recognition criteria for assets, reserves are not captured as asset in the financial statements due to their imprecise nature and difficulty in estimation. They are been subjected to voluntary supplementary disclosures given that the accounting standards and regulations have specified no mandatory disclosure requirements. The UK OIAC only recommends best practices to guide the quality of reserve disclosures. This has placed less importance on reserves which constituted a crucial element in influencing investors' pricing of the firm. (Rai, 2006).

According to Misund (2018), the uncertainty associated with reserves quantities posed a source of confusion for investors whose investment decision largely relied on such information disclosure. Therefore, the central theme for this research was contingent on the premise that should reserve disclosure requirement be value relevant, it followed that the information on reserve quantity disclosures contained in the annual report should reflect on the share price.

Value relevance studies of oil and gas reserves have been explored by previous researchers majorly for USA, Canada and Australia and very few for UK (Bell 1975; Harris and Ohlson 1987; Magliolo, 1986; Spear, 1994; 1996; Ohlson, 1995; Berry, Hasan, & O'Bryan, 1998; Berry and wright, 2001; Boyer & Filion, 2007; Misund & Osmundsen, 2017). Within this narrow band of studies, most have focused on the value relevance of changes in reserves only. Studies on the content of those changes are very sparse and limited to the signalling perspective rather than the measurement perspective (Spear, 1994; Alciatore, 1993; Clinch and Magliolo, 1992; Cormier and Magnan 2002; Coleman, 2005, Olsen et al, 2011, Scholtens and Wagenaar, 2011, Bird et al, 2013; Sabet and Heaney, 2016; Edwin and Thompson, 2016; Misund, 2016; 2017; 2018; Gray et al, 2019, Costabile et al., 2019, Hellstrom, 2006).

Upon this background, a study was conducted with aimed at providing critical evaluation of the measurement of the value relevance of changes in reserves including the contents accounting for such changes as well as the value relevance of the quality of such contents for the UK upstream oil and gas firms listed on the LSE.

The objectives set out to be achieved were as follows:

- To examine the relationship between oil and gas reserves and share price.
- To investigate the value relevance of each individual component of oil and gas reserve quantity adjustments (acquisition announcements, revisions, exploration, production and sales)
- To evaluate the impact of accounting policies and regulations for the UK jurisdiction on reserve disclosure requirements and assess the value relevance of reserve disclosure quality.
- To highlight the implications of adjustments to components of oil and gas reserves for investors, E&P companies and the Petroleum industry.

Incentive theories for oil and gas reserves disclosures such as the information asymmetry theory, agency theory, legitimacy theory, signalling theory and obfuscation hypothesis were explored to better understand the value relevance of oil and gas reserves disclosures. The following principles were derived from these theories:

- An agency relationship exists between the management of E&P firms and shareholders.
- Presence of information asymmetry on reserves disclosure may exist between E&P firm managers and market participants resulting in incorrect pricing of the firm.
- Monitoring and data verification cost to reduce reserve uncertainty constitutes agency costs thus, E&P firms will be driven to disclose reserves information where the perceived benefit of doing so via reductions in risk and contracting cost with agents exceeds the agency cost.
- E&P firms are motivated to voluntarily disclose reserves information as it accords them legitimacy by stakeholders to smoothly continue their operations.
- E&P companies are driven to reduce information asymmetry through an increased level of discretionary disclosure in order to signal their competitive strength through the communication of reserve information to the market so as to improve firm performance.
- E&P managers may, however, be incentivised to obscure, withhold or postpone negative reserves signals while accentuating only positive signals to manipulate market participants' perception of the firm's performance.

With reference to the assumptions of these theories and closely following the Ohlson (1995) and Misund (2018) framework, the multifactor model approach was adopted to formulate the research models. Three research models were derived utilizing statistical modelling technique to express the relationship between average share returns and contents of reserves disclosures as prescribed in research design in section 3.4, in order to achieve the research objectives outlined above. The models speculated a significant relationship between share returns and the predictor variables. Average annual share returns constituted the dependent variable. Change in reserves, exploration, acquisitions, production, revisions and sales of reserves constituted the explanatory variables. The control variables where change in oil price, equity risk, cashflows from operations and the dummy variables (firm size, number of reserve change content disclosed and number of reserves KPI).

A sample of 25 E&P oil and gas companies listed in the AIM and Main Markets of the LSE for the period 2011-2018 constituting 181 firm-year observations was employed. Secondary data was obtained from several sources to empirically test the predominant variables.

To this end, SPSS analysis tool was used to describe the variables, determine trends, correlations and other relationships that either validated or invalidate the predicted relationships in order to provide for a deductive assessment to the research problem. All the statistics were set at 95% significance level.

## 5.3 SUMMARY OF FINDINGS AND CRITICAL DISCUSSION

The presentation of results commenced with the correlation, descriptive and trend analysis of the collected data set. According to the correlation matrix, the independent variables generally were not highly correlated to each other which was desirable to neutralise multicollinearity. The correlation values were below 30%, however, certain pair of independent variables exceeded this level at a range of 30% to 57% given that such variables moved in the same directions and were influenced by similar factors. For instance, change in reserves was correlated more with acquisitions and explorations. Others were, the number of reserves component disclosed and the number of reserves KPI disclosed, change in oil price and equity risk and finally cashflows from operations, number of reserves component disclosed and firm size.

The correlation matrix also revealed small correlations between the dependent variable (Share return) and all the independent variables with the exception of a control variable; change in oil price which had significant correlation suggesting that the share returns of UK listed E&P firms was highly influenced by oil price volatility. (Taamouti et al, 2017; Shaeri, Adaoglu and Katircioglu 2016, Misund 2018). It was also observed that certain correlations were contrary to previous empirical predictions. For instance, Reserve change via acquisitions, the number of reserves component disclosed and the number of reserves KPI disclosed were negatively correlated with share returns while Reserve change via Sales had a positive correlation with share returns. These results were therefore subjected to further rigorous statistical analysis for validation.

The descriptive statistical results revealed that the mean share return which captured the average firm performance for the E&P sector was low at -0.4% for the period 2011-2018. This is in confirmation with the recent findings of Misund (2018) as well as the low trend of returns reported by the Oil and Gas UK business outlook, 2019. A major reason for this has been linked to the changes in oil prices. This study which recorded a mean change in oil price at 13.7% combined with the high correlation between oil price and share return and the trend analysis for oil price changes for the period under study clearly points to the existence of oil price volatility. The decline in oil price is attributed to supply glut caused by booming shale oil production and fall in the demand of oil from emerging economies.

However, the sector witnessed considerable growth in reserves as shown by the mean value of 8.94%. The results highlighted that majority of this growth was due to exploratory activities while production accounted chiefly for the decline in reserves. The trend analysis for the period 2011 to 2018 declared that change in reserves fell by more than 50% from 18.2% to 6.5%, exploratory reserves improved from 2.9% to 6.3%, reserve acquisitions fell by 60% from 20.5 to 7.2%, production increased from 3.7% to 6.8%, revisions to reserves increase from -1.1% to 1% and sales remained unchanged at 0.3%. The mean risk and cash flows from operations for the sector was low at -1.1% and 4.8%.

The descriptive statistics also disclosed that the number of E&P firms listed on the AIM market was more at 64.8% compared to those listed on the Main Market at 35.2%. This may be explained by the firm size. It meant that larger firms which were listed on the main market were less in number and smaller firms listed on the AIM market were more. It could also be argued that aside firm size, the criteria for being listed in the FTSE Main market was more stringent and therefore limited entry of more firms. These criteria are manifested by draconian listing rules and continuing obligations such as the minimum free float set at 25%, minimum market capitalization of £700,000 and high administrative costs constituting a regulatory burden on the firms. The result would be more firms being admitted into the AIM market or being delisted from the Main market into the AIM market.

Finally, the descriptive statistics also affirmed that there was a high level of reserve disclosure for E&P firms in the UK. It also disclosed that AIM-listed firms disclosed as much reserve information as Main listed firms. Generally, 90.9% of the listed firms were found disclose at least two change in reserves component information and similarly 87.3% of them also disclosed reserve KPI information. This result was contrary to the findings of McChelery et al. (2015) who found poor reserve disclosure for UK Oil and Gas firms. However, it is supported by the findings of Odo et al. (2016) who revealed that a high level of reserve disclosures existed for UK oil and gas firms.

Three statistical data analysis for the three models of the study was performed to validate the results of the correlation, descriptive and trend analysis. To achieve the first objective of the study, the first analysis evaluated the relationship between average yearly share returns and changes in oil and gas reserves quantity. The OLS results suggested that change in reserves had a positive relationship with share returns. This implied that an increase in reserves resulted in an increase in share returns. This also supported the Pearson correlation results earlier obtained. However, the t-test statistic was not significant at 5% level suggesting a weak and statistically insignificant relationship leading to an acceptance of the null hypothesis. This corroborates the findings of Harris and Ohlson (1987), Doran et al. (1988), Alciatore (1993) and Spear (1994). According to their findings, the net reserve change is a sum of the effect of several components and these components may offset each other leading to a cumulative insignificant effect on share returns.

Change in oil price and equity risk had a significant positive relationship which was consistent with the correlation, descriptive and trend analysis result previously examined. Market listing proxied as firm size (D1) has a negative significant relationship with share returns suggesting that AIM-listed E&P firms recorded higher returns compared to Main listed firms. This was consistent with the findings from the descriptive statistics previously discussed which proposed that Main listed E&P firm may face more market disincentives compared to AIM-listed E&P firms leading to lower share returns (Nielsson, 2013).

To attain the second objective of the study, the second OLS analysis evaluated the relationship between average yearly share returns and components accounting for changes in oil and gas reserves quantity. The empirical results revealed exploration, acquisition and revisions to have a positive relationship with share returns and production and sales, a negative relationship. This validated the Pearson's correlation statistics. These results are also consistent with the finding of Alciatore (1993), Spear (1994), Berry and wright (2001), Cormier and Magnan (2002), Bird et al. (2013), Sabet and Heaney (2016), Misund (2018). However, the results were all statistically insignificant at 5% level leading to an acceptance of the null hypothesis. The control variables were observed to maintain the same results as with the first analysis.

The insignificance of the results could be attributed to several factors. Firstly, according to Spear (1994), an insignificant relationship may be explained by the fact that oil price has the most significant effect on share returns and also drives the activities for the individual components of changes to reserves. Therefore, the signals conveyed by the components of changes in reserve quantity to the market participants may be so small individually compared to the significant effect of oil price changes.

The second factor is due to the structure of the model. As pointed out in section 2.2, the study adopted the measurement perspective rather than the signalling perspective to determine the value relevance of reserves disclosures. Previous studies as reviewed in section 2.8 adopted the signalling perspective by using share returns accumulated the week following the release of the annual reports or by using the lagged returns which therefore significantly captured the reserve disclosure effects, however the researcher argues that such measure of share returns does not fully capture the average firm performance for the period. This study has therefore employed the average annual share returns to measure the impact of reserve disclosures on the overall firm performance for the period. However, this approach may have resulted in the insignificance of results.

Finally, to accomplish the third objective of the study, the third analysis evaluated the relationship between average yearly share returns and quality of reserves disclosure by including two more dummy variables representing the number of reserves component disclosed and the number of reserves KPI disclosed. Both variables were found to be positive and significantly related to share returns. This was consistent with their descriptive statistics previously analysed. These outcomes are consistent with the findings of Slack et al. (2010), McChelery et al. (2015) and Odo et al. (2016). The null hypothesis is therefore rejected suggesting the quality of reserve disclosures affected share returns. All other variables were still observed to maintain the same results as with the second analysis.

## 5.4 CONCLUSION

Summarily, the empirical findings of all three models suggest that changes in oil and gas reserves, as well as the contents of those changes, have an effect on share results, though, statistically insignificant. However, quality of reserves disclosures had a statistically significant effect on share returns. There was a high level of disclosure for UK listed E&P firms. The AIM-listed firms were found to disclosure as much reserve information as main listed firms. The market listing played a significant explanatory role in the share returns of E&P firms. Oil price and equity risk were also found to have a significant effect on share returns. These findings have been supported by previous empirical studies.

An explanation for the insignificant of the results for the major explanatory variables can be attributed to the significant oil price effect compared with the smaller individual effect of the reserve variables (Alciatore, 1993). It is also due to the limitation of longitudinal effect posed by the model which applies the measurement rather than the signalling approach (Hellstrom, 2006) to capture the effect of the reserves information on average annual share returns over a long period of time.

Other limitations identified for the study include data unavailability for certain firms in certain time periods leading to a consideration of a time frame below 10 years, Small sample size due to restriction to only E&P upstream firms given the nature of the study and the exclusion of certain firms from the sample size due to data unavailability. Also, during data collection, reserve information for certain firms were not explicitly disclosed in the supplementary or additional information sections. However, they were dispersed around the operational and strategic review sections. Also, some reserve component information was not aggregated properly but disclosed as daily rates across different types of reserves. The research had to, therefore, estimate the aggregate values.

An implication of these results raises the question of how beneficial the reserves information disclosed by listed E&P firms in the UK is to financial users. Although the results are insignificant to explain share returns of the firms, it does not isolate the fact that reserves information is a highly useful indicator for the firm's profitability and cashflows, aiding financial users to make informed decisions.

At the firm level, the benefits of reserve information disclosure for the firm is evidenced through the aiding of access to the capital market funding and influencing merger, acquisition and divestiture decisions. Even the national level, the level of reserves disclosure of firms will enable the national regulatory bodies to get first-hand statistics to evaluate its reserves and formulate better policies for the sector.

## 5.5 RECOMMENDATIONS

Given the results of the study, the following recommendations are made:

- The IASB should integrate with UK's FRC to formulate comprehensive accounting standards for reserves disclosure in order to bridge the diversity of reserves reporting among the UK listed upstream oil and gas firms and provide more value relevant information for financial users.

- UK E&P firms should consider the option of diversification given the poor share returns recorded for the sector due to persistent volatility in oil prices.

- Exploratory efforts should be complemented by increasing the acquisition of reserves (which fell during the period under study) in order to boost net reserve growth which is observed to have reduced for the sector by 50% within the last 8 years.

As regards considerations for future research, an extended study could be carried out for the entire population of E&P firms on the LSE, by including all E&P firms listed regardless of the value of the share price. However, this kind of research would be cross-sectional for a very short time period so as not be hindered by data unavailability. Also, more measures of the quality of reserves disclosures as identified by UK SORP (s246-s251) and OFR (p77) may be incorporated into such a study as this current study has examined only two of them.

## APPENDICES

### Appendix 1: Crude oil prices 1861- 2015

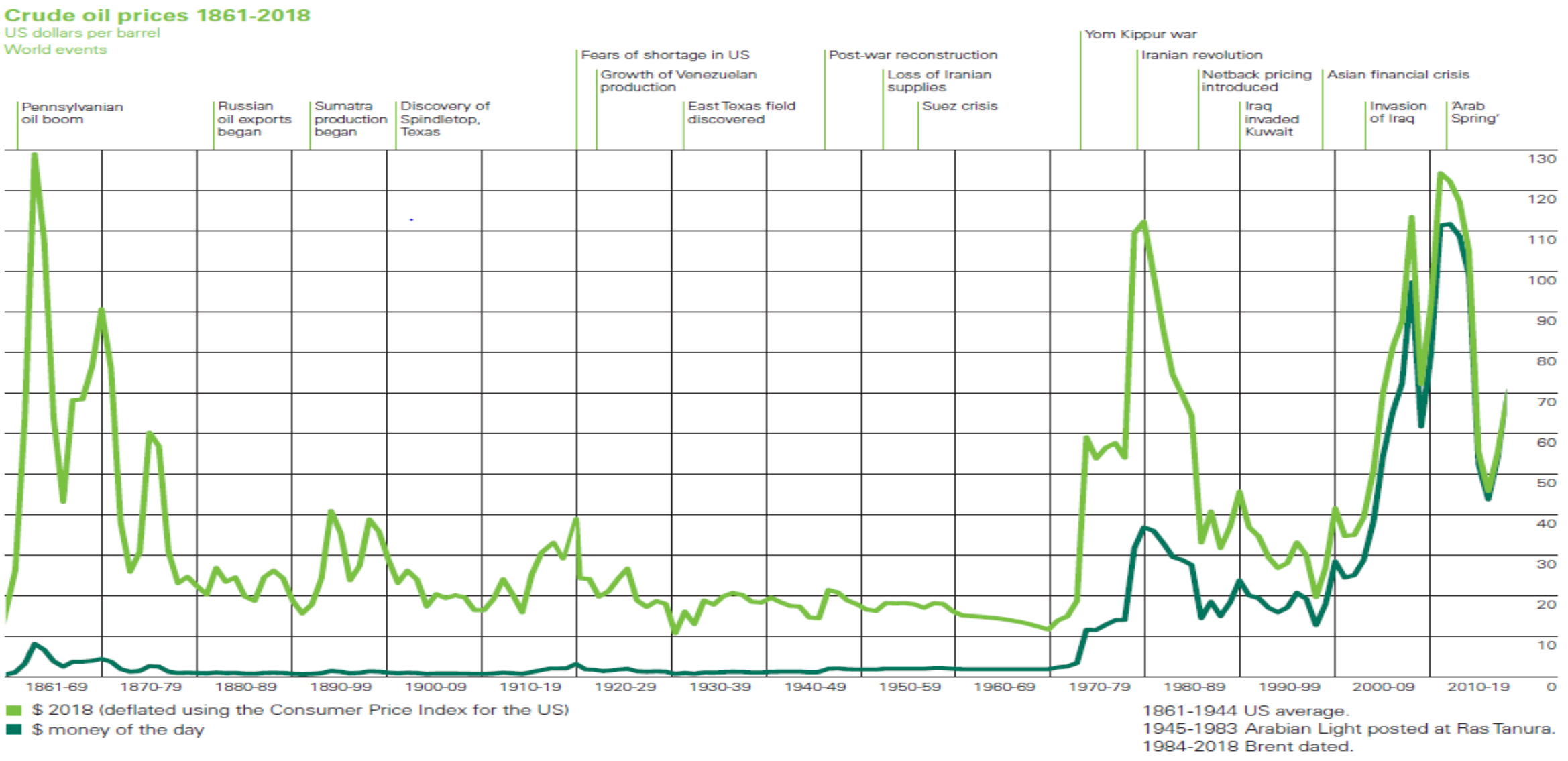

Source: BP Statistical Review of World Energy 2019

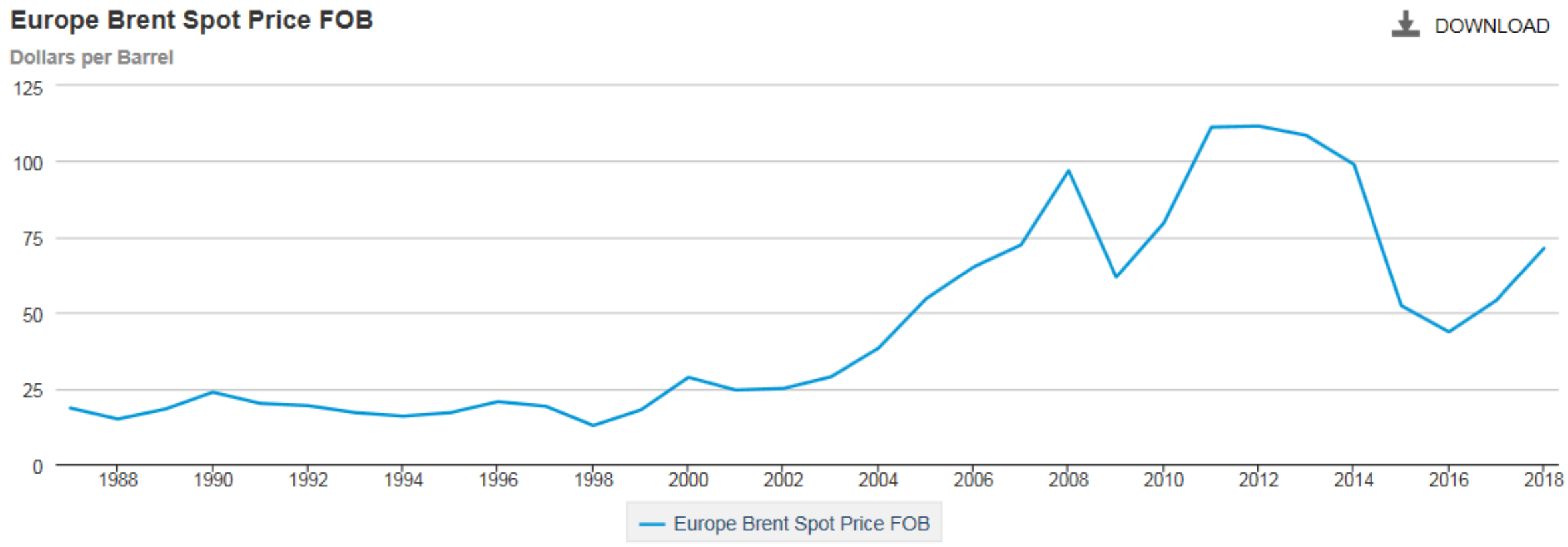

Source: US Energy Information Administration

### Appendix 2a: Average Annual reserve quantity, oil price and risk-free rate for E&P upstream oil and gas companies listed on the London stock exchange

| year | Rit | ΔTR | Exp | Acq | Pro | Rev | Sal | ΔOP | RF |
|---|---|---|---|---|---|---|---|---|---|
| 2018 | 0.042 | 0.065 | 0.063 | 0.072 | -0.068 | 0.010 | -0.003 | 0.3179 | 0.940016 |
| 2017 | 0.267 | 0.075 | 0.070 | 0.048 | -0.075 | 0.050 | -0.016 | 0.2404 | -1.30773 |
| 2016 | -0.140 | 0.032 | 0.062 | 0.036 | -0.088 | 0.028 | -0.001 | -0.1659 | -0.24563 |
| 2015 | -0.393 | 0.098 | 0.033 | 0.009 | -0.064 | 0.035 | -0.013 | -0.4714 | 0.965753 |
| 2014 | 0.101 | 0.081 | 0.052 | 0.168 | -0.069 | -0.079 | -0.003 | -0.0883 | -0.9311 |
| 2013 | -0.015 | 0.074 | 0.180 | 0.043 | -0.070 | -0.050 | -0.017 | -0.0275 | 0.200944 |
| 2012 | -0.061 | 0.140 | 0.203 | 0.005 | -0.059 | -0.081 | -0.007 | 0.0033 | -0.76034 |
| 2011 | 0.123 | 0.182 | 0.029 | 0.205 | -0.037 | -0.011 | -0.003 | 0.3976 | 2.253869 |

## Appendix 2b: Individual Trend

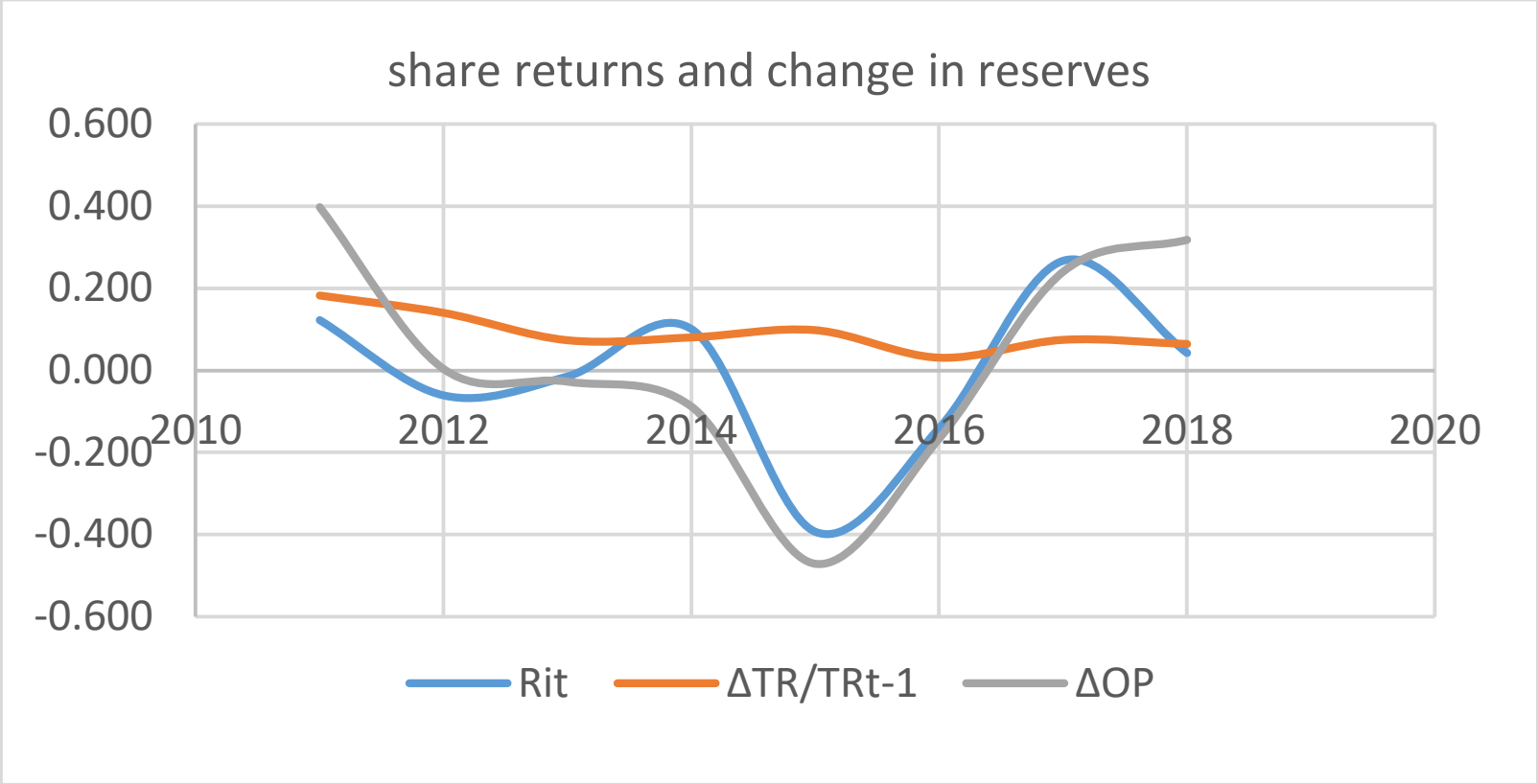

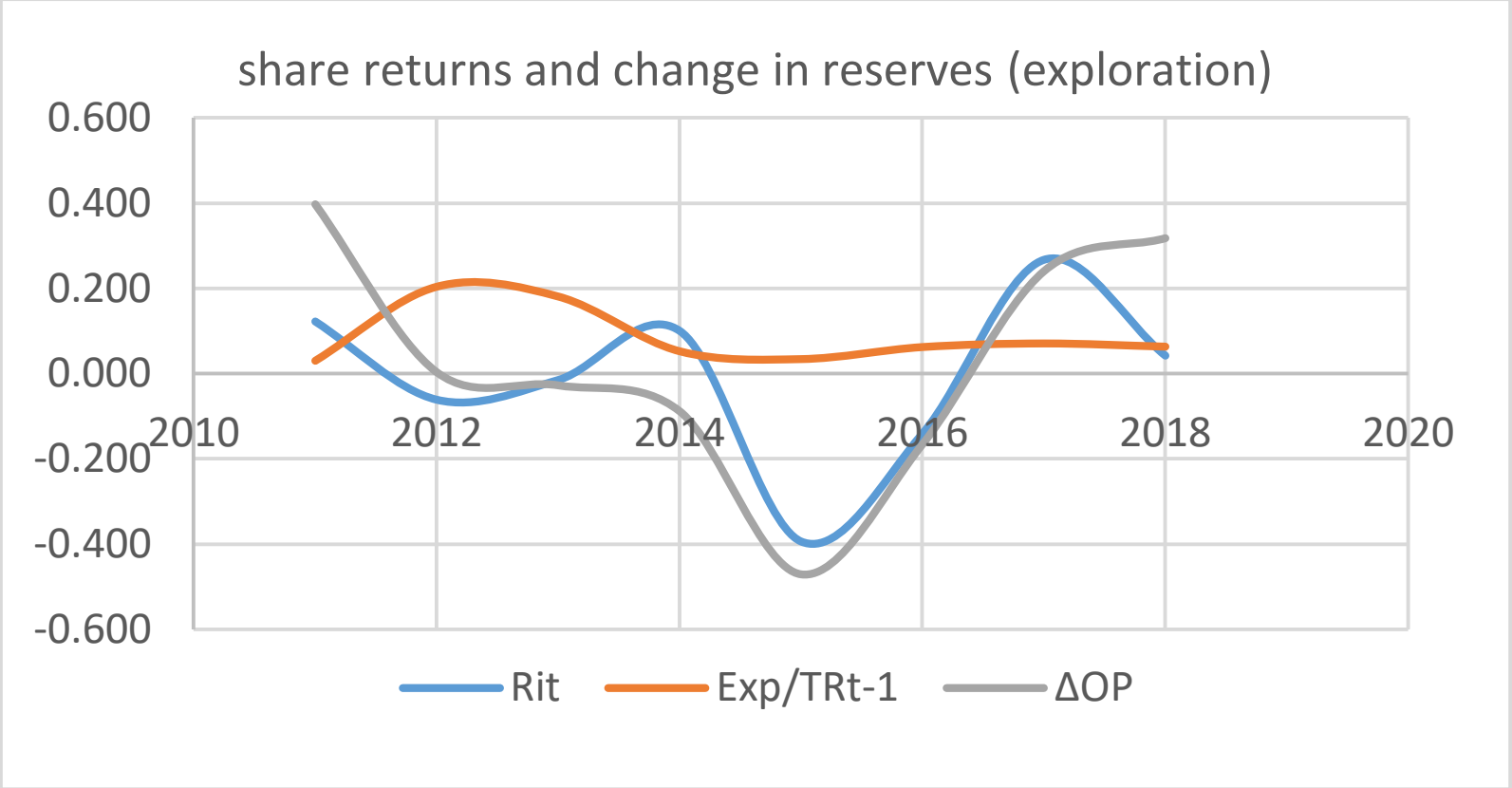

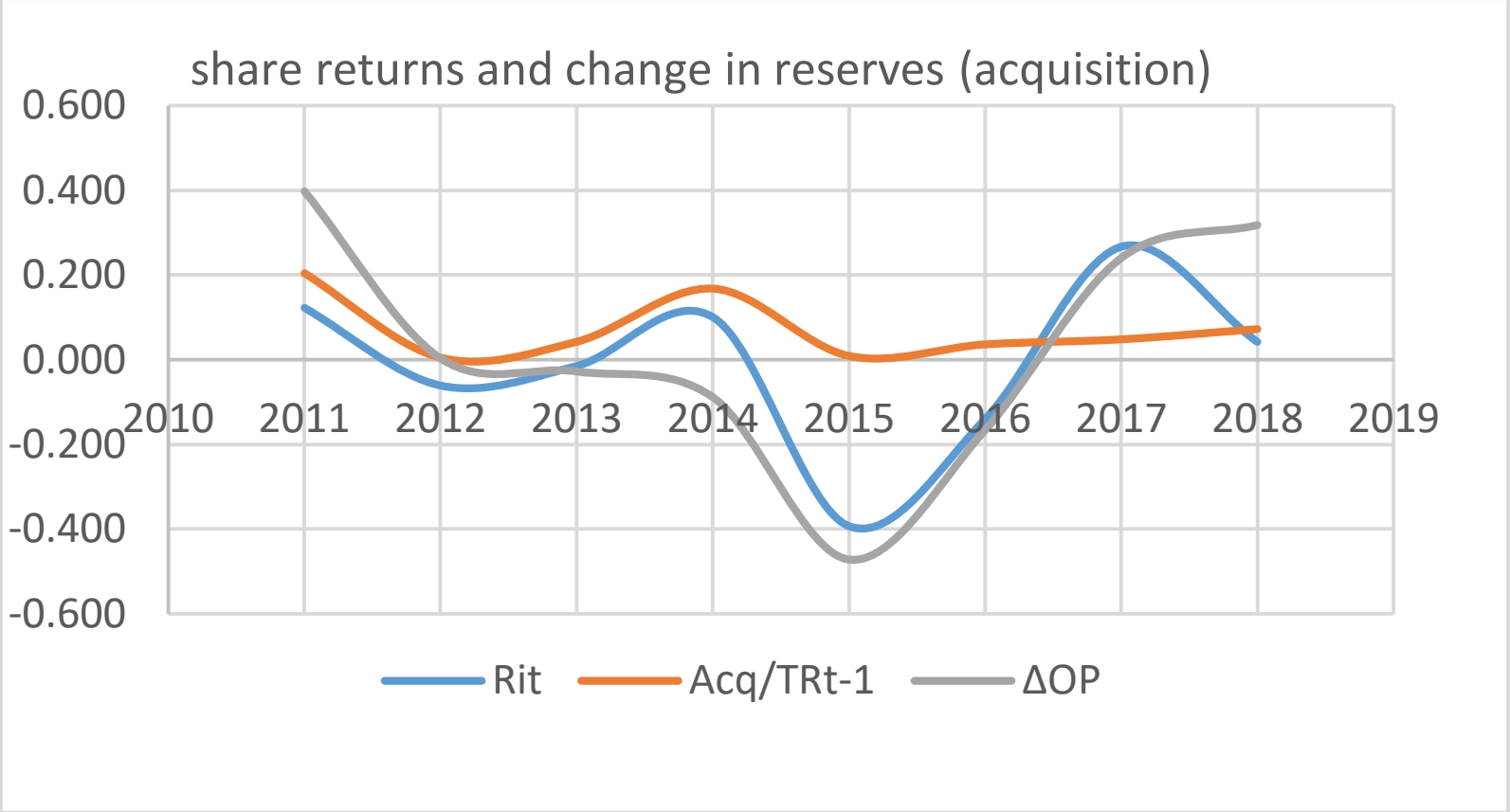

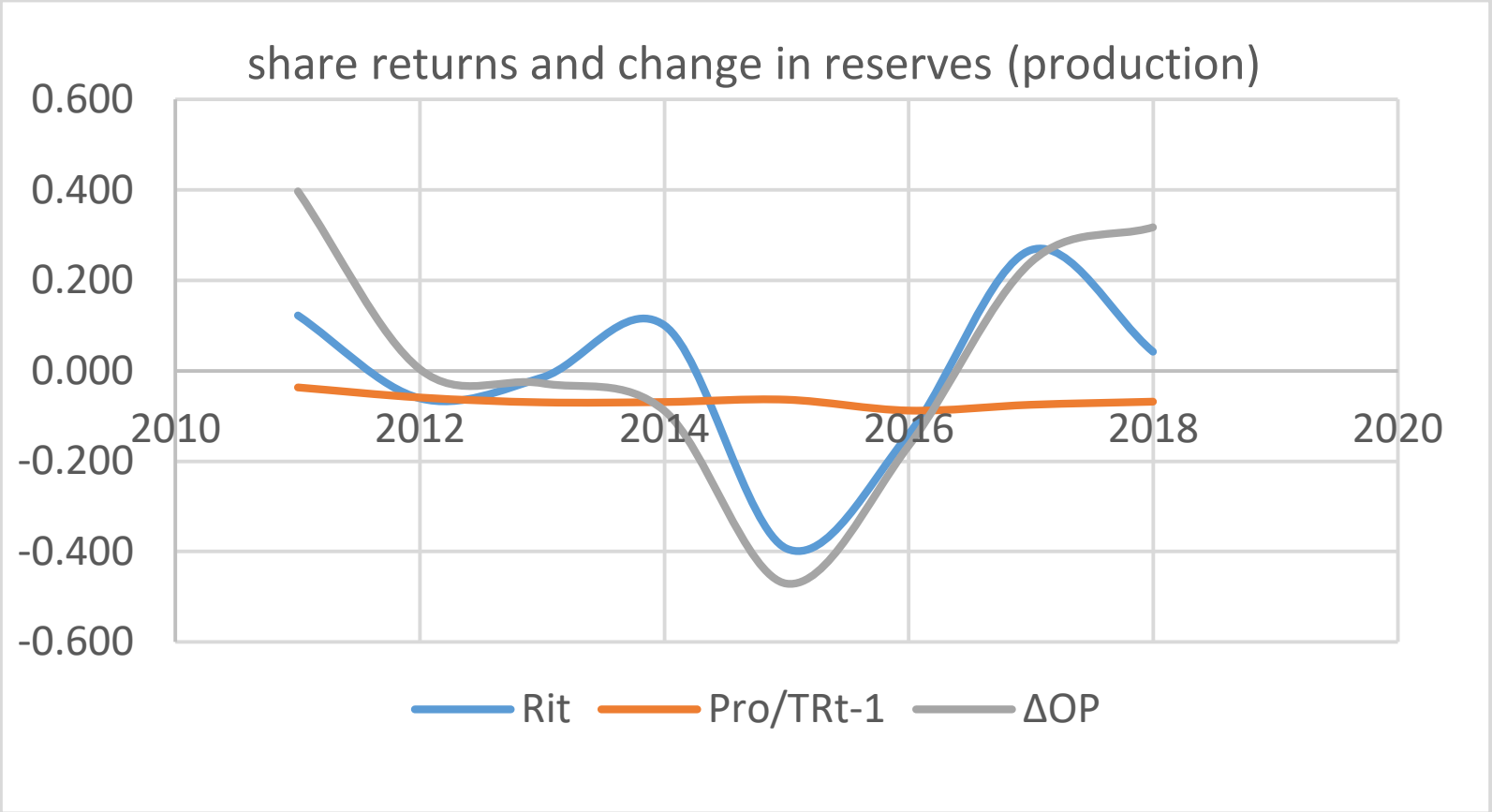

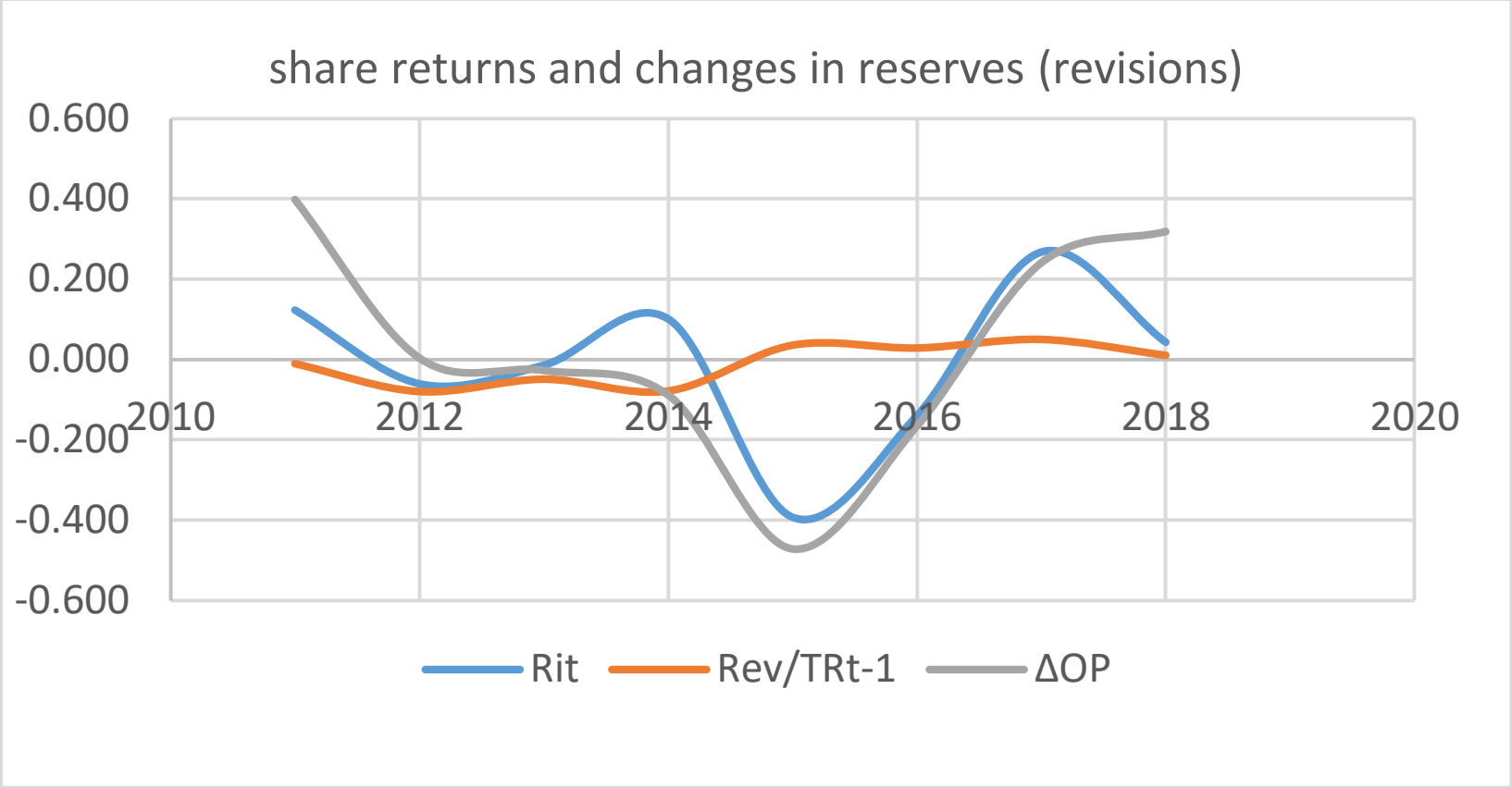

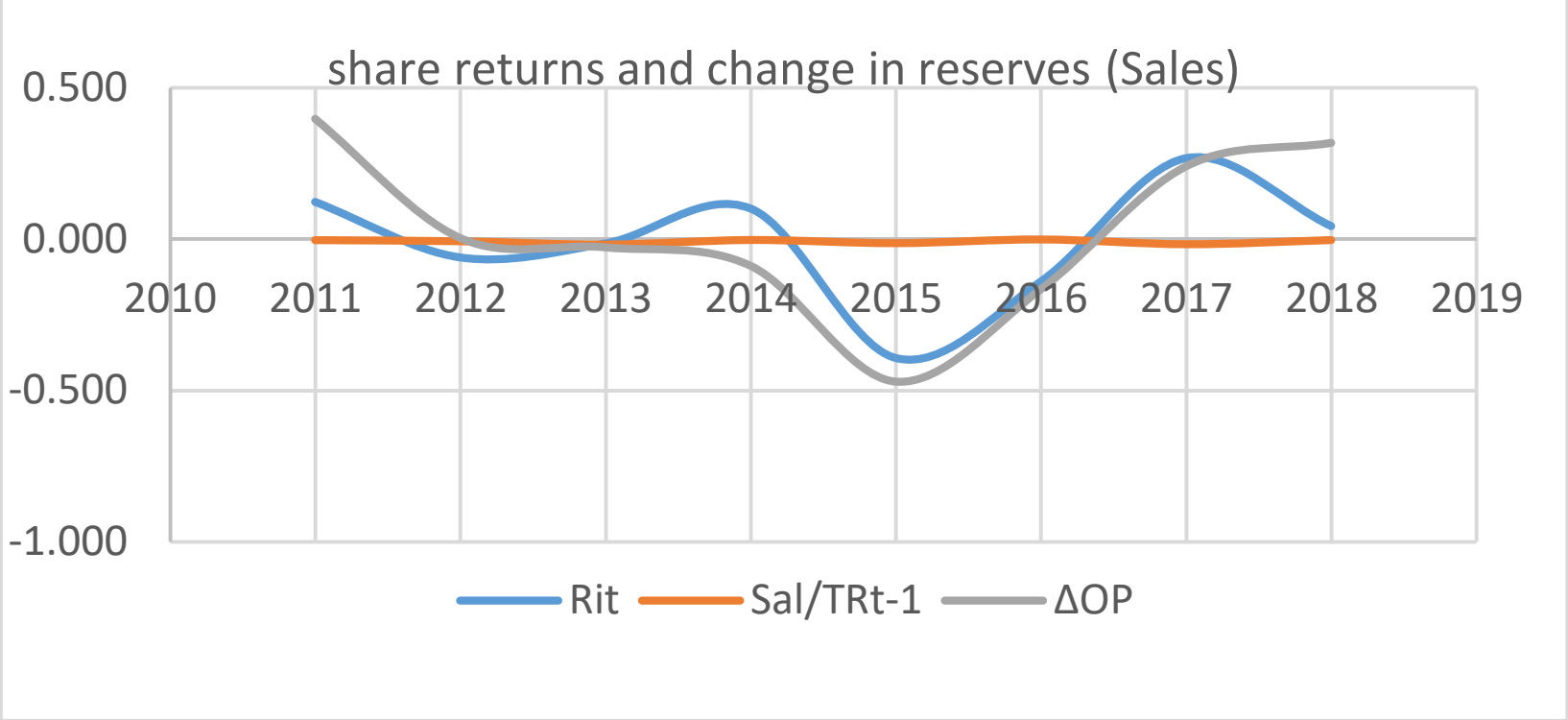

## Appendix 3: Collinearity Diagnosis

| **Coefficients**[a] | | | |
|---|---|---|---|
| | | Collinearity Statistics | |
| Model | | Tolerance | VIF |
| 1 | ResC | 0.965 | 1.037 |
| | OP | 0.898 | 1.114 |
| | risk | 0.876 | 1.142 |
| | CF | 0.804 | 1.243 |
| | D1 | 0.795 | 1.257 |

a. Dependent Variable: Rit

| **Coefficients**[a] | | | |
|---|---|---|---|
| | | Collinearity Statistics | |
| Model | | Tolerance | VIF |
| 2 | Exp | 0.873 | 1.145 |
| | Acq | 0.965 | 1.036 |
| | Pro | 0.933 | 1.072 |
| | Rev | 0.930 | 1.075 |
| | Sal | 0.916 | 1.092 |
| | OP | 0.875 | 1.142 |
| | risk | 0.857 | 1.167 |
| | CF | 0.732 | 1.366 |
| | D1 | 0.700 | 1.429 |

a. Dependent Variable: Rit

| **Coefficients**[a] | | | |
|---|---|---|---|
| | | Collinearity Statistics | |
| Model | | Tolerance | VIF |
| 3 | Exp | 0.873 | 1.146 |
| | Acq | 0.946 | 1.057 |
| | Pro | 0.851 | 1.175 |
| | Rev | 0.924 | 1.082 |
| | Sal | 0.912 | 1.096 |
| | OP | 0.847 | 1.181 |
| | risk | 0.837 | 1.195 |
| | CF | 0.643 | 1.556 |
| | D1 | 0.697 | 1.435 |
| | D2 | 0.690 | 1.449 |
| | D3 | 0.775 | 1.291 |

a. Dependent Variable: Rit

## Appendix 4: Removed cases (Outliers)

| | firm | year | Rit | ResC | Exp | Acq | Pro | Rev | Sal | OP | risk | CF | D1 | D2 | D3 | MAH_1 | Pvalue1 | MAH_2 | Pvalue2 | MAH_3 | Pvalue3 | filter_$ |
|---|---|---|---|---|---|---|---|---|---|---|---|---|---|---|---|---|---|---|---|---|---|---|
| 1 | CASP | 2016... | -.1106 | 23.4167 | 23.8367 | .0000 | -.4200 | .0000 | .0000 | -.1659 | -.1141 | .0003 | 0 | 1 | 1 | 90.9215 | .0000 | 168.3058 | .0000 | 168.3754 | .0000 | 0 |
| 2 | SQZ | 2018... | 1.4474 | 21.1935 | 2.0000 | 18.5484 | -.3226 | .9677 | .0000 | .3179 | 1.4416 | .0027 | 0 | 1 | 1 | 74.5188 | .0000 | 166.8281 | .0000 | 166.8632 | .0000 | 0 |
| 3 | NOG | 2014... | 33.596 | -.0189 | .0000 | .0089 | -.0278 | .0000 | .0000 | -.0883 | 33.5925 | .1706 | 1 | 1 | 0 | 159.9585 | .0000 | 161.2631 | .0000 | 161.8430 | .0000 | 0 |
| 4 | CABC | 2016... | -.4324 | 4.0714 | .0000 | 3.9286 | -.3571 | 2.1429 | 1.6429 | -.1659 | -.2665 | .0354 | 0 | 1 | 1 | 3.2318 | .6643 | 90.2523 | .0000 | 91.1020 | .0000 | 0 |
| 5 | RPT | 2018... | 5.7039 | 2.7037 | .0000 | .0000 | -.0917 | 2.7954 | .0000 | .3179 | 5.6981 | .2947 | 0 | 1 | 1 | 13.6638 | .0179 | 71.6716 | .0000 | 71.6780 | .0000 | 0 |
| 6 | CNE | 2013... | -.0748 | .8813 | .0000 | .0000 | -.0001 | 1.8750 | -.9938 | -.0275 | -.0785 | .0083 | 1 | 1 | 1 | 3.1471 | .6773 | 58.5262 | .0000 | 59.6591 | .0000 | 0 |
| 7 | SOU | 2016... | 1.7098 | -1.0000 | .0000 | .0000 | -1.0000 | .0000 | .0000 | -.1659 | 1.7064 | -.0330 | 0 | 0 | 0 | 1.9634 | .8542 | 49.1090 | .0000 | 58.9250 | .0000 | 0 |
| 8 | SEY | 2017... | -.0209 | -1.0000 | .0000 | .0000 | -1.0000 | .0000 | .0000 | .2404 | -.0234 | .0401 | 0 | 0 | 1 | 1.6854 | .8907 | 48.4938 | .0000 | 55.0557 | .0000 | 0 |
| 9 | CNE | 2011... | -.0916 | -1.0000 | .0000 | .0000 | .0000 | .0000 | -1.0000 | .3976 | -.0974 | .3074 | 1 | 0 | 1 | 7.7769 | .1690 | 29.7960 | .0005 | 39.2905 | .0000 | 0 |
| 10 | CABC | 2014... | -.4258 | 3.8333 | 5.3333 | .0000 | -.6667 | .0000 | -.8333 | -.0883 | -.3375 | .0509 | 0 | 1 | 1 | 2.7365 | .7405 | 37.6347 | .0000 | 38.5877 | .0001 | 0 |
| 11 | SOU | 2012... | -.6605 | -.8142 | -.2702 | .0000 | .0000 | .0000 | -1.0000 | .0033 | -.6652 | -.1718 | 0 | 1 | 0 | 4.0376 | .5440 | 27.9248 | .0010 | 37.0758 | .0001 | 0 |
| 12 | EME | 2016... | .3655 | -1.0000 | .0000 | .0000 | .0000 | .0000 | -1.0000 | -.1659 | .3620 | .0354 | 0 | 0 | 1 | 1.4791 | .9155 | 26.0787 | .0020 | 33.7530 | .0004 | 0 |
| 13 | SEY | 2018... | -.1719 | .0000 | .0000 | .0000 | .0000 | .0000 | .0000 | .3179 | -.1777 | -.5115 | 0 | 0 | 0 | 28.6195 | .0000 | 28.9918 | .0007 | 29.8081 | .0017 | 0 |
| 14 | SEY | 2011... | -.5730 | .5772 | .0000 | .0000 | -.5439 | 1.1211 | .0000 | .3976 | -.5787 | .0369 | 0 | 1 | 1 | 2.6344 | .7561 | 21.2061 | .0118 | 22.1844 | .0230 | 0 |
| 15 | EME | 2017... | 8.1069 | .0000 | .0000 | .0000 | .0000 | .0000 | .0000 | .2404 | 8.1045 | -.2641 | 0 | 0 | 0 | 18.2069 | .0027 | 18.4706 | .0301 | 20.8074 | .0354 | 0 |

## Appendix 5: computed data set

| i | t | Rit | ΔTR | Exp | Acq | Pro | Rev | Sal | ΔOP | Risk | CFo | D1 | D2 | D3 |
|---|---|---|---|---|---|---|---|---|---|---|---|---|---|---|
| BP | 2018 | 0.145 | 0.082 | 0.052 | 0.094 | -0.075 | 0.023 | -0.012 | 0.318 | 0.470 | 0.096 | 1 | 1 | 1 |
| BP | 2017 | 0.124 | 0.035 | 0.056 | 0.012 | -0.075 | 0.052 | 0.010 | 0.240 | 0.778 | 0.077 | 1 | 1 | 1 |
| BP | 2016 | 0.028 | 0.037 | 0.037 | 0.045 | -0.071 | 0.042 | -0.014 | -0.166 | -0.278 | 0.044 | 1 | 1 | 1 |
| BP | 2015 | -0.143 | -0.020 | 0.026 | 0.009 | -0.070 | 0.017 | -0.001 | -0.471 | -0.469 | 0.074 | 1 | 1 | 1 |
| BP | 2014 | 0.022 | -0.026 | 0.031 | 0.004 | -0.065 | 0.010 | -0.006 | -0.088 | -0.055 | 0.124 | 1 | 1 | 1 |
| BP | 2013 | 0.050 | 0.059 | 0.065 | 0.362 | -0.071 | 0.027 | -0.324 | -0.028 | 0.344 | 0.084 | 1 | 1 | 1 |
| BP | 2012 | -0.031 | -0.042 | 0.059 | 0.004 | -0.070 | -0.005 | -0.029 | 0.003 | -0.356 | 0.083 | 1 | 1 | 1 |
| BP | 2011 | -0.044 | -0.018 | 0.034 | 0.012 | -0.071 | 0.039 | -0.032 | 0.398 | 5.039 | 0.085 | 1 | 1 | 1 |
| CNE | 2018 | 0.060 | 0.046 | 0.000 | 0.000 | -0.117 | 0.164 | 0.000 | 0.318 | -2.533 | 0.085 | 1 | 1 | 1 |
| CNE | 2017 | 0.029 | 0.045 | 0.328 | 0.000 | -0.283 | 0.000 | 0.000 | 0.240 | -1.892 | 0.028 | 1 | 1 | 1 |
| CNE | 2016 | 0.182 | 0.040 | 0.000 | 0.000 | 0.000 | 0.040 | 0.000 | -0.166 | -1.707 | -0.012 | 1 | 1 | 1 |
| CNE | 2015 | -0.095 | -0.118 | 0.000 | 0.000 | 0.000 | 0.041 | -0.159 | -0.471 | 15.354 | -0.017 | 1 | 1 | 1 |
| CNE | 2014 | -0.335 | 0.867 | 0.000 | 0.000 | 0.000 | 0.867 | 0.000 | -0.088 | -0.996 | -0.024 | 1 | 1 | 1 |
| CNE | 2013 | -0.075 | 0.881 | 0.000 | 0.000 | 0.000 | 1.875 | -0.994 | -0.028 | 0.245 | 0.008 | 1 | 1 | 1 |
| CNE | 2012 | -0.293 | 0.000 | 0.000 | 0.000 | 0.000 | 0.000 | 0.000 | 0.003 | -0.770 | 0.012 | 1 | 0 | 1 |
| CNE | 2011 | -0.092 | -1.000 | 0.000 | 0.000 | 0.000 | 0.000 | -1.000 | 0.398 | -2.229 | 0.307 | 1 | 0 | 1 |
| ENQ | 2018 | 0.027 | 0.167 | 0.000 | 0.262 | -0.090 | -0.010 | 0.000 | 0.318 | 2.117 | 0.105 | 1 | 1 | 1 |
| ENQ | 2017 | 0.256 | -0.023 | 0.000 | 0.065 | -0.056 | -0.033 | 0.000 | 0.240 | -0.373 | 0.033 | 1 | 1 | 1 |
| ENQ | 2016 | -0.188 | 0.059 | 0.000 | 0.069 | -0.064 | 0.054 | 0.000 | -0.166 | 0.261 | 0.066 | 1 | 1 | 1 |
| ENQ | 2015 | -0.683 | 0.077 | 0.000 | 0.009 | -0.055 | 0.000 | 0.000 | -0.471 | -0.265 | 0.011 | 1 | 1 | 1 |
| ENQ | 2014 | -0.145 | 0.130 | 0.037 | 0.117 | -0.052 | 0.027 | 0.000 | -0.088 | -0.136 | 0.157 | 1 | 1 | 1 |
| ENQ | 2013 | 0.115 | 0.515 | 0.522 | 0.045 | -0.066 | 0.019 | 0.000 | -0.028 | -0.077 | 0.152 | 1 | 1 | 1 |
| ENQ | 2012 | -0.026 | 0.116 | 0.177 | 0.000 | -0.071 | 0.089 | -0.079 | 0.003 | 0.194 | 0.225 | 1 | 1 | 1 |
| ENQ | 2011 | 0.006 | 0.302 | 0.335 | 0.009 | -0.094 | 0.027 | 0.000 | 0.398 | 1.038 | 0.329 | 1 | 1 | 1 |
| PMO | 2018 | 0.537 | -0.358 | 0.000 | 0.000 | -0.098 | -0.228 | -0.033 | 0.318 | 0.435 | 0.084 | 1 | 1 | 1 |
| PMO | 2017 | 0.074 | -0.146 | 0.000 | 0.000 | -0.078 | -0.035 | -0.033 | 0.240 | -0.138 | 0.027 | 1 | 1 | 1 |
| PMO | 2016 | -0.489 | 0.064 | 0.000 | 0.114 | -0.079 | 0.029 | 0.000 | -0.166 | 0.398 | 0.040 | 1 | 1 | 1 |
| PMO | 2015 | -0.595 | 0.364 | 0.000 | 0.000 | -0.087 | 0.577 | -0.126 | -0.471 | -0.316 | 0.147 | 1 | 1 | 1 |
| PMO | 2014 | -0.167 | -0.062 | 0.002 | 0.000 | -0.089 | 0.061 | -0.036 | -0.088 | 0.153 | 0.163 | 1 | 1 | 1 |
| PMO | 2013 | -0.047 | -0.111 | 0.000 | 0.005 | -0.073 | -0.044 | 0.000 | -0.028 | 0.084 | 0.160 | 1 | 1 | 1 |
| PMO | 2012 | -0.134 | 0.025 | 0.018 | 0.039 | -0.074 | 0.042 | 0.000 | 0.003 | 1.226 | 0.193 | 1 | 1 | 1 |
| PMO | 2011 | 0.196 | 0.092 | 0.051 | 0.120 | -0.056 | 0.021 | 0.000 | 0.398 | -0.084 | 0.118 | 1 | 1 | 1 |
| RDS'b | 2018 | 0.131 | 0.033 | 0.074 | 0.001 | -0.125 | 0.098 | -0.014 | 0.318 | 0.347 | 0.113 | 1 | 1 | 1 |
| RDS'b | 2017 | 0.135 | -0.159 | 0.072 | 0.106 | -0.107 | 0.098 | -0.329 | 0.240 | 0.306 | 0.079 | 1 | 1 | 1 |
| RDS'b | 2016 | -0.151 | 0.180 | 0.028 | 0.228 | -0.127 | 0.054 | -0.002 | -0.166 | 0.406 | 0.038 | 1 | 1 | 1 |
| RDS'b | 2015 | 0.026 | -0.135 | 0.008 | 0.000 | -0.090 | -0.041 | -0.012 | -0.471 | -0.444 | 0.072 | 1 | 1 | 1 |
| RDS'b | 2014 | 0.022 | -0.074 | 0.012 | 0.000 | -0.082 | 0.009 | -0.001 | -0.088 | 0.022 | 0.106 | 1 | 1 | 1 |
| RDS'b | 2013 | -0.012 | 0.069 | 0.096 | 0.008 | -0.091 | 0.057 | -0.001 | -0.028 | -0.165 | 0.095 | 1 | 1 | 1 |
| RDS'b | 2012 | 0.138 | 0.024 | 0.017 | 0.014 | -0.099 | 0.104 | -0.011 | 0.003 | -0.093 | 0.093 | 1 | 1 | 1 |
| RDS'b | 2011 | 0.141 | -0.016 | 0.059 | 0.000 | -0.099 | 0.031 | -0.006 | 0.398 | 0.295 | 0.076 | 1 | 1 | 1 |
| SIA | 2018 | -0.266 | -0.181 | 0.000 | 0.000 | -0.096 | -0.085 | -0.003 | 0.318 | -2.995 | 0.122 | 1 | 1 | 1 |
| SIA | 2017 | -0.145 | -0.156 | 0.000 | 0.000 | -0.090 | -0.066 | -0.002 | 0.240 | -1.760 | 0.124 | 1 | 1 | 1 |
| SIA | 2016 | -0.189 | -0.161 | 0.000 | 0.000 | -0.097 | -0.011 | 0.000 | -0.166 | 0.031 | 0.077 | 1 | 1 | 1 |
| SIA | 2015 | -0.519 | -0.086 | 0.000 | 0.000 | -0.105 | 0.020 | 0.000 | -0.471 | -0.594 | 0.120 | 1 | 1 | 1 |
| SIA | 2014 | 0.106 | -0.686 | 0.000 | 0.000 | -0.038 | -0.648 | -0.005 | -0.088 | -0.406 | 0.298 | 1 | 1 | 1 |
| SIA | 2013 | 0.237 | 0.012 | 0.059 | 0.000 | -0.047 | 0.000 | 0.000 | -0.028 | -0.064 | 0.349 | 1 | 1 | 1 |
| SIA | 2012 | -0.079 | -0.014 | 0.000 | 0.028 | -0.041 | 0.000 | 0.000 | 0.003 | 1.675 | 0.343 | 1 | 1 | 1 |
| SIA | 2011 | -0.111 | -0.017 | 0.000 | 0.000 | -0.017 | 0.000 | 0.000 | 0.398 | 3.566 | 0.108 | 1 | 1 | 1 |
| TLW | 2018 | 0.137 | -0.038 | 0.009 | 0.000 | -0.102 | 0.062 | -0.006 | 0.318 | -0.004 | 0.098 | 1 | 1 | 1 |
| TLW | 2017 | -0.054 | -0.043 | 0.043 | 0.000 | -0.105 | 0.018 | 0.000 | 0.240 | 21.674 | 0.077 | 1 | 1 | 1 |
| TLW | 2016 | -0.158 | -0.056 | 0.000 | 0.000 | -0.076 | 0.020 | 0.000 | -0.166 | -0.915 | 0.039 | 1 | 1 | 1 |
| TLW | 2015 | -0.594 | -0.068 | 0.003 | 0.000 | -0.078 | 0.012 | -0.005 | -0.471 | -0.886 | 0.103 | 1 | 1 | 1 |
| TLW | 2014 | -0.325 | -0.097 | 0.000 | 0.021 | -0.072 | -0.046 | -0.006 | -0.088 | 2.753 | 0.121 | 1 | 1 | 1 |
| TLW | 2013 | -0.267 | -0.014 | 0.068 | 0.002 | -0.079 | 0.030 | -0.035 | -0.028 | -0.262 | 0.169 | 1 | 1 | 1 |
| TLW | 2012 | 0.053 | 0.304 | 0.378 | 0.001 | -0.097 | 0.023 | 0.000 | 0.003 | 0.287 | 0.185 | 1 | 1 | 1 |
| TLW | 2011 | 0.131 | 0.011 | 0.000 | 0.076 | -0.097 | 0.032 | 0.000 | 0.398 | 1.410 | 0.171 | 1 | 1 | 1 |
| AMER | 2018 | -0.274 | 0.237 | 0.386 | 0.000 | -0.151 | 0.000 | 0.000 | 0.318 | 0.278 | 0.131 | 0 | 1 | 1 |
| AMER | 2017 | -0.238 | -0.154 | 0.000 | 0.000 | -0.072 | -0.082 | 0.000 | 0.240 | -14.347 | 0.068 | 0 | 1 | 1 |
| AMER | 2016 | -0.102 | 0.032 | 0.000 | 0.000 | -0.048 | 0.080 | 0.000 | -0.166 | -23.217 | 0.041 | 0 | 1 | 1 |
| AMER | 2015 | -0.454 | -0.033 | 0.000 | 0.000 | -0.066 | 0.099 | 0.000 | -0.471 | -0.999 | 0.038 | 0 | 1 | 1 |
| AMER | 2014 | 0.158 | -0.253 | 0.000 | 0.000 | -0.069 | -0.184 | 0.000 | -0.088 | 0.099 | 0.073 | 0 | 1 | 1 |
| AMER | 2013 | 0.438 | 0.097 | 0.154 | 0.000 | -0.057 | 0.000 | 0.000 | -0.028 | 3.203 | 0.108 | 0 | 1 | 1 |
| AMER | 2012 | 0.673 | 2.883 | 3.013 | 0.000 | -0.130 | 0.000 | 0.000 | 0.003 | 2.720 | 0.217 | 0 | 1 | 1 |
| AMER | 2011 | 0.313 | 0.140 | 0.000 | 0.000 | 0.000 | 0.140 | 0.000 | 0.398 | 0.241 | -0.309 | 0 | 0 | 0 |
| CASP | 2018 | -0.070 | 0.000 | 0.000 | 0.000 | -0.024 | 0.024 | 0.000 | 0.318 | -0.176 | 0.078 | 0 | 1 | 1 |
| CASP | 2017 | -0.033 | -0.017 | 0.000 | 0.000 | -0.028 | 0.010 | 0.000 | 0.240 | -0.159 | 0.099 | 0 | 1 | 1 |
| CASP | 2016 | -0.111 | 23.417 | 23.837 | 0.000 | -0.420 | 0.000 | 0.000 | -0.166 | 2.872 | 0.000 | 0 | 1 | 1 |
| CASP | 2015 | 0.148 | 0.000 | 0.000 | 0.000 | 0.000 | 0.000 | 0.000 | -0.471 | -1.057 | -0.093 | 0 | 1 | 1 |
| CASP | 2014 | 1.327 | 0.000 | 0.000 | 0.000 | -0.006 | 0.006 | 0.000 | -0.088 | -4.477 | 0.013 | 0 | 1 | 1 |
| CASP | 2013 | 0.259 | 0.960 | 1.101 | 0.000 | -0.141 | 0.000 | 0.000 | -0.028 | -0.136 | -0.011 | 0 | 1 | 1 |
| CASP | 2012 | -0.223 | 0.000 | 0.000 | 0.000 | -0.074 | 0.074 | 0.000 | 0.003 | -1.132 | -0.012 | 0 | 1 | 1 |
| CASP | 2011 | -0.420 | 0.000 | 0.000 | 0.000 | -0.012 | 0.012 | 0.000 | 0.398 | -2.093 | -0.031 | 0 | 1 | 1 |
| ELA | 2018 | 0.824 | 0.066 | 0.142 | 0.000 | -0.076 | 0.000 | 0.000 | 0.318 | 27.104 | 0.109 | 0 | 1 | 1 |
| ELA | 2017 | 0.833 | 0.002 | 0.040 | 0.000 | -0.037 | 0.000 | 0.000 | 0.240 | -1.014 | 0.038 | 0 | 1 | 1 |
| ELA | 2016 | -0.378 | 0.000 | 0.012 | 0.000 | -0.012 | 0.000 | 0.000 | -0.166 | 14.432 | -0.035 | 0 | 1 | 1 |
| ELA | 2015 | -0.499 | 0.052 | 0.065 | 0.000 | -0.013 | 0.000 | 0.000 | -0.471 | -0.984 | -0.032 | 0 | 1 | 1 |
| ELA | 2014 | -0.137 | -0.005 | 0.000 | 0.000 | -0.005 | 0.000 | 0.000 | -0.088 | -0.425 | -0.031 | 0 | 0 | 1 |
| ELA | 2013 | 0.018 | 0.139 | 0.139 | 0.000 | 0.000 | 0.000 | 0.000 | -0.028 | 0.883 | -0.078 | 0 | 0 | 1 |
| EME | 2018 | 0.059 | 0.000 | 0.000 | 0.000 | 0.000 | 0.000 | 0.000 | 0.318 | -0.953 | -0.245 | 0 | 0 | 0 |
| EME | 2017 | 8.107 | 0.000 | 0.000 | 0.000 | 0.000 | 0.000 | 0.000 | 0.240 | -35.381 | -0.264 | 0 | 0 | 0 |
| EME | 2016 | 0.365 | -1.000 | 0.000 | 0.000 | 0.000 | 0.000 | -1.000 | -0.166 | -0.951 | 0.035 | 0 | 0 | 1 |
| EME | 2015 | -0.563 | 0.116 | 0.145 | 0.000 | -0.029 | 0.000 | 0.000 | -0.471 | 0.260 | 0.154 | 0 | 1 | 1 |
| EME | 2014 | 1.207 | 0.939 | 0.990 | 0.000 | -0.051 | 0.000 | 0.000 | -0.088 | 0.720 | 0.229 | 0 | 1 | 1 |
| EME | 2013 | -0.100 | 1.038 | 1.143 | 0.000 | -0.114 | 0.000 | 0.000 | -0.028 | 1.699 | 0.146 | 0 | 1 | 1 |
| EME | 2012 | 0.196 | 1.417 | 0.000 | 0.000 | 0.000 | 0.000 | 0.000 | 0.003 | -4.903 | 0.133 | 0 | 0 | 0 |
| EME | 2011 | -0.095 | 0.000 | 0.000 | 0.000 | 0.000 | 0.000 | 0.000 | 0.398 | 0.290 | -0.013 | 0 | 0 | 0 |
| HUR | 2018 | 0.124 | 0.000 | 0.000 | 0.000 | 0.000 | 0.000 | 0.000 | 0.318 | -0.277 | -0.004 | 0 | 0 | 1 |
| HUR | 2017 | 0.589 | 0.000 | 0.000 | 0.000 | 0.000 | 0.000 | 0.000 | 0.240 | 1.121 | -0.015 | 0 | 0 | 1 |
| HUR | 2016 | 0.691 | 0.000 | 0.000 | 0.000 | 0.000 | 0.000 | 0.000 | -0.166 | 0.213 | -0.014 | 0 | 0 | 1 |
| HUR | 2015 | -0.531 | 0.000 | 0.000 | 0.000 | 0.000 | 0.000 | 0.000 | -0.471 | -0.373 | -0.014 | 0 | 0 | 1 |
| IGAS | 2018 | -0.005 | 0.067 | 0.178 | 0.000 | -0.060 | -0.051 | 0.000 | 0.318 | -3.412 | 0.037 | 0 | 1 | 1 |
| IGAS | 2017 | -0.642 | 0.020 | 0.000 | 0.000 | -0.067 | 0.087 | 0.000 | 0.240 | -1.625 | 0.004 | 0 | 1 | 1 |
| IGAS | 2016 | -0.413 | 0.003 | 0.000 | 0.000 | -0.064 | 0.067 | 0.000 | -0.166 | -1.465 | -0.064 | 0 | 1 | 1 |
| IGAS | 2015 | -0.778 | 0.055 | 0.012 | 0.000 | -0.056 | 0.100 | 0.000 | -0.471 | -0.295 | -0.027 | 0 | 1 | 1 |
| IGAS | 2014 | 0.075 | -0.190 | 0.000 | 0.018 | -0.061 | -0.148 | 0.000 | -0.088 | -0.126 | 0.054 | 0 | 1 | 1 |
| IGAS | 2013 | 0.446 | 0.444 | 0.000 | 0.474 | -0.079 | 0.049 | 0.000 | -0.028 | -3.595 | 0.003 | 0 | 1 | 1 |
| IGAS | 2012 | 0.134 | -0.021 | 0.000 | 0.000 | -0.021 | 0.000 | 0.000 | 0.003 | -0.074 | -0.010 | 0 | 1 | 1 |
| IGAS | 2011 | -0.170 | 0.000 | 0.000 | 0.000 | 0.000 | 0.000 | 0.000 | 0.398 | 7.132 | -0.003 | 0 | 1 | 1 |
| PMG | 2018 | 0.305 | 0.671 | 0.000 | 0.800 | -0.129 | 0.000 | 0.000 | 0.318 | 0.522 | 0.038 | 0 | 1 | 1 |
| PMG | 2017 | -0.225 | -0.007 | 0.000 | 0.000 | -0.088 | 0.096 | 0.000 | 0.240 | 0.025 | -0.006 | 0 | 1 | 1 |
| PMG | 2016 | -0.423 | 0.187 | 0.000 | 0.270 | -0.083 | 0.000 | 0.000 | -0.166 | -0.769 | -0.121 | 0 | 1 | 1 |
| PMG | 2015 | -0.557 | -0.100 | 0.000 | 0.000 | -0.070 | -0.030 | 0.000 | -0.471 | -2.205 | -0.017 | 0 | 1 | 1 |
| PMG | 2014 | 0.067 | 0.214 | 0.000 | 0.245 | -0.031 | 0.000 | 0.000 | -0.088 | -3.365 | 0.055 | 0 | 1 | 1 |
| PMG | 2013 | -0.087 | -0.143 | 0.000 | 0.000 | -0.016 | -0.127 | 0.000 | -0.028 | 0.018 | -0.088 | 0 | 1 | 1 |
| PMG | 2012 | -0.122 | 0.000 | 0.000 | 0.000 | 0.000 | 0.000 | 0.000 | 0.003 | 0.339 | -0.102 | 0 | 1 | 1 |
| PMG | 2011 | 2.404 | 0.000 | 0.000 | 0.000 | 0.000 | 0.000 | 0.000 | 0.398 | 1.707 | -0.088 | 0 | 0 | 0 |
| RPT | 2018 | 5.704 | 2.704 | 0.000 | 0.000 | -0.092 | 2.795 | 0.000 | 0.318 | 3.602 | 0.295 | 0 | 1 | 1 |
| RPT | 2017 | 0.535 | 0.000 | 0.000 | 0.000 | -0.060 | 0.060 | 0.000 | 0.240 | 0.385 | 0.264 | 0 | 1 | 1 |
| RPT | 2016 | -0.253 | 0.000 | 0.000 | 0.000 | -0.046 | 0.046 | 0.000 | -0.166 | 0.868 | 0.155 | 0 | 1 | 1 |
| RPT | 2015 | -0.487 | 0.154 | 0.000 | 0.194 | -0.040 | 0.000 | 0.000 | -0.471 | -0.487 | 0.126 | 0 | 1 | 1 |
| RPT | 2014 | -0.586 | 0.000 | 0.000 | 0.000 | -0.043 | 0.043 | 0.000 | -0.088 | 0.210 | 0.207 | 0 | 1 | 1 |
| RPT | 2013 | -0.122 | -0.630 | 0.000 | 0.000 | -0.016 | -0.613 | 0.000 | -0.028 | -0.390 | 0.170 | 0 | 1 | 1 |
| RPT | 2012 | -0.426 | -0.791 | 0.000 | 0.000 | -0.004 | -0.787 | 0.000 | 0.003 | -7.844 | 0.110 | 0 | 1 | 1 |
| RPT | 2011 | 0.018 | 0.000 | 0.000 | 0.000 | 0.000 | 0.000 | 0.000 | 0.398 | -0.765 | -0.075 | 0 | 0 | 0 |
| SDX | 2018 | 0.035 | -0.033 | 0.185 | 0.000 | -0.153 | -0.065 | 0.000 | 0.318 | 1.293 | 0.262 | 0 | 1 | 1 |
| SDX | 2017 | 1.111 | 1.200 | 0.794 | 0.499 | -0.276 | 0.183 | 0.000 | 0.240 | -1.127 | 0.153 | 0 | 1 | 1 |
| SQZ | 2018 | 1.447 | 21.194 | 2.000 | 18.548 | -0.323 | 0.968 | 0.000 | 0.318 | -0.710 | 0.003 | 0 | 1 | 1 |
| SQZ | 2017 | 1.901 | -0.184 | 0.000 | 0.000 | -0.184 | 0.000 | 0.000 | 0.240 | 2.060 | 0.205 | 0 | 1 | 1 |
| SQZ | 2016 | 1.019 | -0.095 | 0.000 | 0.000 | -0.143 | 0.048 | 0.000 | -0.166 | 5.359 | 0.078 | 0 | 1 | 1 |
| SQZ | 2015 | -0.454 | 0.000 | 0.000 | 0.000 | 0.000 | 0.000 | 0.000 | -0.471 | -1.052 | 0.110 | 0 | 1 | 1 |
| SQZ | 2014 | -0.495 | -1.000 | 0.000 | 0.000 | 0.000 | -1.000 | 0.000 | -0.088 | 2.775 | -0.043 | 0 | 1 | 1 |
| SQZ | 2013 | -0.222 | -0.055 | 0.000 | 0.000 | -0.018 | -0.036 | 0.000 | -0.028 | -3.113 | -0.026 | 0 | 1 | 1 |
| SQZ | 2012 | 0.076 | -0.191 | 0.000 | 0.000 | -0.059 | -0.132 | 0.000 | 0.003 | -0.782 | 0.036 | 0 | 1 | 1 |
| SQZ | 2011 | -0.576 | -0.181 | 0.000 | 0.000 | -0.096 | -0.084 | 0.000 | 0.398 | -0.107 | 0.070 | 0 | 1 | 1 |
| SEY | 2018 | -0.172 | 0.000 | 0.000 | 0.000 | 0.000 | 0.000 | 0.000 | 0.318 | -0.769 | -0.511 | 0 | 0 | 0 |
| SEY | 2017 | -0.021 | -1.000 | 0.000 | 0.000 | -1.000 | 0.000 | 0.000 | 0.240 | 0.029 | 0.040 | 0 | 0 | 1 |
| SEY | 2016 | -0.051 | -0.578 | 0.000 | 0.000 | -0.590 | 0.012 | 0.000 | -0.166 | -0.453 | -0.086 | 0 | 1 | 1 |
| SEY | 2015 | -0.492 | -0.408 | 0.000 | 0.000 | -0.387 | -0.021 | 0.000 | -0.471 | 12.009 | -0.039 | 0 | 1 | 1 |
| SEY | 2014 | -0.144 | -0.478 | 0.000 | 0.000 | -0.283 | -0.195 | 0.000 | -0.088 | -1.151 | -0.010 | 0 | 1 | 1 |
| SEY | 2013 | -0.037 | 0.177 | 0.000 | 0.000 | -0.404 | 0.581 | 0.000 | -0.028 | 0.334 | 0.041 | 0 | 1 | 1 |
| SEY | 2012 | -0.184 | -0.285 | 0.000 | 0.000 | -0.285 | 0.000 | 0.000 | 0.003 | -0.651 | 0.055 | 0 | 1 | 1 |
| SEY | 2011 | -0.573 | 0.577 | 0.000 | 0.000 | -0.544 | 1.121 | 0.000 | 0.398 | 1.243 | 0.037 | 0 | 1 | 1 |
| SOU | 2018 | -0.406 | 0.000 | 0.000 | 0.000 | 0.000 | 0.000 | 0.000 | 0.318 | -1.779 | 0.005 | 0 | 0 | 0 |
| SOU | 2017 | 0.431 | 0.000 | 0.000 | 0.000 | 0.000 | 0.000 | 0.000 | 0.240 | -2.858 | -0.058 | 0 | 0 | 0 |
| SOU | 2016 | 1.710 | -1.000 | 0.000 | 0.000 | -1.000 | 0.000 | 0.000 | -0.166 | 1.011 | -0.033 | 0 | 0 | 0 |
| SOU | 2015 | 0.733 | 2.667 | 0.188 | 0.000 | -0.111 | 0.000 | 0.000 | -0.471 | 0.295 | -0.176 | 0 | 1 | 0 |
| SOU | 2014 | 0.123 | 0.000 | 0.000 | 0.000 | -0.250 | 0.000 | 0.000 | -0.088 | -0.633 | -0.113 | 0 | 1 | 0 |
| SOU | 2013 | -0.145 | 0.091 | 0.440 | 0.000 | -0.045 | -0.045 | 0.000 | -0.028 | 0.995 | -0.117 | 0 | 1 | 0 |
| SOU | 2012 | -0.661 | -0.814 | -0.270 | 0.000 | 0.000 | 0.000 | -1.000 | 0.003 | -0.256 | -0.172 | 0 | 1 | 0 |
| SOU | 2011 | 0.723 | 0.000 | 0.000 | 0.000 | 0.000 | 0.000 | 0.000 | 0.398 | 1.149 | -0.082 | 0 | 0 | 0 |
| TRIN | 2018 | 0.271 | 0.055 | 0.000 | 0.000 | -0.045 | 0.100 | 0.000 | 0.318 | -1.077 | 0.028 | 0 | 1 | 1 |
| TRIN | 2017 | 2.652 | 0.092 | 0.000 | 0.000 | -0.043 | 0.136 | 0.000 | 0.240 | 2.601 | -0.051 | 0 | 1 | 1 |
| TRIN | 2016 | -0.884 | -0.025 | 0.000 | 0.000 | -0.042 | 0.067 | 0.000 | -0.166 | -1.627 | 0.055 | 0 | 1 | 1 |
| TRIN | 2015 | -0.720 | -0.138 | 0.000 | 0.000 | -0.043 | -0.095 | 0.000 | -0.471 | -0.832 | -0.032 | 0 | 1 | 1 |
| TRIN | 2014 | -0.139 | -0.470 | 0.000 | 0.000 | -0.027 | -0.442 | 0.000 | -0.088 | -2.247 | 0.051 | 0 | 1 | 1 |
| TRIN | 2013 | -0.729 | 0.340 | 0.000 | 0.000 | -0.039 | 0.379 | 0.000 | -0.028 | 1.113 | 0.107 | 0 | 1 | 1 |
| ZOL | 2018 | -0.079 | 0.000 | 0.000 | 0.000 | -0.010 | 0.010 | 0.000 | 0.318 | -1.384 | 0.097 | 0 | 1 | 0 |
| ZOL | 2017 | -0.142 | 0.000 | 0.000 | 0.000 | -0.012 | 0.012 | 0.000 | 0.240 | -1.908 | 0.089 | 0 | 1 | 0 |
| ZOL | 2016 | -0.497 | 0.000 | 0.000 | 0.000 | -0.014 | 0.014 | 0.000 | -0.166 | 1.313 | 0.071 | 0 | 1 | 0 |
| ZOL | 2015 | -0.542 | 0.000 | 0.000 | 0.000 | -0.013 | 0.013 | 0.000 | -0.471 | -0.761 | 0.024 | 0 | 1 | 0 |
| ZOL | 2014 | -0.127 | 1.939 | 0.000 | 1.976 | -0.037 | 0.000 | 0.000 | -0.088 | -8.682 | -0.029 | 0 | 1 | 0 |
| ZOL | 2013 | 0.566 | 0.000 | 0.000 | 0.000 | 0.000 | 0.000 | 0.000 | -0.028 | 0.221 | -0.108 | 0 | 1 | 0 |
| NOG | 2018 | -0.468 | -0.160 | 0.000 | 0.000 | -0.022 | -0.138 | 0.000 | 0.318 | 4.745 | 0.079 | 1 | 1 | 0 |
| NOG | 2017 | 0.376 | 0.047 | 0.077 | 0.000 | -0.030 | 0.000 | 0.000 | 0.240 | 0.998 | 0.061 | 1 | 1 | 0 |
| NOG | 2016 | -0.401 | -0.009 | 0.022 | 0.000 | -0.030 | 0.000 | 0.000 | -0.166 | -0.481 | 0.082 | 1 | 1 | 0 |
| NOG | 2015 | 0.297 | -0.177 | 0.000 | 0.000 | -0.025 | 0.152 | 0.000 | -0.471 | -0.708 | 0.068 | 1 | 1 | 0 |
| NOG | 2014 | 33.596 | -0.019 | 0.000 | 0.009 | -0.028 | 0.000 | 0.000 | -0.088 | 0.749 | 0.171 | 1 | 1 | 0 |
| CABC | 2018 | -0.312 | 0.689 | 0.292 | 0.335 | -0.123 | 0.184 | 0.000 | 0.318 | 0.603 | 0.027 | 0 | 1 | 1 |
| CABC | 2017 | 0.455 | 0.493 | 0.000 | 0.120 | -0.077 | 0.451 | 0.000 | 0.240 | 0.718 | 0.029 | 0 | 1 | 1 |
| CABC | 2016 | -0.432 | 4.071 | 0.000 | 3.929 | -0.357 | 2.143 | 1.643 | -0.166 | -0.784 | 0.035 | 0 | 1 | 1 |
| CABC | 2015 | -0.755 | -0.034 | 0.000 | 0.000 | -0.034 | 0.000 | 0.000 | -0.471 | -0.825 | -0.108 | 0 | 1 | 1 |
| CABC | 2014 | -0.426 | 3.833 | 5.333 | 0.000 | -0.667 | 0.000 | -0.833 | -0.088 | 0.618 | 0.051 | 0 | 1 | 1 |
| CABC | 2013 | -0.451 | -0.999 | 0.000 | 0.000 | 0.000 | -0.999 | 0.000 | -0.028 | 4.346 | 0.026 | 0 | 1 | 1 |
| CABC | 2012 | -0.269 | -0.239 | 0.000 | 0.000 | -0.001 | -0.238 | 0.000 | 0.003 | -1.604 | -0.023 | 0 | 1 | 1 |
| CABC | 2011 | -0.135 | -0.155 | 0.000 | 0.000 | -0.006 | -0.134 | -0.014 | 0.398 | 6.314 | -0.019 | 0 | 1 | 1 |
| EGRE | 2018 | -0.118 | 0.020 | 0.061 | 0.000 | -0.041 | 0.000 | 0.000 | 0.318 | 0.169 | -0.049 | 0 | 1 | 1 |
| EGRE | 2017 | -0.214 | 0.078 | 0.132 | 0.000 | -0.054 | 0.000 | 0.000 | 0.240 | -0.375 | -0.013 | 0 | 1 | 1 |
| EGRE | 2016 | 0.127 | 0.847 | 1.018 | 0.000 | -0.171 | 0.000 | 0.000 | -0.166 | -0.406 | -0.006 | 0 | 1 | 1 |
| EGRE | 2015 | -0.531 | -0.080 | 0.073 | 0.000 | -0.153 | 0.000 | 0.000 | -0.471 | 10.986 | -0.041 | 0 | 1 | 1 |
| EGRE | 2014 | 1.306 | -0.195 | 0.000 | 0.000 | -0.170 | -0.025 | 0.000 | -0.088 | -0.370 | 0.012 | 0 | 1 | 1 |
| EGRE | 2013 | 0.054 | -0.416 | 0.000 | 0.000 | -0.091 | -0.325 | 0.000 | -0.028 | -0.786 | -0.025 | 0 | 1 | 1 |
| EGRE | 2012 | -0.448 | -0.616 | 0.000 | 0.000 | -0.022 | -0.594 | 0.000 | 0.003 | -1.665 | 0.036 | 0 | 1 | 1 |
| EGRE | 2011 | 0.115 | 0.000 | 0.022 | 0.000 | -0.022 | 0.000 | 0.000 | 0.398 | 16.658 | 0.005 | 0 | 1 | 1 |
| PPTC | 2018 | 0.275 | 0.063 | 0.000 | 0.102 | 0.031 | 0.128 | 0.007 | 0.318 | -2.115 | 0.057 | 0 | 1 | 1 |
| PPTC | 2017 | -0.028 | 0.340 | 0.000 | 0.255 | -0.020 | 0.104 | 0.000 | 0.240 | -0.429 | -0.038 | 0 | 1 | 1 |
| PPTC | 2016 | -0.275 | 0.106 | 0.117 | 0.000 | -0.010 | 0.000 | 0.000 | -0.166 | -0.007 | -0.001 | 0 | 1 | 1 |
| PPTC | 2015 | -0.647 | 0.270 | 0.284 | 0.000 | -0.012 | -0.001 | 0.000 | -0.471 | -2.172 | -0.019 | 0 | 1 | 1 |
| PPTC | 2014 | 0.301 | 1.145 | 0.017 | 1.151 | -0.023 | 0.000 | 0.000 | -0.088 | -6.446 | -0.012 | 0 | 1 | 1 |
| PPTC | 2013 | -0.272 | -0.020 | 0.003 | 0.000 | -0.023 | 0.000 | 0.000 | -0.028 | -0.475 | 0.050 | 0 | 1 | 1 |
| PPTC | 2012 | -0.128 | -0.047 | 0.000 | 0.000 | -0.018 | -0.028 | 0.000 | 0.003 | -0.757 | -0.045 | 0 | 1 | 1 |
| PPTC | 2011 | -0.412 | 2.941 | 0.000 | 3.263 | -0.050 | -0.272 | 0.000 | 0.398 | 2.021 | -0.088 | 0 | 1 | 1 |

## Appendix 4: Pooled (raw) data set

| E&P | t | price | TRt | TRt-1 | ΔTR | Exp | Acq | Pro | Rev | Sales | Cfo | TA | Earnings |
|---|---|---|---|---|---|---|---|---|---|---|---|---|---|
| | | £ | (mmboe) | (mmboe) | (mmboe) | (mmboe) | (mmboe) | (mmboe) | (mmboe) | (mmboe) | (£m) | (£m) | (£m) |
| BP | 2018 | 536.771 | 19945.00 | 18441.00 | 1505.00 | 956.00 | 1727.00 | -1375.00 | 425.00 | -229.00 | 20723.00 | 216474.00 | 28306.0 |
| BP | 2017 | 468.850 | 18441.00 | 17810.00 | 631.00 | 1004.00 | 222.00 | -1342.00 | 921.00 | 175.00 | 15609.00 | 201410.00 | 19260.0 |
| BP | 2016 | 417.029 | 17810.00 | 17180.00 | 630.00 | 634.00 | 765.00 | -1226.00 | 720.00 | -245.00 | 9267.00 | 212946.00 | 10835.0 |
| BP | 2015 | 405.483 | 17180.00 | 17523.00 | -343.00 | 450.00 | 151.00 | -1225.00 | 301.00 | -20.00 | 13055.00 | 175564.00 | 15000.0 |
| BP | 2014 | 473.233 | 17523.00 | 17996.00 | -473.00 | 560.00 | 74.00 | -1177.00 | 182.00 | -114.00 | 22638.00 | 182308.00 | 28248.0 |
| BP | 2013 | 462.838 | 17996.00 | 17000.00 | 996.00 | 1098.00 | 6152.00 | -1209.00 | 466.00 | -5511.00 | 15512.00 | 183731.00 | 29897.0 |
| BP | 2012 | 440.854 | 17000.00 | 17748.00 | -748.00 | 1048.00 | 68.00 | -1246.00 | -95.00 | -523.00 | 15246.00 | 184670.00 | 22248.0 |
| BP | 2011 | 454.775 | 17748.00 | 18071.00 | -323.00 | 614.00 | 211.00 | -1282.00 | 706.00 | -572.00 | 16033.00 | 188566.00 | 34548.0 |
| CNE | 2018 | 212.150 | 56.30 | 53.80 | 2.50 | 0.00 | 0.00 | -6.30 | 8.80 | 0.00 | 133.23 | 1568.57 | 121.8 |
| CNE | 2017 | 200.175 | 53.80 | 51.50 | 2.30 | 16.90 | 0.00 | -14.60 | 0.00 | 0.00 | 66.78 | 2407.56 | -79.4 |
| CNE | 2016 | 194.500 | 51.50 | 49.50 | 2.00 | 0.00 | 0.00 | 0.00 | 2.00 | 0.00 | -24.88 | 1992.22 | 89.0 |
| CNE | 2015 | 164.550 | 49.50 | 56.10 | -6.60 | 0.00 | 0.00 | 0.00 | 2.30 | -8.90 | -26.69 | 1563.50 | -125.9 |
| CNE | 2014 | 181.800 | 56.10 | 30.10 | 26.10 | 0.00 | 0.00 | 0.00 | 26.10 | 0.00 | -45.66 | 1935.10 | -7.7 |
| CNE | 2013 | 273.192 | 30.10 | 16.00 | 14.10 | 0.00 | 0.00 | 0.00 | 30.00 | -15.90 | 18.05 | 2177.20 | -183.2 |
| CNE | 2012 | 295.278 | 16.00 | 0.00 | 0.00 | 0.00 | 16.00 | 0.00 | 0.00 | 0.00 | 33.22 | 2662.38 | -147.2 |
| CNE | 2011 | 417.833 | 0.00 | 225.00 | -225.00 | 0.00 | 0.00 | 0.00 | 0.00 | -225.00 | 1457.76 | 4742.68 | -638.7 |
| ENQ | 2018 | 29.011 | 245.00 | 210.00 | 35.00 | 0.00 | 55.00 | -19.00 | -2.00 | 0.00 | 464.76 | 4437.24 | 701.9 |
| ENQ | 2017 | 28.260 | 210.00 | 215.00 | -5.00 | 0.00 | 14.00 | -12.00 | -7.00 | 0.00 | 124.36 | 3726.13 | 225.2 |
| ENQ | 2016 | 22.495 | 215.00 | 203.00 | 12.00 | 0.00 | 14.00 | -13.00 | 11.00 | 0.00 | 210.34 | 3182.03 | 359.3 |
| ENQ | 2015 | 27.700 | 203.00 | 220.00 | 17.00 | 0.00 | 2.00 | -12.00 | 0.00 | 0.00 | 28.76 | 2558.65 | 284.9 |
| ENQ | 2014 | 87.253 | 220.00 | 194.76 | 25.24 | 7.25 | 22.75 | -10.08 | 5.33 | 0.00 | 411.44 | 2626.74 | 387.7 |
| ENQ | 2013 | 102.048 | 194.76 | 128.52 | 66.24 | 67.04 | 5.76 | -8.49 | 2.43 | 0.00 | 326.16 | 2143.61 | 449.0 |
| ENQ | 2012 | 91.498 | 128.52 | 115.21 | 13.31 | 20.40 | 0.00 | -8.21 | 10.23 | -9.11 | 351.52 | 1565.54 | 486.7 |
| ENQ | 2011 | 93.944 | 115.21 | 88.51 | 26.70 | 29.69 | 0.82 | -8.36 | 2.43 | 0.00 | 412.00 | 1253.91 | 407.6 |
| PMO | 2018 | 100.308 | 193.68 | 301.84 | -108.16 | 0.00 | 0.00 | -29.55 | -68.74 | -9.87 | 369.36 | 4400.39 | 752.2 |
| PMO | 2017 | 65.250 | 301.80 | 353.30 | -51.50 | 0.00 | 0.00 | -27.40 | -12.40 | -11.70 | 121.58 | 4538.23 | 524.3 |
| PMO | 2016 | 60.771 | 353.30 | 331.90 | 21.40 | 0.00 | 37.80 | -26.10 | 9.70 | 0.00 | 198.74 | 4911.33 | 608.4 |
| PMO | 2015 | 118.858 | 331.90 | 243.30 | 88.60 | 0.00 | 0.00 | -21.10 | 140.40 | -30.60 | 529.70 | 3593.71 | 435.3 |
| PMO | 2014 | 293.467 | 243.30 | 259.40 | -16.10 | 0.60 | 0.00 | -23.10 | 15.80 | -9.40 | 636.38 | 3903.55 | 636.6 |
| PMO | 2013 | 352.408 | 259.40 | 291.90 | -32.50 | 0.10 | 1.40 | -21.20 | -12.80 | 0.00 | 560.23 | 3509.57 | 552.1 |
| PMO | 2012 | 369.875 | 291.90 | 284.80 | 7.10 | 5.10 | 11.10 | -21.10 | 12.00 | 0.00 | 573.24 | 2977.18 | 509.4 |
| PMO | 2011 | 427.317 | 284.80 | 260.80 | 24.00 | 13.40 | 31.20 | -14.70 | 5.60 | 0.00 | 295.86 | 2502.67 | 228.9 |
| RDS'b | 2018 | 2523.958 | 5437.00 | 5262.00 | 175.00 | 388.00 | 3.00 | -658.00 | 517.00 | -76.00 | 34875.00 | 308106.00 | 48198.0 |
| RDS'b | 2017 | 2231.542 | 5262.00 | 6258.00 | -996.00 | 451.00 | 666.00 | -667.00 | 612.00 | -2058.00 | 23666.00 | 298993.00 | 35790.0 |
| RDS'b | 2016 | 1966.542 | 6258.00 | 5303.00 | 955.00 | 151.00 | 1207.00 | -675.00 | 284.00 | -12.00 | 12762.00 | 332161.00 | 27397.0 |
| RDS'b | 2015 | 2315.625 | 5303.00 | 6130.00 | -827.00 | 52.00 | 2.00 | -553.00 | -252.00 | -76.00 | 16439.00 | 227457.00 | 19483.0 |
| RDS'b | 2014 | 2256.667 | 6130.00 | 6621.00 | -491.00 | 77.00 | 0.00 | -544.00 | 62.00 | -8.00 | 23784.00 | 225379.00 | 35027.0 |
| RDS'b | 2013 | 2208.167 | 6621.00 | 6196.00 | 425.00 | 594.00 | 48.00 | -564.00 | 351.00 | -4.00 | 20287.00 | 213692.00 | 34259.0 |
| RDS'b | 2012 | 2234.458 | 6196.00 | 6048.00 | 148.00 | 101.00 | 82.00 | -598.00 | 629.00 | -66.00 | 19890.00 | 213934.00 | 41009.0 |
| RDS'b | 2011 | 1963.792 | 6048.00 | 6146.00 | -98.00 | 360.00 | 0.00 | -611.00 | 190.00 | -37.00 | 16433.00 | 214818.00 | 45210.0 |
| SIA | 2018 | 92.125 | 23.00 | 28.10 | -5.10 | 0.00 | 0.00 | -2.70 | -2.40 | 0.00 | 78.14 | 641.07 | 107.1 |
| SIA | 2017 | 125.521 | 28.10 | 33.30 | -5.20 | 0.00 | 0.00 | -3.00 | -2.20 | 0.00 | 64.93 | 524.92 | -53.7 |
| SIA | 2016 | 146.833 | 33.30 | 37.30 | -6.00 | 0.00 | 0.00 | -3.60 | -0.40 | 0.00 | 69.38 | 905.09 | 70.7 |
| SIA | 2015 | 180.988 | 37.30 | 40.80 | -3.50 | 0.00 | 0.00 | -4.30 | 0.80 | 0.00 | 96.39 | 800.04 | 68.6 |
| SIA | 2014 | 375.994 | 40.80 | 130.10 | -89.30 | 0.00 | 0.00 | -5.00 | -84.30 | 0.00 | 244.98 | 822.10 | 168.9 |
| SIA | 2013 | 340.038 | 130.10 | 128.50 | 1.60 | 7.60 | 0.00 | -6.00 | 0.00 | 0.00 | 286.97 | 822.62 | 284.1 |
| SIA | 2012 | 274.990 | 128.50 | 130.30 | -1.80 | 0.00 | 3.60 | -5.40 | 0.00 | 0.00 | 302.86 | 884.10 | 303.6 |
| SIA | 2011 | 298.506 | 130.30 | 132.60 | -2.30 | 0.00 | 0.00 | -2.30 | 0.00 | 0.00 | 89.11 | 822.29 | 113.5 |
| TLW | 2018 | 217.896 | 279.50 | 290.50 | -11.00 | 2.50 | 0.00 | -29.70 | 18.00 | -1.80 | 812.70 | 8334.95 | 915.6 |
| TLW | 2017 | 191.608 | 290.50 | 303.70 | -13.20 | 13.20 | 0.00 | -31.90 | 5.50 | 0.00 | 627.13 | 8155.97 | 918.9 |
| TLW | 2016 | 202.534 | 303.70 | 321.80 | -18.10 | 0.00 | 0.00 | -24.50 | 6.40 | 0.00 | 344.63 | 8754.82 | 40.5 |
| TLW | 2015 | 240.403 | 321.80 | 345.30 | -23.50 | 0.90 | 0.00 | -26.80 | 4.10 | -1.70 | 793.40 | 7686.65 | 475.9 |
| TLW | 2014 | 592.110 | 345.30 | 382.40 | -37.10 | 0.00 | 8.20 | -27.50 | -17.40 | -2.30 | 886.55 | 7324.89 | 4176.0 |
| TLW | 2013 | 877.675 | 382.40 | 388.00 | -5.60 | 26.30 | 0.60 | -30.60 | 11.70 | -13.60 | 1177.14 | 6948.38 | 1112.7 |
| TLW | 2012 | 1198.174 | 388.00 | 297.60 | 90.40 | 112.40 | 0.20 | -29.00 | 6.80 | 0.00 | 1067.43 | 5771.64 | 1506.8 |
| TLW | 2011 | 1138.134 | 297.60 | 294.40 | 3.20 | 0.00 | 22.30 | -28.50 | 9.40 | 0.00 | 1169.04 | 6842.61 | 1171.1 |
| AMER | 2018 | 15.127 | 25.60 | 20.70 | 4.90 | 8.00 | 0.00 | -3.12 | 0.00 | 0.00 | 28.49 | 217.59 | 17.4 |

| | | | | | | | | | | | |
|---|---|---|---|---|---|---|---|---|---|---|---|
| AMER | 2017 | 20.833 | 20.70 | 24.47 | -3.77 | 0.00 | 0.00 | -1.76 | -2.01 | 0.00 | 13.30 | 195.38 | 13.6 |
| AMER | 2016 | 27.341 | 24.47 | 23.70 | 0.77 | 0.00 | 0.00 | -1.13 | 1.90 | 0.00 | 8.08 | 196.97 | -1.0 |
| AMER | 2015 | 30.458 | 23.70 | 24.50 | -0.80 | 0.00 | 0.00 | -1.62 | 2.42 | 0.00 | 5.96 | 157.02 | 0.0 |
| AMER | 2014 | 55.750 | 24.50 | 32.80 | -8.30 | 0.00 | 0.00 | -2.28 | -6.02 | 0.00 | 12.55 | 172.01 | 56.9 |
| AMER | 2013 | 48.125 | 32.80 | 29.90 | 2.90 | 4.60 | 0.00 | -1.70 | 0.00 | 0.00 | 15.99 | 147.89 | 51.8 |
| AMER | 2012 | 33.458 | 29.20 | 7.70 | 22.20 | 23.20 | 0.00 | -1.00 | 0.00 | 0.00 | 20.38 | 94.08 | 12.3 |
| AMER | 2011 | 20.000 | 7.70 | 6.75 | 0.95 | 0.00 | 0.00 | 0.00 | 0.95 | 0.00 | -13.75 | 44.56 | 3.3 |
| CASP | 2018 | 8.965 | 28.80 | 28.80 | 0.00 | 0.00 | 0.00 | -0.69 | 0.69 | 0.00 | 3.99 | 51.35 | -2.0 |
| CASP | 2017 | 9.635 | 28.80 | 29.30 | -0.50 | 0.00 | 0.00 | -0.81 | 0.31 | 0.00 | 5.96 | 60.43 | -2.5 |
| CASP | 2016 | 9.969 | 29.30 | 1.20 | 28.10 | 28.60 | 0.00 | -0.50 | 0.00 | 0.00 | 0.02 | 69.08 | -2.9 |
| CASP | 2015 | 11.208 | 1.20 | 0.00 | 1.20 | 1.53 | 0.00 | -0.33 | 0.00 | 0.00 | -5.36 | 57.58 | -0.8 |
| CASP | 2014 | 9.767 | 14.70 | 14.70 | 0.00 | 0.00 | 0.00 | -0.08 | 0.08 | 0.00 | 1.26 | 95.20 | 13.2 |
| CASP | 2013 | 4.198 | 14.70 | 7.50 | 7.20 | 8.26 | 0.00 | -1.06 | 0.00 | 0.00 | -1.14 | 105.02 | -3.8 |
| CASP | 2012 | 3.333 | 7.50 | 7.50 | 0.00 | 0.00 | 0.00 | -0.56 | 0.56 | 0.00 | -1.23 | 104.65 | -4.4 |
| CASP | 2011 | 4.292 | 7.50 | 7.50 | 0.00 | 0.00 | 0.00 | -0.09 | 0.09 | 0.00 | -2.71 | 87.08 | 33.4 |
| ELA | 2018 | 105.258 | 91.50 | 85.80 | 5.70 | 12.20 | 0.00 | -6.50 | 0.00 | 0.00 | 40.57 | 371.24 | 82.3 |
| ELA | 2017 | 57.708 | 85.80 | 85.60 | 0.20 | 3.39 | 0.00 | -3.19 | 0.00 | 0.00 | 7.56 | 198.64 | 2.9 |
| ELA | 2016 | 31.489 | 85.60 | 85.60 | 0.00 | 1.06 | 0.00 | -1.06 | 0.00 | 0.00 | -6.15 | 177.94 | -21.1 |
| ELA | 2015 | 50.625 | 85.60 | 81.40 | 4.20 | 5.26 | 0.00 | -1.06 | 0.00 | 0.00 | -4.66 | 143.64 | -1.4 |
| ELA | 2014 | 101.042 | 81.40 | 81.80 | -0.40 | 0.00 | 0.00 | -0.40 | 0.00 | 0.00 | -4.39 | 141.12 | -8.6 |
| ELA | 2013 | 117.083 | 81.80 | 71.80 | 10.00 | 10.00 | 0.00 | 0.00 | 0.00 | 0.00 | -8.61 | 110.58 | -14.9 |
| EME | 2018 | 10.327 | 0.00 | 0.00 | 0.00 | 0.00 | 0.00 | 0.00 | 0.00 | 0.00 | -2.14 | 8.76 | -0.6 |
| EME | 2017 | 9.748 | 0.00 | 0.00 | 0.00 | 0.00 | 0.00 | 0.00 | 0.00 | 0.00 | -1.56 | 5.91 | -13.6 |
| EME | 2016 | 1.070 | 0.00 | 14.11 | -14.11 | 0.00 | 0.00 | 0.00 | 0.00 | -14.11 | 1.03 | 29.14 | 0.4 |
| EME | 2015 | 0.784 | 14.11 | 12.64 | 1.47 | 1.84 | 0.00 | -0.37 | 0.00 | 0.00 | 6.69 | 43.53 | 8.0 |
| EME | 2014 | 1.794 | 12.64 | 6.52 | 6.12 | 6.46 | 0.00 | -0.34 | 0.00 | 0.00 | 6.26 | 27.38 | 6.4 |
| EME | 2013 | 0.813 | 6.52 | 3.20 | 3.32 | 3.66 | 0.00 | -0.37 | 0.00 | 0.00 | 3.41 | 23.39 | 3.7 |
| EME | 2012 | 0.903 | 3.20 | 1.32 | 1.88 | 0.00 | 0.00 | 0.00 | 0.00 | 0.00 | 2.03 | 15.25 | 1.4 |
| EME | 2011 | 0.755 | 1.32 | 0.00 | 1.32 | 0.00 | 0.00 | 0.00 | 0.00 | 0.00 | -0.18 | 13.22 | -0.4 |
| HUR | 2018 | 43.853 | 37.00 | 37.00 | 0.00 | 0.00 | 0.00 | 0.00 | 0.00 | 0.00 | -3.48 | 776.01 | -9.9 |
| HUR | 2017 | 39.000 | 37.00 | 37.00 | 0.00 | 0.00 | 0.00 | 0.00 | 0.00 | 0.00 | -10.32 | 693.72 | -13.7 |
| HUR | 2016 | 24.548 | 37.00 | 37.00 | 0.00 | 0.00 | 0.00 | 0.00 | 0.00 | 0.00 | -4.79 | 333.77 | -6.5 |
| HUR | 2015 | 14.517 | 37.00 | 0.00 | 0.00 | 0.00 | 0.00 | 0.00 | 0.00 | 0.00 | -2.55 | 187.00 | -5.3 |
| IGAS | 2018 | 99.558 | 14.56 | 13.64 | 0.92 | 2.43 | 0.00 | -0.82 | -0.69 | 0.00 | 9.07 | 244.85 | -14.3 |
| IGAS | 2017 | 100.010 | 13.64 | 13.37 | 0.27 | 0.00 | 0.00 | -0.89 | 1.16 | 0.00 | 1.09 | 254.93 | 5.9 |
| IGAS | 2016 | 279.467 | 13.37 | 13.33 | 0.04 | 0.00 | 0.00 | -0.85 | 0.89 | 0.00 | -15.80 | 248.19 | -9.5 |
| IGAS | 2015 | 476.054 | 13.33 | 12.63 | 0.70 | 0.15 | 0.00 | -0.71 | 1.26 | 0.00 | -6.86 | 255.24 | 20.4 |
| IGAS | 2014 | 2143.693 | 13.25 | 16.36 | -3.11 | 0.00 | 0.30 | -0.99 | -2.42 | 0.00 | 15.63 | 286.75 | 29.0 |
| IGAS | 2013 | 1993.631 | 16.36 | 11.33 | 5.03 | 0.00 | 5.37 | -0.90 | 0.56 | 0.00 | 0.96 | 353.71 | 33.2 |
| IGAS | 2012 | 1378.354 | 11.33 | 11.57 | -0.24 | 0.00 | 0.00 | -0.24 | 0.00 | 0.00 | -1.97 | 199.82 | -12.8 |
| IGAS | 2011 | 1215.510 | 11.57 | 0.00 | 11.57 | 0.00 | 10.70 | -0.04 | 0.91 | 0.00 | -0.63 | 199.83 | -13.8 |
| PMG | 2018 | 55.558 | 46.30 | 27.70 | 18.60 | 0.00 | 22.17 | -3.57 | 0.00 | 0.00 | 2.97 | 78.89 | -4.8 |
| PMG | 2017 | 42.563 | 27.70 | 27.90 | -0.20 | 0.00 | 0.00 | -2.47 | 2.67 | 0.00 | -0.46 | 82.16 | -3.2 |
| PMG | 2016 | 54.907 | 27.90 | 23.50 | 4.40 | 0.00 | 6.35 | -1.95 | 0.00 | 0.00 | -10.58 | 87.48 | -3.1 |
| PMG | 2015 | 95.125 | 23.50 | 26.10 | -2.60 | 0.00 | 0.00 | -1.83 | -0.78 | 0.00 | -1.76 | 105.56 | -13.4 |
| PMG | 2014 | 214.885 | 26.10 | 21.50 | 4.60 | 0.00 | 5.27 | -0.67 | 0.00 | 0.00 | 7.01 | 127.44 | 11.1 |
| PMG | 2013 | 201.428 | 21.50 | 25.10 | -3.60 | 0.00 | 0.00 | -0.41 | -3.19 | 0.00 | -4.69 | 53.38 | -4.7 |
| PMG | 2012 | 220.624 | 25.10 | 0.00 | 25.10 | 0.00 | 25.21 | -0.11 | 0.00 | 0.00 | -2.33 | 22.91 | -4.6 |
| PMG | 2011 | 251.251 | 0.00 | 0.00 | 0.00 | 0.00 | 0.00 | 0.00 | 0.00 | 0.00 | -1.09 | 12.33 | -3.4 |
| RPT | 2018 | 32.967 | 50.00 | 13.50 | 36.50 | 0.00 | 0.00 | -1.24 | 37.74 | 0.00 | 28.49 | 96.66 | 59.5 |
| RPT | 2017 | 4.918 | 13.50 | 13.50 | 0.00 | 0.00 | 0.00 | -0.82 | 0.82 | 0.00 | 13.30 | 50.31 | 12.9 |
| RPT | 2016 | 3.204 | 13.50 | 13.50 | 0.00 | 0.00 | 0.00 | -0.62 | 0.62 | 0.00 | 8.08 | 52.25 | 9.3 |
| RPT | 2015 | 4.290 | 13.50 | 11.70 | 1.80 | 0.00 | 2.27 | -0.47 | 0.00 | 0.00 | 5.96 | 47.21 | 5.0 |
| RPT | 2014 | 8.365 | 11.70 | 11.70 | 0.00 | 0.00 | 0.00 | -0.50 | 0.50 | 0.00 | 12.55 | 60.74 | 9.7 |
| RPT | 2013 | 20.208 | 11.70 | 31.60 | -19.90 | 0.00 | 0.00 | -0.52 | -19.38 | 0.00 | 15.99 | 93.87 | 8.0 |
| RPT | 2012 | 23.021 | 31.60 | 151.30 | -119.70 | 0.00 | 0.00 | -0.56 | -119.14 | 0.00 | 20.38 | 184.41 | 13.2 |
| RPT | 2011 | 40.135 | 151.30 | 151.30 | 0.00 | 0.00 | 0.00 | 0.00 | 0.00 | 0.00 | -13.75 | 182.71 | -1.9 |
| SDX | 2018 | 52.908 | 12.85 | 13.30 | -0.44 | 2.45 | 0.00 | -2.03 | -0.86 | 0.00 | 36.24 | 138.11 | 7.8 |
| SDX | 2017 | 51.113 | 13.30 | 6.04 | 7.25 | 4.80 | 3.02 | -1.67 | 1.10 | 0.00 | 21.62 | 141.06 | 3.4 |
| SQZ | 2018 | 85.967 | 68.80 | 3.10 | 65.70 | 6.20 | 57.50 | -1.00 | 3.00 | 0.00 | 1.29 | 476.85 | 3.4 |
| SQZ | 2017 | 35.125 | 3.10 | 3.80 | -0.70 | 0.00 | 0.00 | -0.70 | 0.00 | 0.00 | 17.67 | 86.26 | 11.7 |
| SQZ | 2016 | 12.106 | 3.80 | 4.20 | -0.40 | 0.00 | 0.00 | -0.60 | 0.20 | 0.00 | 6.11 | 77.85 | 3.8 |

| | | | | | | | | | | | | |
|---|---|---|---|---|---|---|---|---|---|---|---|---|
| SQZ | 2015 | 5.998 | 4.20 | 0.00 | 4.20 | 0.00 | 4.80 | -0.60 | 0.00 | 0.00 | 6.66 | 60.53 | 0.6 |
| SQZ | 2014 | 10.989 | 0.00 | 5.20 | -5.20 | 0.00 | 0.00 | 0.00 | -5.20 | 0.00 | -1.94 | 45.11 | -11.6 |
| SQZ | 2013 | 21.754 | 5.20 | 5.50 | -0.30 | 0.00 | 0.00 | -0.10 | -0.20 | 0.00 | -1.66 | 64.30 | -3.1 |
| SQZ | 2012 | 27.957 | 5.50 | 6.80 | -1.30 | 0.00 | 0.00 | -0.40 | -0.90 | 0.00 | 2.27 | 62.69 | 1.5 |
| SQZ | 2011 | 25.992 | 6.80 | 8.30 | -1.50 | 0.00 | 0.00 | -0.80 | -0.70 | 0.00 | 5.54 | 79.35 | 6.7 |
| SEY | 2018 | 12.517 | 0.00 | 0.00 | 0.00 | 0.00 | 0.00 | 0.00 | 0.00 | 0.00 | -34.68 | 67.80 | -1.9 |
| SEY | 2017 | 15.115 | 0.00 | 0.07 | -0.07 | 0.00 | 0.00 | -0.07 | 0.00 | 0.00 | 4.15 | 103.65 | -8.4 |
| SEY | 2016 | 15.438 | 0.07 | 0.17 | -0.10 | 0.00 | 0.00 | -0.10 | 0.00 | 0.00 | -9.93 | 115.41 | -8.1 |
| SEY | 2015 | 16.271 | 0.17 | 0.29 | -0.12 | 0.00 | 0.00 | -0.11 | -0.01 | 0.00 | -4.92 | 125.63 | -14.9 |
| SEY | 2014 | 32.031 | 0.29 | 0.56 | -0.27 | 0.00 | 0.00 | -0.16 | -0.11 | 0.00 | -1.35 | 142.16 | -1.1 |
| SEY | 2013 | 37.417 | 0.56 | 0.48 | 0.08 | 0.00 | 0.00 | -0.19 | 0.28 | 0.00 | 6.27 | 151.06 | 7.6 |
| SEY | 2012 | 38.865 | 0.48 | 0.66 | -0.19 | 0.00 | 0.00 | -0.19 | 0.00 | 0.00 | 7.80 | 141.28 | 5.7 |
| SEY | 2011 | 47.646 | 0.66 | 0.42 | 0.24 | 0.00 | 0.00 | -0.23 | 0.47 | 0.00 | 5.57 | 150.94 | 16.3 |
| SOU | 2018 | 37.351 | 0.00 | 0.00 | 0.00 | 0.00 | 0.00 | 0.00 | 0.00 | 0.00 | 0.99 | 209.64 | -8.7 |
| SOU | 2017 | 62.875 | 0.00 | 0.00 | 0.00 | 0.00 | 0.00 | 0.00 | 0.00 | 0.00 | -11.75 | 202.15 | 11.2 |
| SOU | 2016 | 43.948 | 0.00 | 0.66 | -0.66 | 0.00 | 0.00 | -0.66 | 0.00 | 0.00 | -2.97 | 90.11 | -6.0 |
| SOU | 2015 | 16.218 | 0.66 | 0.18 | 0.48 | 0.50 | 0.00 | -0.02 | 0.00 | 0.00 | -6.02 | 34.29 | -3.0 |
| SOU | 2014 | 9.361 | 0.24 | 0.24 | 0.00 | 0.06 | 0.00 | -0.06 | 0.00 | 0.00 | -4.34 | 38.46 | -2.3 |
| SOU | 2013 | 8.339 | 0.24 | 0.22 | 0.02 | 0.04 | 0.00 | -0.01 | -0.01 | 0.00 | -2.77 | 23.68 | -6.3 |
| SOU | 2012 | 9.753 | 0.22 | 1.18 | -0.96 | 0.22 | 0.00 | 0.00 | 0.00 | -1.18 | -4.32 | 25.12 | -3.2 |
| SOU | 2011 | 28.730 | 1.18 | 0.00 | 1.18 | 1.18 | 0.00 | 0.00 | 0.00 | 0.00 | -2.97 | 36.04 | -4.2 |
| TRIN | 2018 | 17.279 | 24.49 | 23.21 | 1.28 | 0.00 | 0.00 | -1.04 | 2.32 | 0.00 | 2.57 | 90.79 | -1.6 |
| TRIN | 2017 | 13.594 | 23.21 | 21.25 | 1.96 | 0.00 | 0.00 | -0.92 | 2.88 | 0.00 | -3.92 | 77.53 | 20.4 |
| TRIN | 2016 | 3.722 | 21.25 | 21.80 | -0.55 | 0.00 | 0.00 | -0.92 | 1.47 | 0.00 | 4.83 | 87.91 | 5.7 |
| TRIN | 2015 | 31.990 | 21.80 | 25.30 | -3.50 | 0.00 | 0.00 | -1.10 | -2.40 | 0.00 | -2.76 | 87.24 | -9.1 |
| TRIN | 2014 | 114.323 | 25.30 | 47.70 | -22.40 | 0.00 | 0.00 | -1.30 | -21.10 | 0.00 | 6.75 | 132.86 | -53.8 |
| TRIN | 2013 | 132.771 | 47.70 | 35.60 | 12.10 | 0.00 | 0.00 | -1.40 | 13.50 | 0.00 | 24.18 | 226.89 | 43.1 |
| ZOL | 2018 | 18.375 | 210.40 | 210.40 | 0.00 | 0.00 | 0.00 | -2.11 | 2.11 | 0.00 | 9.21 | 95.30 | 3.9 |
| ZOL | 2017 | 19.958 | 210.40 | 210.40 | 0.00 | 0.00 | 0.00 | -2.62 | 2.62 | 0.00 | 8.87 | 99.43 | -10.2 |
| ZOL | 2016 | 23.250 | 210.40 | 210.40 | 0.00 | 0.00 | 0.00 | -3.00 | 3.00 | 0.00 | 8.97 | 125.77 | 11.3 |
| ZOL | 2015 | 46.250 | 210.40 | 210.40 | 0.00 | 0.00 | 0.00 | -2.80 | 2.80 | 0.00 | 2.11 | 89.78 | 4.9 |
| ZOL | 2014 | 100.938 | 210.40 | 75.00 | 145.40 | 0.00 | 148.19 | -2.79 | 0.00 | 0.00 | -3.33 | 115.72 | 20.4 |
| ZOL | 2013 | 115.583 | 75.00 | 0.00 | 75.00 | 0.00 | 77.92 | -2.92 | 0.00 | 0.00 | -3.03 | 28.08 | -2.7 |
| CABC | 2018 | 305.667 | 3.58 | 2.12 | 1.46 | 0.62 | 0.71 | -0.26 | 0.39 | 0.00 | 1.94 | 70.86 | 966.8 |
| CABC | 2017 | 444.533 | 2.12 | 1.42 | 0.70 | 0.00 | 0.17 | -0.11 | 0.64 | 0.00 | 1.75 | 60.80 | 168.3 |
| CABC | 2016 | 305.420 | 1.42 | 0.28 | 1.14 | 0.00 | 1.10 | -0.10 | 0.60 | 0.46 | 1.71 | 48.48 | 84.2 |
| CABC | 2015 | 538.121 | 0.28 | 0.29 | -0.01 | 0.00 | 0.00 | -0.01 | 0.00 | 0.00 | -3.54 | 32.88 | 162.4 |
| CABC | 2014 | 2195.077 | 0.29 | 0.06 | 0.23 | 0.32 | 0.00 | -0.04 | 0.00 | -0.05 | 2.55 | 50.06 | 557.0 |
| CABC | 2013 | 3822.945 | 0.06 | 57.50 | -57.44 | 0.00 | 0.00 | -0.01 | -57.43 | 0.00 | 2.13 | 81.34 | -6.0 |
| CABC | 2012 | 6958.175 | 57.50 | 75.55 | -18.05 | 0.00 | 0.00 | -0.04 | -18.01 | 0.00 | -2.86 | 123.64 | -3.7 |
| CABC | 2011 | 9517.886 | 75.55 | 89.45 | -13.90 | 0.00 | 0.00 | -0.58 | -12.03 | -1.29 | -2.34 | 120.88 | -2.2 |
| EGRE | 2018 | 7.729 | 0.77 | 0.76 | 0.02 | 0.05 | 0.00 | -0.03 | 0.00 | 0.00 | -1.67 | 34.13 | -10.1 |
| EGRE | 2017 | 8.760 | 0.76 | 0.70 | 0.06 | 0.09 | 0.00 | -0.04 | 0.00 | 0.00 | -0.46 | 36.11 | -57.3 |
| EGRE | 2016 | 11.138 | 0.70 | 0.38 | 0.32 | 0.39 | 0.00 | -0.07 | 0.00 | 0.00 | -0.19 | 32.32 | -35.4 |
| EGRE | 2015 | 9.883 | 0.38 | 0.41 | -0.03 | 0.03 | 0.00 | -0.06 | 0.00 | 0.00 | -1.44 | 34.82 | -6.6 |
| EGRE | 2014 | 21.062 | 0.41 | 0.51 | -0.10 | 0.00 | 0.00 | -0.09 | -0.01 | 0.00 | 0.49 | 42.06 | 11.0 |
| EGRE | 2013 | 9.132 | 0.51 | 0.88 | -0.37 | 0.00 | 0.00 | -0.08 | -0.29 | 0.00 | -0.51 | 20.48 | -1.9 |
| EGRE | 2012 | 8.663 | 0.82 | 2.13 | -1.31 | 0.00 | 0.00 | -0.05 | -1.26 | 0.00 | 0.74 | 20.48 | -1.7 |
| EGRE | 2011 | 15.704 | 2.13 | 2.13 | 0.00 | 0.05 | 0.00 | -0.05 | 0.00 | 0.00 | 0.12 | 23.84 | -2.7 |
| PPTC | 2018 | 9.663 | 25.43 | 27.15 | 1.71 | 0.00 | 2.78 | 0.83 | 3.48 | 0.18 | 9.43 | 165.86 | -4.5 |
| PPTC | 2017 | 7.578 | 27.15 | 20.27 | 6.88 | 0.00 | 5.17 | -0.41 | 2.12 | 0.00 | -5.32 | 141.46 | -0.4 |
| PPTC | 2016 | 7.798 | 20.27 | 18.32 | 1.95 | 2.13 | 0.00 | -0.19 | 0.00 | 0.00 | -0.19 | 144.44 | -0.6 |
| PPTC | 2015 | 10.762 | 18.32 | 14.42 | 3.89 | 4.09 | 0.00 | -0.18 | -0.02 | 0.00 | -2.20 | 113.20 | -2.8 |
| PPTC | 2014 | 30.521 | 14.42 | 6.73 | 7.70 | 0.12 | 7.74 | -0.16 | 0.00 | 0.00 | -1.55 | 132.70 | 4.2 |
| PPTC | 2013 | 23.455 | 6.73 | 6.86 | -0.14 | 0.02 | 0.00 | -0.16 | 0.00 | 0.00 | 3.37 | 67.21 | 11.3 |
| PPTC | 2012 | 32.211 | 6.86 | 7.20 | -0.34 | 0.00 | 0.00 | -0.13 | -0.21 | 0.00 | -2.76 | 61.12 | -10.1 |
| PPTC | 2011 | 36.939 | 7.20 | 1.83 | 5.37 | 0.00 | 5.96 | -0.09 | -0.50 | 0.00 | -3.70 | 42.11 | -17.7 |